\documentclass[a4paper]{article}

\usepackage[a4paper,top=3cm,bottom=2cm,left=3cm,right=3cm,marginparwidth=1.75cm]{geometry}
\usepackage[utf8]{inputenc} 
\usepackage[T1]{fontenc}    
\usepackage{url}            
\usepackage{booktabs}       
\usepackage{amsfonts}       
\usepackage{nicefrac}       
\usepackage[round]{natbib}
\usepackage{amsmath}
\usepackage{amssymb}
\usepackage{graphicx}

\usepackage{filecontents}

\title{Effects of Stochastic Parametrization on Extreme Value Statistics}
\author{Guannan Hu,$^{1,2}$ Tam{\'a}s B{\'o}dai,$^3$ and Valerio Lucarini$^{3,4,2}$ \\
    \small $^1$School of Integrated Climate System Sciences (SICSS), University of Hamburg, Hamburg, Germany\\
    \small $^2$CEN, University of Hamburg, Hamburg, Germany\\
    \small$^3$Department of Mathematics and Statistics, University of Reading, Reading, UK\\
    \small $^4$Centre for the Mathematics of Planet Earth, University of Reading, Reading, UK\\
}

\date{} 

\begin{document}
\maketitle

\begin{abstract}
Extreme geophysical events are of crucial relevance to our daily life: they threaten human lives and cause property damage. To assess the risk and reduce losses, we need to model and probabilistically predict these events. Parametrizations are computational tools used in Earth system models, which are aimed at reproducing the impact of unresolved scales on resolved scales. The performance of parametrizations has usually been examined on typical events rather than on extreme events. In this paper we consider a modified version of the two-level Lorenz'96 model and investigate how two parametrizations of the fast degrees of freedom perform in terms of the representation of extreme events. One parametrization is constructed following Wilks (2005) and is constructed through an empirical fitting procedure; the other parametrization is constructed through the statistical mechanical approach proposed by Wouters and Lucarini (2012, 2013). The two strategies show different advantages and disadvantages. We discover that the agreement between parametrized models and true model is in general worse when looking at extremes rather than at the bulk of the statistics. The results suggest that stochastic parametrizations should be accurately and specifically tested against their performance on extreme events, as usual optimization procedures might neglect them.
\end{abstract}

\section{Introduction}

Weather and climate models are mathematical representations of the physical processes in the Earth system. These physical processes operate on different temporal and spatial scales. For instance, atmospheric convection acts from minutes to hours on a spatial scale of kilometers, while ocean circulation acts from years to multiple thousands of years on a spatial scale of thousands of kilometers. Since weather and climate models have certain temporal and spatial resolutions, the processes that happen on shorter timescales or smaller spatial scales cannot be resolved by the model. However, the processes on different scales interact with each other, and the unresolved processes will influence the resolved processes. Parametrizations are computational tools aimed at reproducing the impact of the unresolved scales on the scales that can be directly resolved. Different parametrization schemes strike different balance between computational efficiency and various aspects of accuracy. While traditionally parametrizations are formulated as empirical deterministic formulas, in recent years the scientific community has advocated the need for using more general parametrization methods comprising also of stochastic components; see Palmer and Williams (2008), Franzke \textit{et al}. (2015) and Berner \textit{et al}. (2017) for a summary of recent developments in this field. The idea of using a stochastic climate model has been proposed by Hasselmann (1976). The goal of this paper is to address, in a simple yet informative case, the problem of how well parametrizations perform, but, instead of looking at the bulk of the statistics, as usually done, we look at extreme value statistics.\\
\indent The understanding, modeling and probabilistic prediction of extremes are of key interest to the financial market, the insurance sector, and also to civil defence services concerning natural catastrophes, e.g. hurricanes, storms, floods, etc. A mature statistical framework, known as extreme value theory (EVT), is widely applied to analyze extremes. The theory was developed in the course of the $20$th century by
Fisher and Tippett (1928), Gnedenko (1943), Balkema and de Haan (1974), and Pickands (1975); an excellent summary of the main results can be found in, e.g. Embrechts, Klüppelberg, and Mikosch (1997), and Coles (2001). Two approaches are often used to analyze extremes: the block maxima (BM) and the peak over threshold (POT) approaches. Extremes are defined in different ways by these two approaches: the first one takes the maximal values in blocks or batches of sample data as extremes, while the second one selects the exceedances above a given high threshold. Under rather general conditions the block maxima are distributed according to the Generalized Extreme Value (GEV) distribution, and the threshold exceedances follow the Generalized Pareto (GP) distribution. That is in the limit of large block sizes and thresholds, respectively, upon some suitable normalisation. Holland \textit{et al}. (2012) applied the BM method to the extremes of a number of chaotic deterministic dynamical systems and found that for sufficiently smooth so-called physical observables, a parameter of the limiting GEV distribution, the shape parameter, is determined by the dimensions of the stable and unstable manifolds of the chaotic attractor. Later Lucarini \textit{et al}. (2014) applied the POT method and suggested by a heuristic argument that the formula of Holland \textit{et al.} (2012) should apply generically. These two papers together with other earlier publications \cite{collet01,fre10} establish a connection between extreme value statistics and the geometrical properties of the attractor. A comprehensive summary of the main results of extreme events of observables of deterministic systems with examples and applications can be found in Lucarini \textit{et al}. (2016). This link has been reexamined by G{\'a}lfi, B{\'o}dai, and Lucarini (2017), in which the authors presented the convergence of shape parameter estimates to the theoretical value in a two-level quasi-geostrophic atmospheric model, or the lack of it, as this convergence could be observed only in the model with a strong forcing. Furthermore, 
B{\'o}dai (2017) argued that the convergence of the shape parameter can be observed typically for high-dimensional systems, and in low-dimensional systems, such as the Lorenz'84 and one-level Lorenz'96 models that he studied, the shape parameter estimates can increase nonmonotonically with the block size, or fluctuate owing to the fractality of the natural measure, in which latter case no extreme value law exists in a strict sense.\\
\indent We use a conceptual model of the atmosphere, the two-level Lorenz'96 model (L96) \cite{L96} and two parametrized versions of this model, which are constructed by resolving the large-scale variables and parametrizing the influence of the unresolved small-scale processes on the evolution of the resolved variables in the two-level L96. We consider two stochastic parametrization schemes for the unresolved processes: an empirical parametrization and another one based on response theory. The first one was proposed by Wilks (2005), using multivariate regression and a simple first-order autoregressive model. This parametrization has been widely applied to the two-level L96 and showed very good performance in reproducing the large-scale variables of the full dynamic model \cite{arn13,Christensen15,Harlim:2017}. The second parametrization scheme was recently introduced by 
Vissio and Lucarini (2018), using the methodology proposed by Wouters and Lucarini (2012, 2013, 2016). This is a scale-adaptive parametrization, such that one can derive a universal expression for the parametrization that explicitly represents a scale parameter to do with the coupling strength between the resolved and unresolved processes. The two parametrizations show comparable skills in reproducing the probability density, the spatial correlation, and the temporal autocorrelation of the large-scale variables of the full model \cite{vis18}, with the Wilks approach providing marginally better results at the expense of the fact of being ad-hoc and not adaptive. In this paper, we evaluate how well the two parametrization schemes perform in terms of reproducing the extreme value statistics of the full model. The Wilks parametrizations is constructed in such a way that typical events are represented as well as possible but there is no a-priori control on how well extreme events are represented. The Wouters-Lucarini parametrization allows, in principle to control the error in the expectation value of any observable, including those relevant for defining the properties of the tails of the distributions. Conversely, the latter approach suffers from being a perturbative one, with the risk of lack of convergence. In an earlier study, Franzke (2012) showed that a reduced order model constructed by systematic stochastic mode reduction strategy has similar extreme value statistics as the full dynamical model for a wide range of time-scale separations. As a contribution to the main goal of this paper, like \cite{fra12,gal17,Bodai:2017}, we also examine the convergence/approximation of the shape parameter estimates of the said different models to the theoretical value of the two-level L96. Note that the shape parameter of a stochastic model is most probably not the same as that of a deterministic model. \\
\indent In the following section, we will introduce the two-level L96 and the two parametrization schemes. In Sec.~\ref{sec:EVT}, we present the two approaches from EVT and the mathematical expression for the shape parameter in dynamical systems. In Sec.~\ref{sec:comparison_EVT}, we apply the two approaches of EVT to the parametrized and full models and compare the estimates of the GEV and GP parameters obtained for the different models. In Sec.~\ref{sec:comparison_direct}, we provide an empirical comparison of the extremes in the different models. In Sec.~\ref{sec:return_time}, we compare the mean return periods of the extremes of the same magnitude in the different models. In Sec.~\ref{sec:coupling_strength}, we repeat the analysis for a different coupling strength in the two-level L96. We close the paper with a summary and conclusions in Sec.~\ref{sec:summary}.  

\section{The two-level Lorenz'96 Model}\label{sec:model}

The two-level L96 was introduced by Lorenz (1995), the governing equations of which describe the dynamics of a lattice with periodic boundary conditions and represent, in a very conceptual way, the main processes occurring in the atmosphere: advection, forcing, and dissipation. In order to apply the Wouters-Lucarini (W-L) parametrization to the two-level L96, Vissio and Lucarini (2018) made two changes to the original model: 1) introduced a forcing term in the equations for the small-scale variables, and 2) restricted the periodic boundary conditions of the small-scale variables within the corresponding large-scale sectors. The first change was aimed at fulfilling a basic requirement for the W-L parametrization: the presence of chaos in the uncoupled dynamics, so that the autocorrelation of the variables decays fast. This requires, in physical terms, an external forcing providing energy to the small-scale variables. The second change was implemented in order for the small-scale variables to represent subgrid-scale phenomena of the sectors they belong to. Additionally, the latter made the implementation of the W-L parametrization easier. The modified governing equations of the two-level L96 are given as: 
\begin{align} 
\frac{dX_{k}}{dt} &= -X_{k-1}(X_{k-2} - X_{k+1}) - X_{k} + F_x -\frac{hc}{b}\sum_{j=1}^{J} Y_{j,k}, \label{eqn:L96b1}\\
\frac{dY_{j,k}}{dt}&= -cbY_{j+1,k}(Y_{j+2,k}-Y_{j-1,k}) - cY_{j,k} + \frac{c}{b}F_y + \frac{hc}{b}X_{k}, \label{eqn:L96b2}
\end{align} 
where the variables $X_k$ and variables $Y_{j,k}$ are defined for $k=1, ..., K$ and $j=1, ..., J$. We consider the variables $X_k$ to be large-scale variables, while the variables $Y_{j,k}$ to be small-scale variables. The boundary conditions are defined as:
\begin{align*}
&X_{-1}=X_{K-1}, \\
&X_{0}=X_{K}, \\
&X_{K+1}=X_{1},\\
&Y_{J+1,k}=Y_{1,k},\\
&Y_{J+2,k}=Y_{2,k},\\
&Y_{0,k}=Y_{J,k}.
\end{align*}
The parameters $F_x$ and $F_y$ represent forcing terms in the equations for the variables $X_k$ and $Y_{j,k}$, respectively. The parameter $h$ is a coupling coefficient, and the parameters $c$ and $b$ can be thought of as time-scale ratio and spatial-scale ratio, respectively. These three parameters determine the coupling strength of the system: $\epsilon=hc/b$. As we keep $b$ and $c$ equal, we have $\epsilon=h$. The variables $X_k$ can represent some atmospheric quantity in $K$ sectors of a latitude cycle, while the variables $Y_{j,k}$ can represent some other atmospheric quantity in smaller $JK$ sectors. There are $J$ smaller sectors in each larger sector. In our computations we set the parameter values of the model as follows: $K=10$, $J=10$, $F_x=10.0$, $F_y=6.0$, $h=1.0$, $b=10.0$, and $c=10.0$. \\
\indent In many practical applications, small-scale processes are too expensive to resolve and their impacts on the evolution of large-scale processes are parametrized by deterministic terms, or stochastic terms, or both of them. In the two-level L96 we can parametrize the effects of the evolution of the $Y_{j,k}$ variables on the evolution of the $X_k$ variables, then the evolution equations of $X_k$ variables are given as
\begin{equation}\label{eqn:reduced}
\frac{dX_{k}}{dt} = -X_{k-1}(X_{k-2} - X_{k+1}) - X_{k} + F_x + U,
\end{equation}
where $U$ denotes the parametrization of the effects of the unresolved processes, which has to represent the model error when only the large-scale variables $X_k$ are resolved in place of the full dynamics. We call Eq.~(\ref{eqn:reduced}) the parametrized model, contrasting with the full model given by Eqs. (\ref{eqn:L96b1}) and (\ref{eqn:L96b2}). We now introduce two parametrization schemes.

\subsection{Wilks parametrization: \\multivariate regression and autoregressive process}

The first parametrization scheme was proposed by Wilks (2005), which used a polynomial equation and a correlated noise term to represent the unresolved processes:
\begin{equation}\label{eqn:polynomial_and_noise}
U(t)=P_k(X_{k}(t))+e_{k}(t),
\end{equation}
where the polynomial equation is a function of the $X_k$ variables:
\begin{equation}
P_k(X_{k}(t))=a_0+a_1X_k(t)+a_2X_{k}^{2}(t)+a_3X_{k}^{3}(t)+a_4X_{k}^{4}(t),
\end{equation}
and the noise term is a simple first-order autoregressive model:
\begin{equation}\label{eqn:noise}
e_{k,i}=\phi e_{k,i-1}+\sigma_e(1-\phi^2)^{1/2}z_{k,i},
\end{equation}
where $e_{k,i} = e_k(t_i)$, $t_i = dt\times i$, $i = 1,\dots, I$, $dt = 0.005$ [MTU], where MTU is the abbreviation of `model time unit', and $z_{k,i}$ are unit variance Gaussian random variables, independent wrt. both $k$ and $i$.  The autoregressive parameter $\phi$ is equal to the autocorrelation of the time series $e_{k}(t_i)$ with a time lag of $dt$, and $\sigma_e$ denotes the standard deviation of the time series. We first estimate the polynomial coefficients by regressing $P_k(X_k(t_i))$ against 
\begin{equation*}
-\frac{hc}{b}\sum_{j=1}^{J} Y_{j,k}(t_i).
\end{equation*}
The estimated coefficients are $a_0= -1.81$, $a_1=-0.1467$, $a_2=0.001357 $, $a_3=-0.001446$, and $a_4=0.0001313$. Next we fit the residual of the polynomial fitting
\begin{equation*}
- \frac{hc}{b}\sum_{j=1}^{J} Y_{j,k}(t_i) - P_k(X_k(t_i);a_0,a_1,a_2,a_3,a_4)
\end{equation*}
with the first-order autoregressive model (Eq.~\ref{eqn:noise}). In our computation, we get $\phi= 0.9997$ and $\sigma_e =0.8965$; we refer to \citet{neu01} for the estimation of parameters of autoregressive models. The Wilks parametrization is an empirical parametrization which is constructed based on the fact that in the two-level L96 the unresolved tendency is strongly and nonlinearly dependent on the value of the resolved variable \citep{wilks05}. The implementation of the two modifications to the original model do not change this characteristic, therefore, the Wilks parametrization is still valid. A weakness of the empirical parametrizations is that their parameters need to be recalculated if the configuration of the full model is changed.

\subsection{Wouters–Lucarini parametrization:\\ averaging, correlations and memory}

The application of the second parametrization scheme to the two-level L96 was first demonstrated by Vissio and Lucarini (2018). The W-L parametrization is based on Ruelle's response theory \citep{Ruelle1997,Ruelle2009}, and it was proposed by Wouters and Lucarini (2012). Later Wouters and Lucarini (2013) showed that the parametrization scheme can also be obtained through the Mori-Zwanzig approach \cite{mor74,zwa60,zwa61}. Note that this point of view, as well as the more sophisticated results by Chekroun, Liu, and Wang (2015), provide the mathematical foundations of the theory of stochastic parametrizations. In the W-L parametrization, the coupling of the variables to be parametrized is considered as a small perturbation to the variables of interest. The coupling is decomposed into three terms: an averaging term, a correlation term, and a memory term. Therefore, the formula of the W-L parametrization is given as:
\begin{equation}
U=D+S_{k}+M_{k}.
\end{equation}
In the above the averaging term $D$ is a constant, accounting for the "averaged influence" of the $Y_{j,k}$ variables on the long-term statistics of the $X_k$ variables, and it is calculated by 
\begin{equation}
D=-\frac{1}{b}\lim_{T \rightarrow \infty}\frac{1}{T}\int_{0}^{T}\sum_{j=1}^{J}\Tilde{Y}_{j,k}(\tau)d\tau,
\end{equation}
where $k=1,...,K$. Due to symmetry, this term has the same value for all $X_k$ variables. The time series of $\Tilde{Y}_{j,k}$ are obtained by integrating the equations given as: 
\begin{equation}\label{eqn:rescaled}
\frac{d\Tilde{Y}_{j,k}}{d\tau}= -\Tilde{Y}_{j+1,k}(\Tilde{Y}_{j+2,k}-\Tilde{Y}_{j-1,k}) - \Tilde{Y}_{j,k} + F_y,
\end{equation}
where the $\Tilde{Y}_{j,k}$ denote the rescaled small-scale variables:
\begin{equation}
\Tilde{Y}_{j,k}=bY_{j,k},
\end{equation}
and time is rescaled as
\begin{equation}
\tau = ct.
\end{equation}
The correlation term $S_k$ is a stochastic term which accounts for the fluctuations of the influence of the $Y_{j,k}$ variables on the long-term statistics of the $X_k$ variables. It is constructed as an additive noise which reproduces the temporal correlation of the fluctuations. The autocovariance of the fluctuations is given as:
\begin{equation}
R_{k}(\tau)=\lim_{T \rightarrow \infty}\frac{1}{T}\int_{0}^{T}p_{k}(\tau_1)p_{k}(\tau_1 + \tau)d\tau_1,
\end{equation}
where 
\begin{equation}
p_{k}(\tau)=-\sum_{j=1}^{J}\frac{\Tilde{Y}_{j,k}(\tau)}{b} - D.
\end{equation}
Following Vissio and Lucarini (2018), we generate the stochastic term $S_{k}$ using a simple autoregressive model. The $M_k$ is a non-Markovian term, accounting for the memory effects, which is important for the parametrization of the small-scale processes of the two-level L96; see the comparison of the first- and second-order parametrizations in Vissio and Lucarini (2018), where the first one contains only the averaging term, while the second one contains also the correlation term and the memory term. $M_{k}$ describes the influence of the past values on the present values of the $X_k$ variables through the coupling of the $Y_{j,k}$ variables, and an explicit expression for $M_k$ was provided by Vissio and Lucarini (2018):
\begin{equation}
M_{k}=-\frac{1}{b}\int_{0}^{\infty}X_{k}(t-\tau_1)H(\tau_1)d\tau_{1},
\end{equation}
where
\begin{equation}
H(\tau_1)=\frac{1}{\Omega}\lim_{\Omega \rightarrow \infty}\int_{0}^{\Omega}\sum_{j=1}^{J}\frac{\partial \Tilde{Y}_{j,k}(\tau_1 + \omega)}{\partial \Tilde{Y}_{j,k}(\omega)}d\omega.
\end{equation}
The W-L parametrization is a scale-adaptive parametrization; in the two-level L96, it can be analytically adapted to the changes in the values of $h$, $b$, and $c$ by simply rescaling the three terms $D$, $S_k$ and $M_k$. However, at the expense of being flexible, it may be somewhat less accurate than the empirical parametrization. A limitation of the W-L parametrization is that it is only valid for weakly coupled systems.  

\subsection{Local and Global Observables}
\begin{figure}[!h]
    \begin{center}
	\includegraphics[width=2.5in]{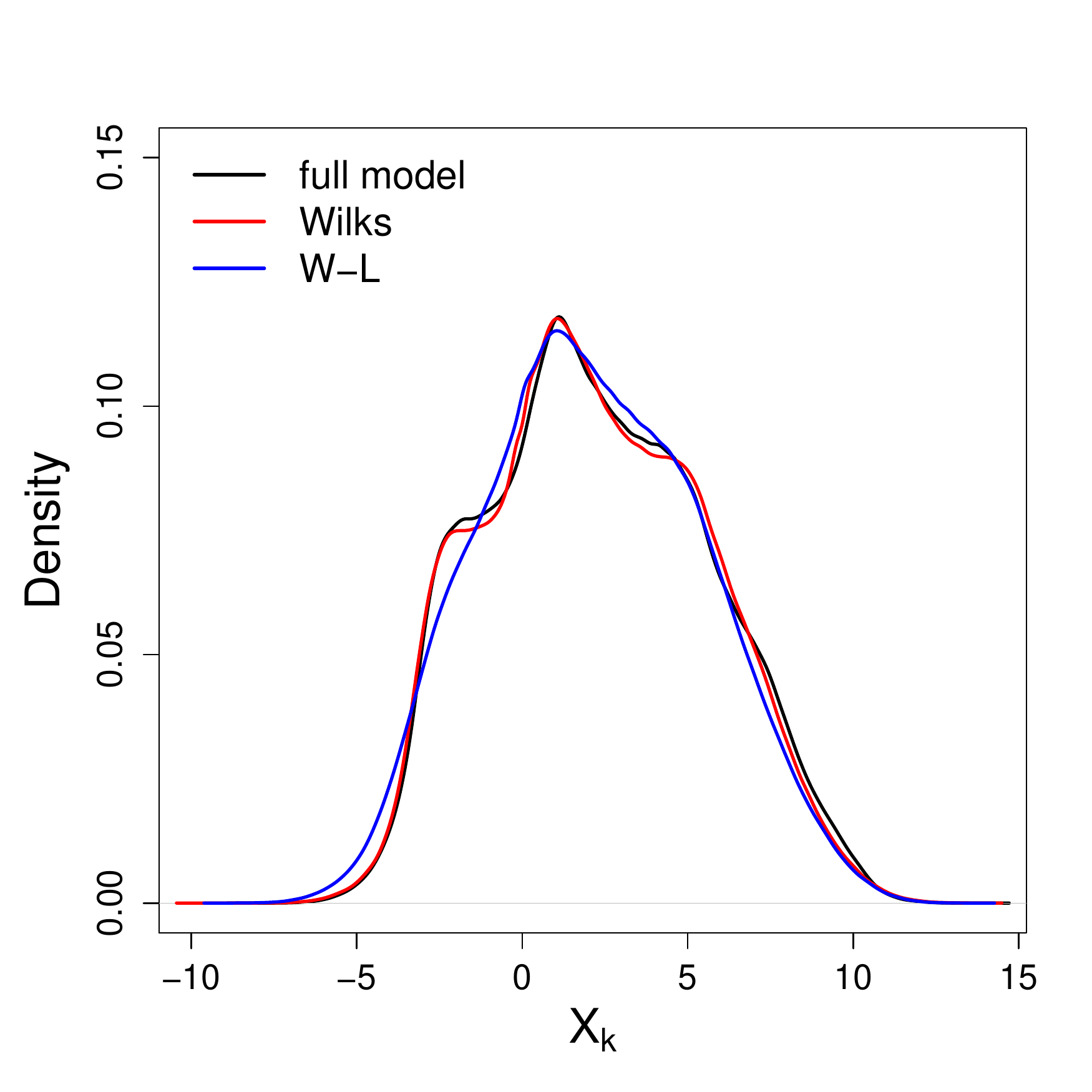}
	\caption{\label{fig:compare_pdf_x}Probability density functions of the observable $A_x$ from the full and parametrized models.}
	\end{center}
\end{figure}

\begin{figure*}[!t]
    \begin{center}
	\includegraphics[width=2.5in]{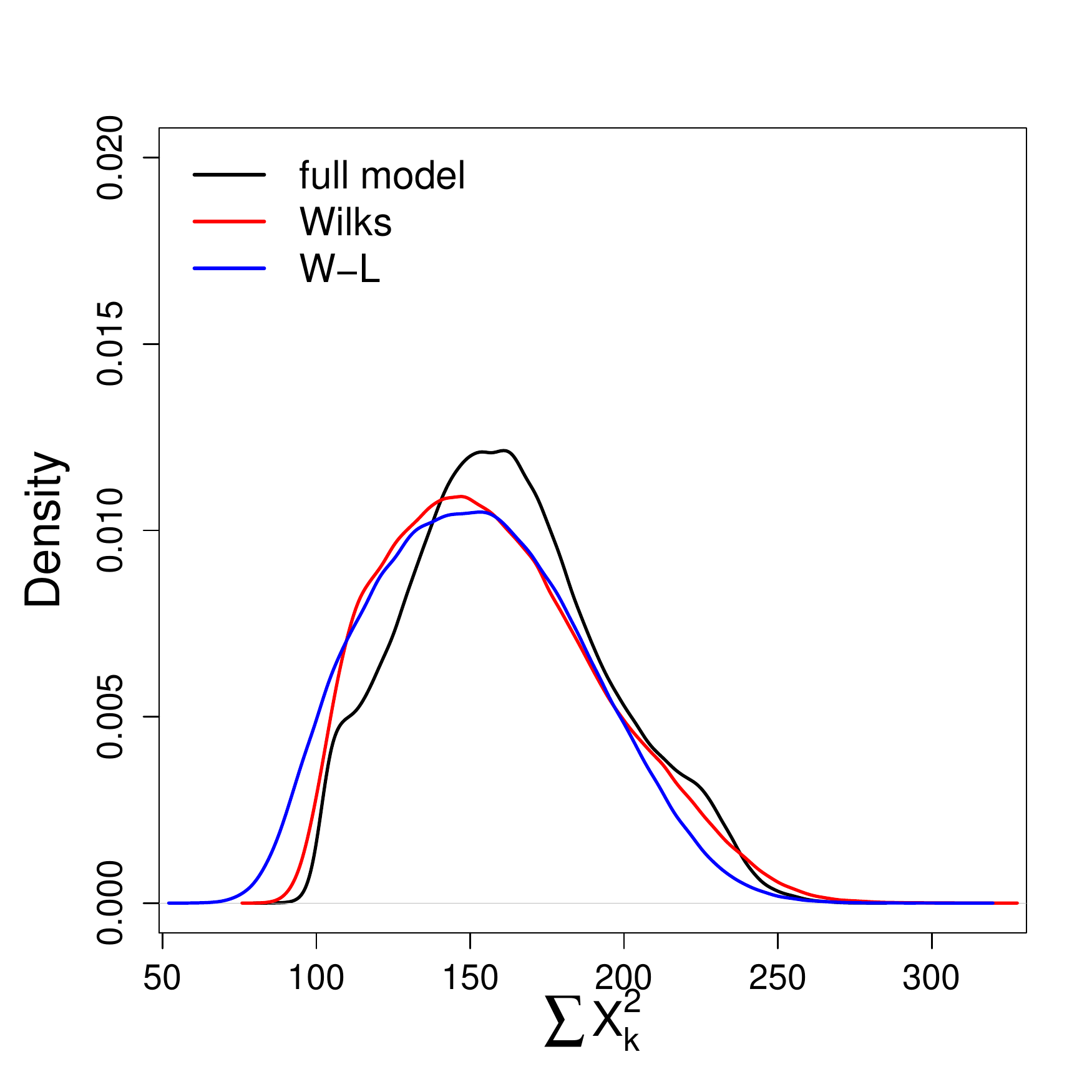}
	\includegraphics[width=2.5in]{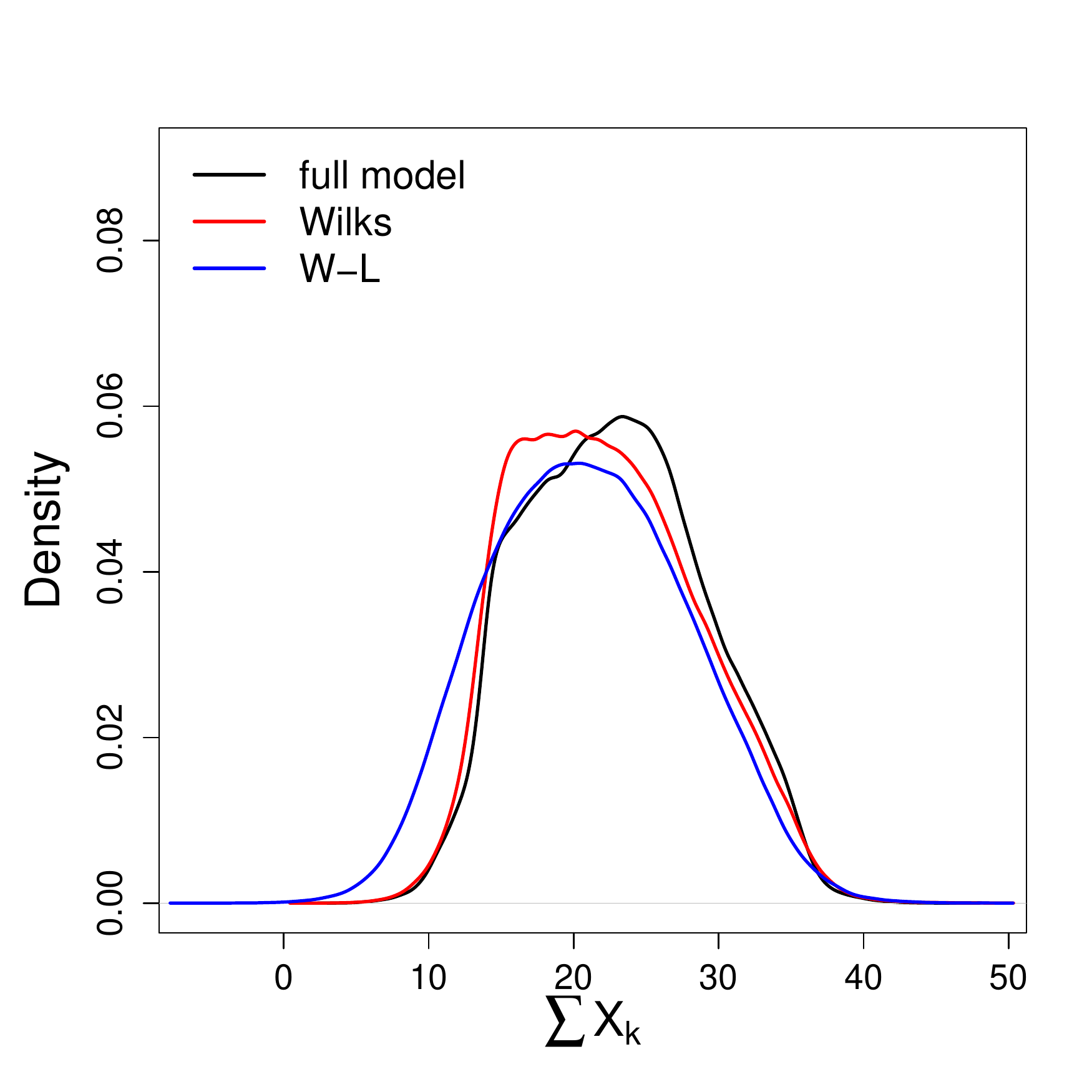}
	\caption{\label{fig:compare_pdf_energy}Probability density functions of the observables $A_E$ (left) and $A_p$ (right) from the full and parametrized models.}
	\end{center}
\end{figure*}

In physics an observable is a variable that can be measured. We consider a local observable and two global observables of the full and parametrized models; we define the local observable as $A_x = X_k$, for $k=1,...,K$ (the $X_k$ variables have the same statistics due the the symmetry of the model), and the global observables as $A_E = \sum_{k=1}^{K}X_{k}^{2}$ and $A_p = \sum_{k=1}^{K}X_{k}$, representing the total energy and the total momentum of the $X_k$ variables, respectively. Fig.~\ref{fig:compare_pdf_x} compares the probability density functions (PDF) of the local observable between the full and parametrized models, and
Fig.~\ref{fig:compare_pdf_energy} compares the PDFs of the global observables between the full and parametrized models. 
The sample data of the local and global observables are produced by integrating the models for $3.2 \times 10^5$ time steps with adaptive stepsizes. We record the three observables at each time step and since the $10$ different $X_k$ variables are statistically equivalent, we get $3.2 \times 10^6$ samples for the local observable. The two parametrized models can capture the statistics of the considered observables of the full model relatively well, but they produce more precise statistics for the local observable than the global observables. Moreover, compared to the W-L parametrized model, the Wilks parametrized model better reproduces the statistics of the local observables of the full model. However, it must be noted that what we look at here is the bulk of the statistics, and these figures do not allow us to compare the extreme value statistics.

\section{Extreme Value Analysis}\label{sec:EVT}
\subsection{Two Approaches}

Two straightforward approaches exist for extreme value analysis: the BM approach and the POT approach; we refer the reader to Embrechts, Klüppelberg, and Mikosch (1997), Coles (2001), and Lucarini \textit{et al.} (2016) for more details. According to the BM approach the data is divided into sufficiently large blocks of equal length and the maximal value of each block is retained. Then the GEV distribution is fitted to these maxima. The cumulative distribution function (CDF) of the GEV distribution is given as:
\begin{equation}
P(X\leqslant x)=F(x;\mu,\sigma,\xi)=\text{exp}\left\{1- \left[1+ \xi\left(\frac{x-\mu}{\sigma}\right)\right]^{-1/\xi}\right\},
\label{eqn:gev}
\end{equation}
which holds for $1+ \xi(x-\mu)/ \sigma > 0$, where $ -\infty< \mu < \infty$ is referred to as location parameter, $\sigma > 0$ is called scale parameter and $ -\infty< \xi < \infty$ denotes the so-called shape parameter. When $\xi =0$, we take the limit of (\ref{eqn:gev}) as $\xi \to 0$, which is 
\begin{equation}
F(x;\mu,\sigma)=\text{exp}\left\{-\text{exp} \left[-\left(\frac{x-\mu}{\sigma}\right)\right]\right\}.
\label{eqn:gev_0}
\end{equation}
The location and scale parameters are scaling constants used to normalize the random variable $X$ of the maximal values of blocks. The shape parameter characterises the tail behavior: when $\xi = 0$, the tail decays exponentially; when $\xi > 0$, a heavy tail occurs, which decays following a power law; and when $\xi <0$, the tail is bounded, i.e., there is an upper limit of the domain of the distribution. The POT approach selects data whose magnitude is above a high threshold. The threshold exceedances are fitted by the GP distribution, which is defined by the CDF:
\begin{equation}
P(\hat{X} \leqslant \hat{x})=F(\hat{x};\hat{\sigma},\hat{\xi}) = 1-\left(1+\hat{\xi} \frac{\hat{x}}{\hat{\sigma}}\right)^{-1/\hat{\xi}},
\label{eqn:gp}
\end{equation}
where $\hat{X} >0$ denotes the exceedances above a threshold $u$, e.g. $\hat{X}=X_k-u$. The scale parameter $\hat{\sigma} > 1$, and $1+\hat{\xi}\hat{x}/\hat{\sigma} > 0$. The shape parameter $\hat{\xi}$ again characterises the tail behaviour like it does in the GEV distributions. As with the GEV distribution, for $\hat{\xi}=0$, the CDF is given as:
\begin{equation}
F(\hat{x};\hat{\sigma}) = 1-\text{exp}\left(-\frac{\hat{x}}{\hat{\sigma}}\right).
\label{eqn:gp2}
\end{equation}
It should be noted that the shape parameter is the same for corresponding limit GEV and GP distributions, i.e., $\xi = \hat{\xi}$ \cite{lucarini2016extremes}. A simple functional relation connects the two distributions; the natural logarithm of Eq.~(\ref{eqn:gev}) plus one equals Eq.~(\ref{eqn:gp}). Under general conditions, while the two approaches define extremes differently, they are fundamentally equivalent. 

\subsection{The Theoretical Value of Shape Parameter}

For general classes of smooth physical observables, Holland \textit{et al.} (2012) and Lucarini \textit{et al.} (2014) provided an explicit expression for the shape parameter:
\begin{equation}\label{eqn:theo1}
    \xi_{theo}=-\frac{1}{\delta},
\end{equation}
where 
\begin{equation}\label{eqn:theo2}
    \delta=d_s+\frac{(d_u+d_n)}{2},
\end{equation}
for continuous-time flows. In the above, $d_u$ is equal to the number of positive Lyapunov exponents, $d_n$ is equal to the number of zero exponents, which is at least one for Axiom A systems \citep{luc14}, and $d_s$ equals the number of stable directions and is given by 
\begin{equation}\label{eqn:theo3}
    d_s=d_{KY}-d_u-d_n.
\end{equation}
In the above, the Kaplan-Yorke dimension is defined as \cite{kap79}
\begin{equation}\label{eqn:theo4}
    d_{KY}=n+\frac{\sum_{i=1}^{n}\lambda_{i}}{\left|{\lambda_{n+1}}\right|},
\end{equation}
where $\lambda_i$ denote the Lyapunov exponents of the system, arranged in a descending order, and $n$ is the number when $\sum_{i=1}^n \lambda_i$ is larger than zero while $\sum_{i=1}^{n+1} \lambda_i$ is smaller than zero.\\
\indent Eqs.~(\ref{eqn:theo1}-\ref{eqn:theo4}) give an estimator for the shape parameter via the estimators of the dimensions, or via the estimators of the Lyapunov exponents, instead of estimating it by fitting a GEV or a GP distribution. Clearly, the value of $\xi_{theo}$ is always negative; this can be seen from the formulae, where we always have $\delta > 0$. A negative shape parameter indicates that the distribution of extremes has an upper bound. This is not surprising as smooth physical observables are considered and the dynamical attractor is compact \cite{luc14}. Finite-size positive estimates are of course possible. We should think that whenever positive shape parameters are and have been encountered, a deterministic dynamical system is still an appropriate mathematical representation of the physical process. When we consider a high-dimensional chaotic system, we have a large Kaplan-Yorker dimension of the attractor and so the value of $\xi_{theo}$ is close to zero. It means that the occurrence of very large extreme events becomes more likely. 

\subsection{Return Period}

The mean or expected return period is the reciprocal of the probability of an extreme value of a given magnitude. The magnitude of the extreme value in this context is commonly referred to as the return level. The fitted GEV and GP distributions provide estimates of mean return periods for different return levels via inverting those distributions. As an extrapolation, one can choose higher return levels than ever observed, however, the corresponding mean return time estimates are in general biased. We can calculate the empirical return periods for the data up to reasonable return levels, which can be compared -- in order to facilitate a goodness-of-fit check -- to the said estimates by fitting a GEV distribution and inverting it. Using the block maxima data, the empirical mean return period for return level $m$ is calculated as:
\begin{equation}
\tau_{BM}(m) = \left(\frac{n_p(m)}{n_b}\right)^{-1},
\end{equation}
where $n_p$ is the number of block maxima which have a value greater than or equal to $m$, and $n_b$ is the number of blocks, or the size of data. We construct mean return time vs return level plots to compare the full and parametrized models yet another way. We create these diagrams using the package "extRemes 2.0" \cite{extRemes} of the software environment R.

\section{Comparison of GEV and GP Parameters}\label{sec:comparison_EVT}

We apply the BM and POT methods to the local and global observables of the full and parametrized models. We consider a range of block sizes $B$, exponentially increasing from the smallest block size considered $B_0 = 1000$ MTU, and a range of exceedance ratios $E$, exponentially decreasing from the highest ratio considered $E_0 = 0.1\%$, for the BM method and POT method, respectively. In order to compare fairly the GP and GEV parameters estimated from the samples of the same size, we match choices of $B$ with choices of $E = 1/B$. The number of the block maxima given by the block size $B_0$, equaling the number of the threshold exceedances given by the exceedance ratio $E_0$, is $4 \times 10^6$ for the local observable ($4 \times 10^5$ for each $X_k$ variable) and $4 \times 10^5$ for the global observables. We take advantage of the package "extRemes 2.0" \cite{extRemes} of the software environment R to estimate the GEV and GP parameters; we use the method of L-moments \cite{hos90} for estimation. For the local observable, we plot the averaged estimates of the parameters over ten $X_k$ variables, and present the estimator variance as the sample standard deviation for the ten estimates given by error bars. For the global observables on the other hand, we show the $95\%$ confidence intervals of each estimate, where the confidence interval is calculated by a parametric bootstrap method \cite{extRemes}. 

\subsection{The GEV Parameters}\label{sec:comparison_EVT_a}

\begin{figure*}
    \begin{center}
	\includegraphics[width=0.30\textwidth]{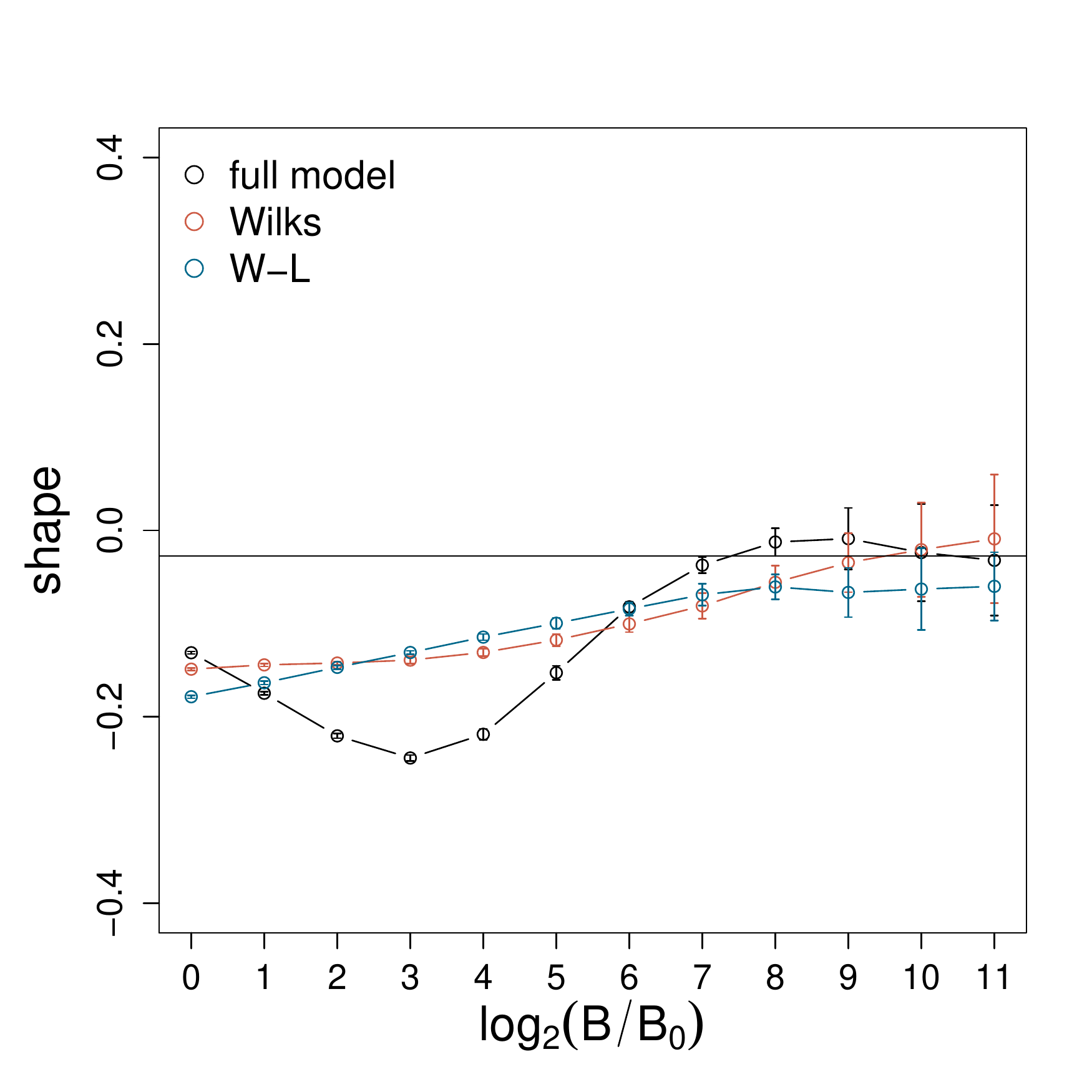}
	\includegraphics[width=0.30\textwidth]{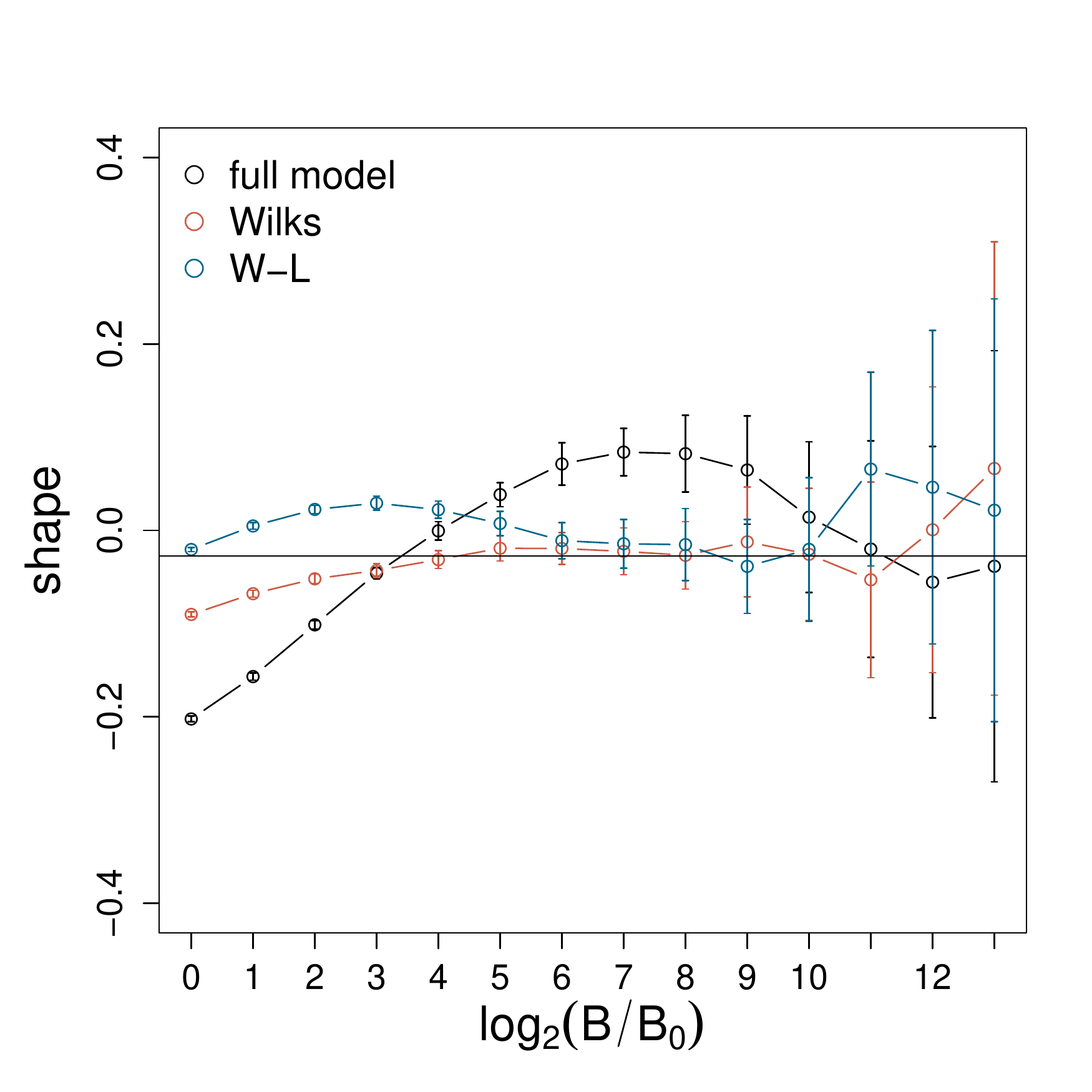}
	\includegraphics[width=0.30\textwidth]{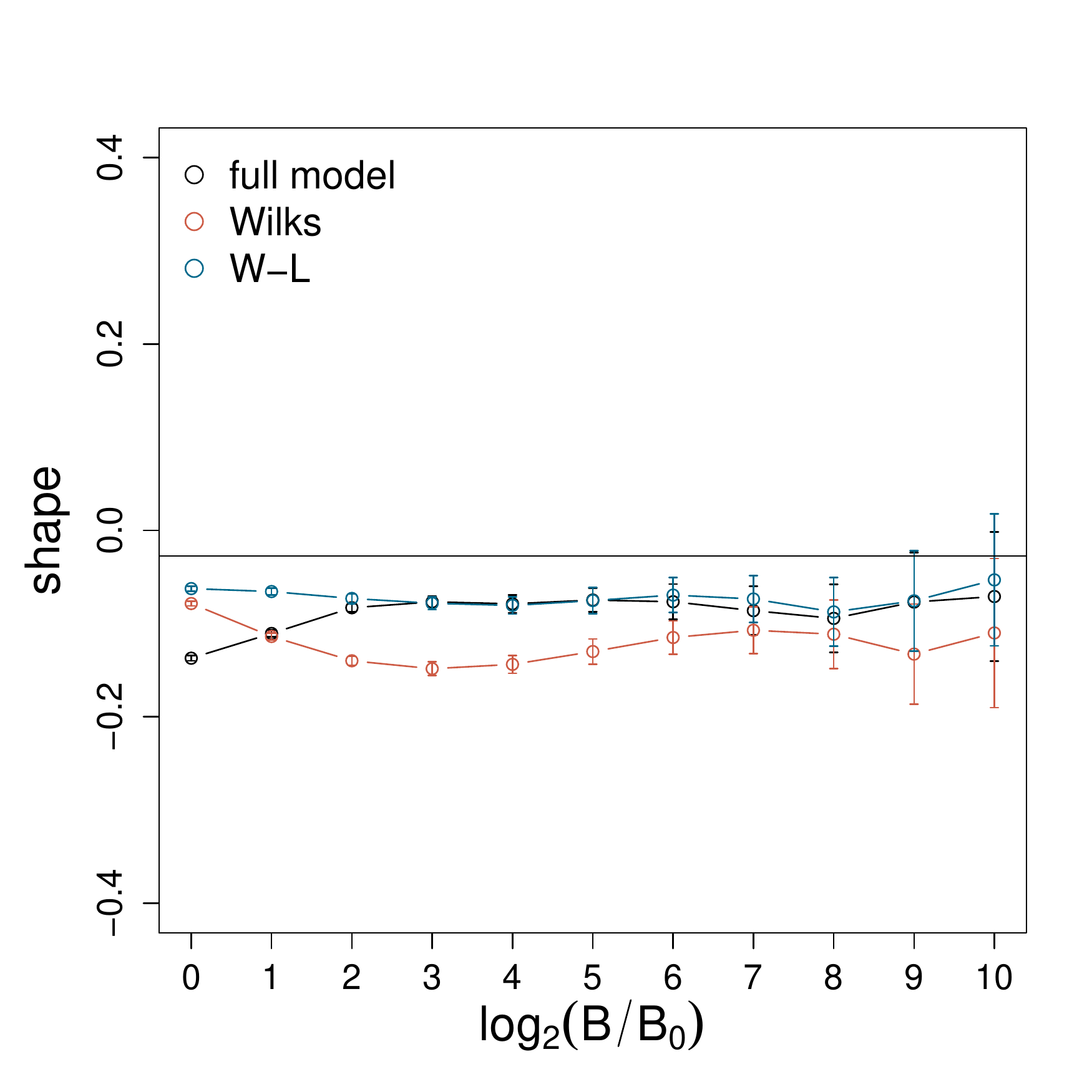}
 	\caption[]{\label{fig:compare_shape}Comparison of the estimates of the GEV shape parameter for the observable $A_x$ (left), $A_E$ (middle), and $A_p$ (right) over a range of block sizes. The horizontal line shows the theoretical value of the shape parameter for the full model.}
 	\end{center}
\end{figure*}
\begin{figure*}
    \begin{center}
	\includegraphics[width=0.30\textwidth]{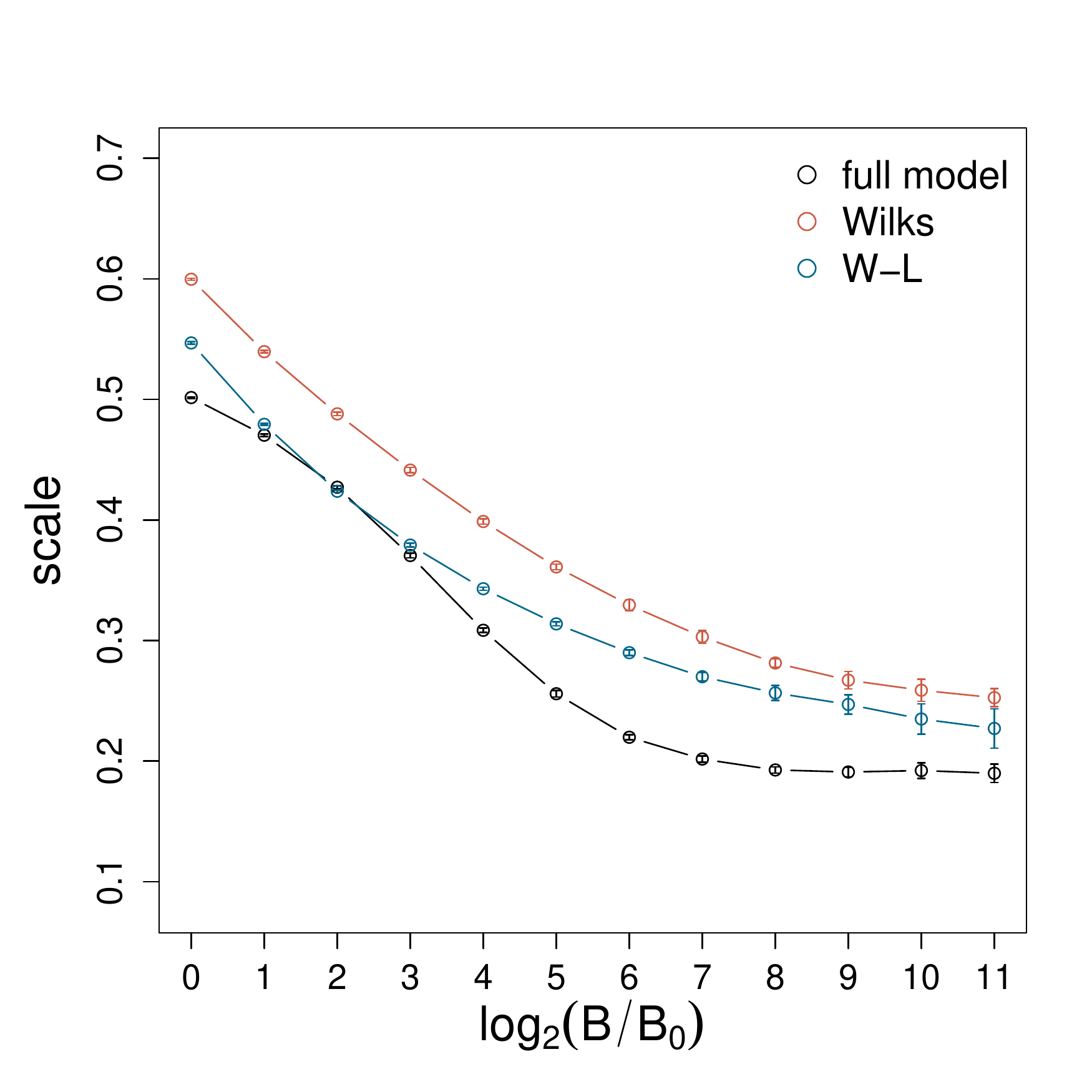}
	\includegraphics[width=0.30\textwidth]{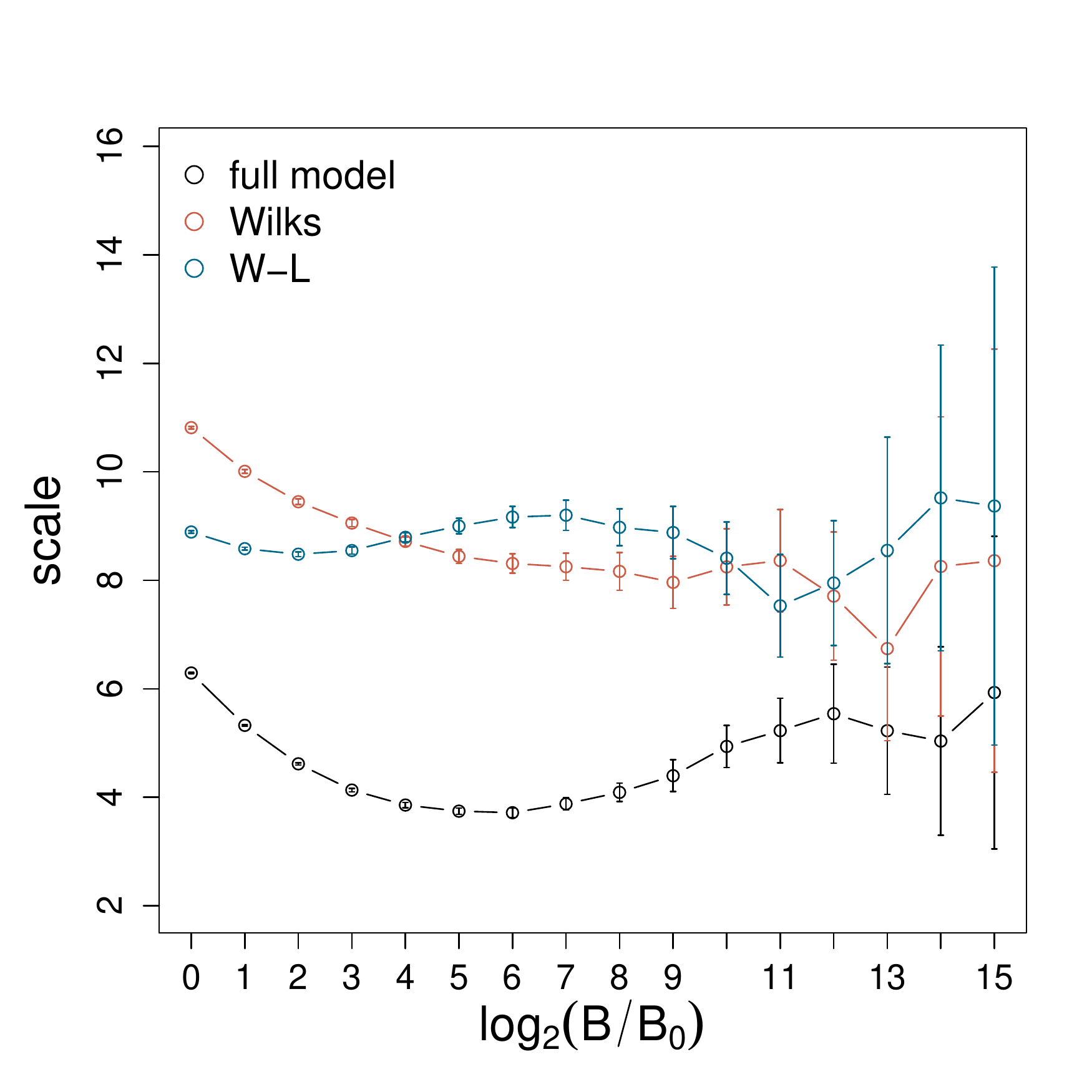}
	\includegraphics[width=0.30\textwidth]{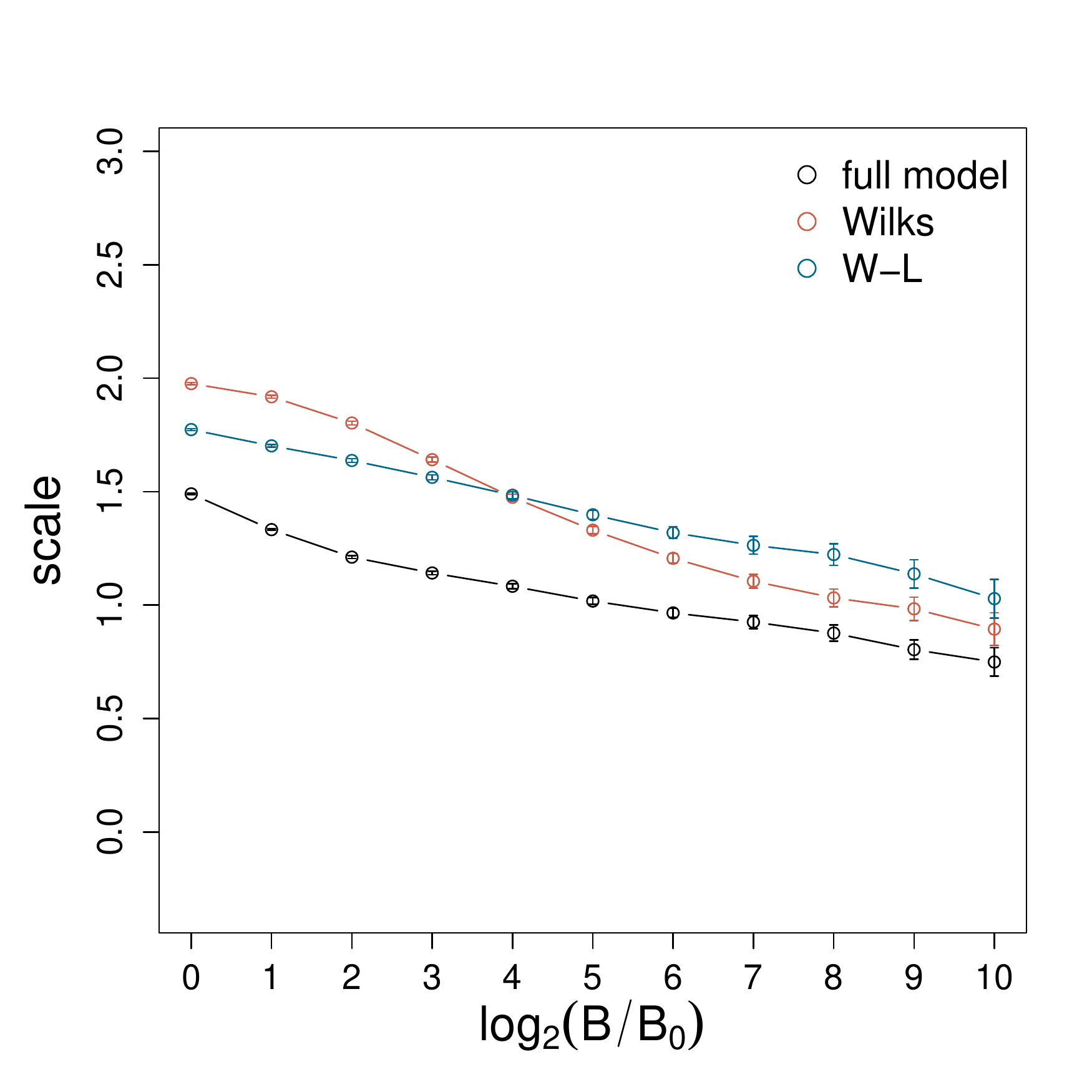}
 	\caption[]{\label{fig:compare_scale}Comparison of the estimates of the GEV scale parameter for the observable $A_x$ (left), $A_E$ (middle), and $A_p$ (right) over a range of block sizes.}
 	 \end{center}
\end{figure*}
\begin{figure*}
    \begin{center}
	\includegraphics[width=0.30\textwidth]{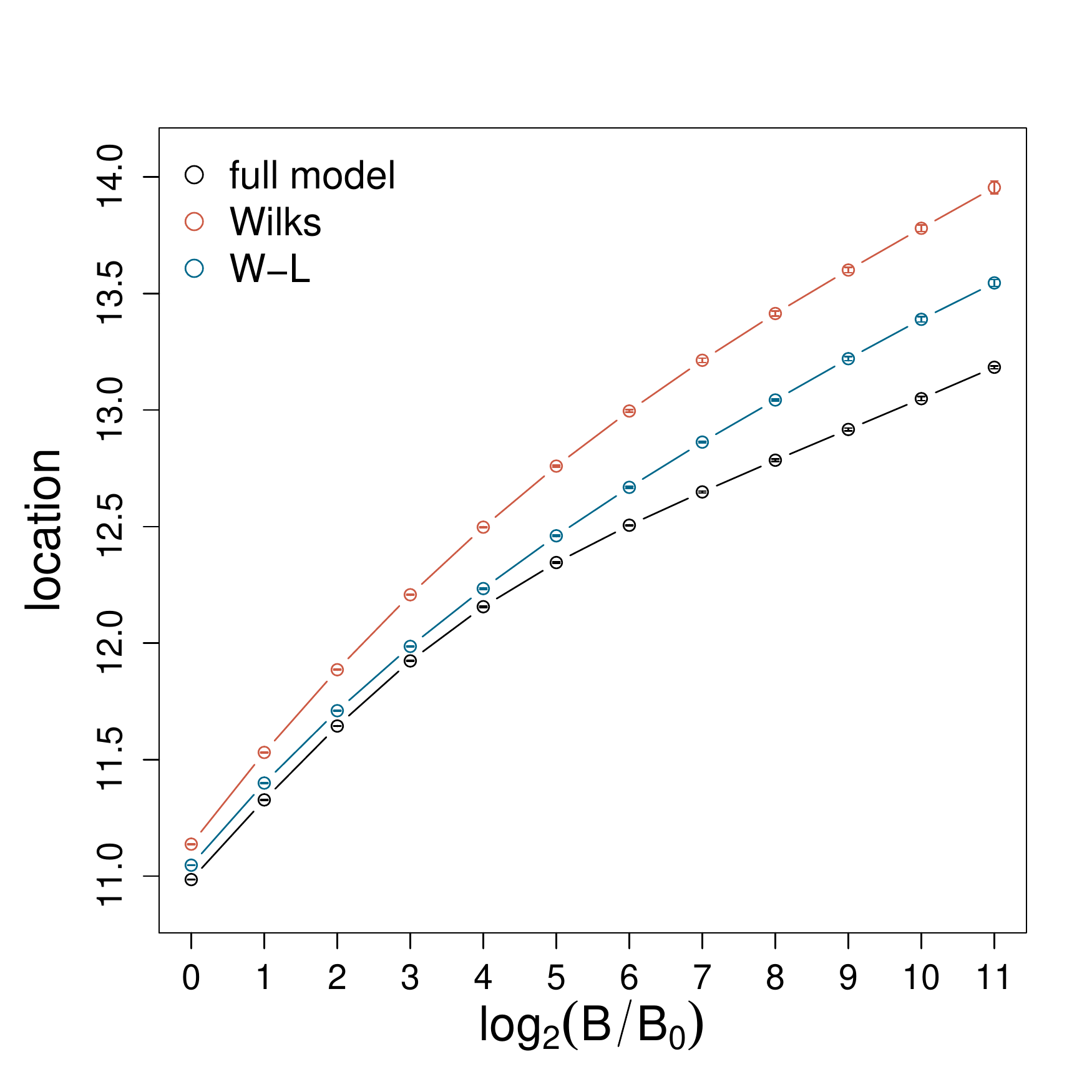}
    \includegraphics[width=0.30\textwidth]{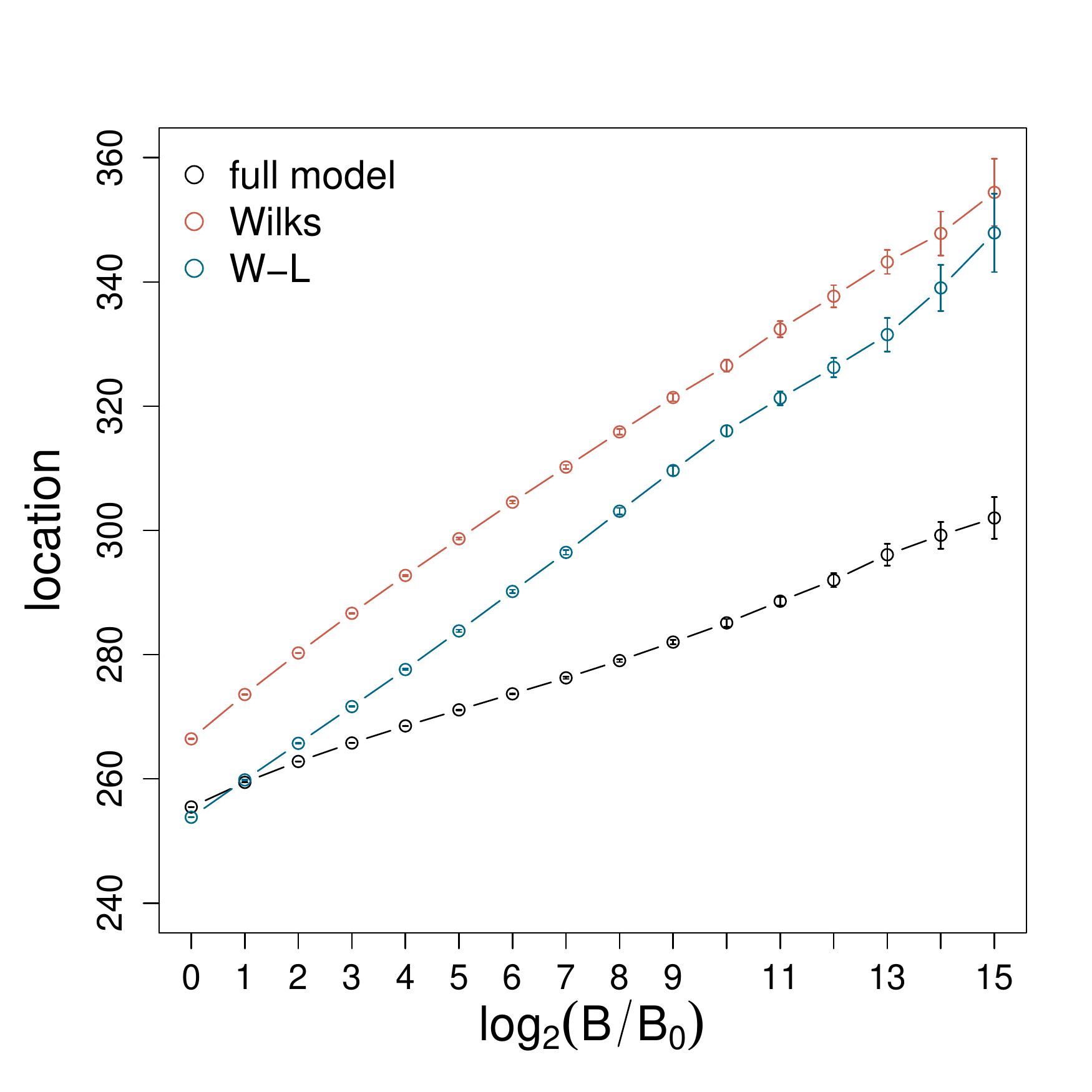}
	\includegraphics[width=0.30\textwidth]{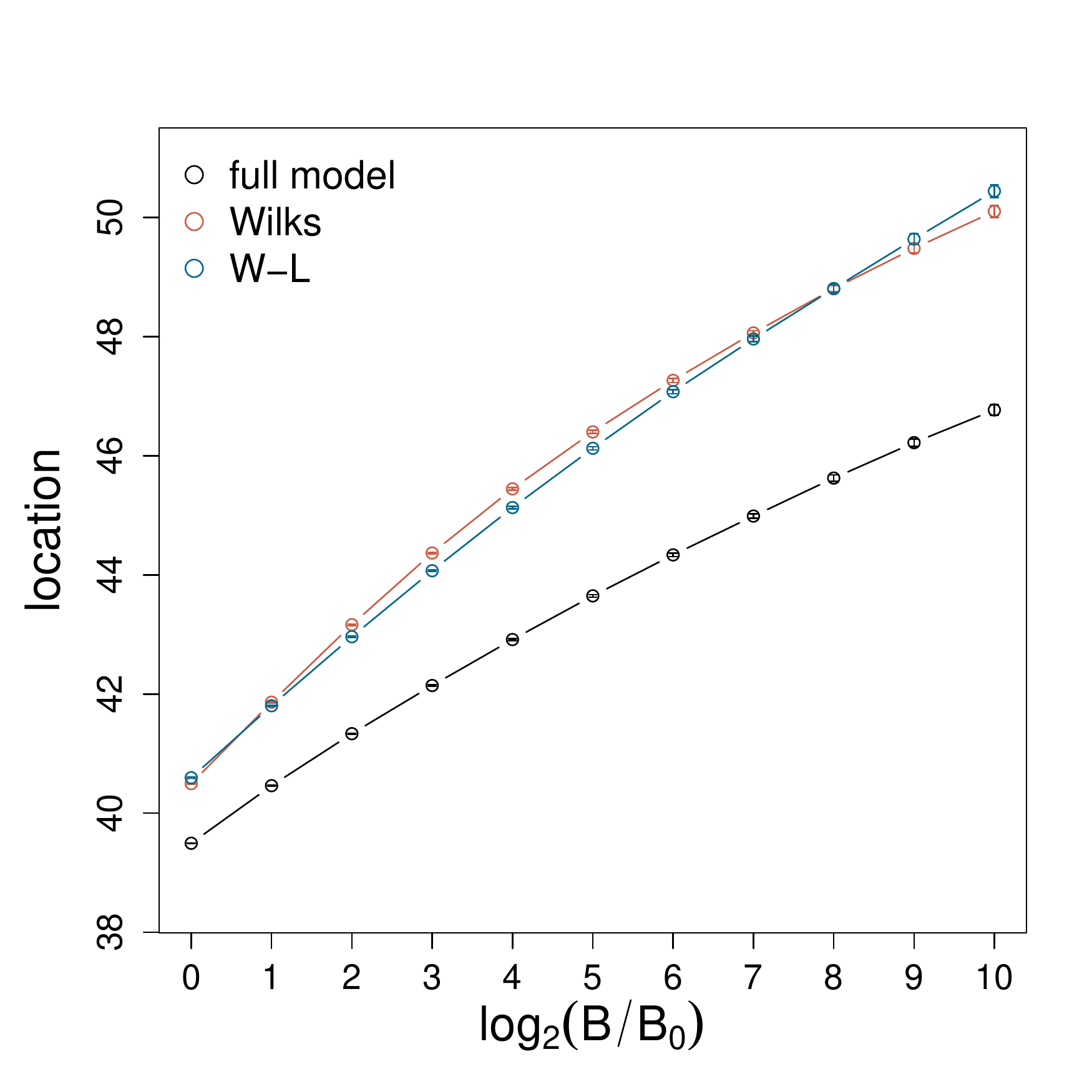}
 	\caption[]{\label{fig:compare_location}Comparison of the estimates of the location parameter for the observable $A_x$ (left), $A_E$ (middle), and $A_p$ (right) over a range of block sizes.}
 	\end{center}
\end{figure*}
\begin{figure*}
    \begin{center}
	\includegraphics[width=0.30\textwidth]{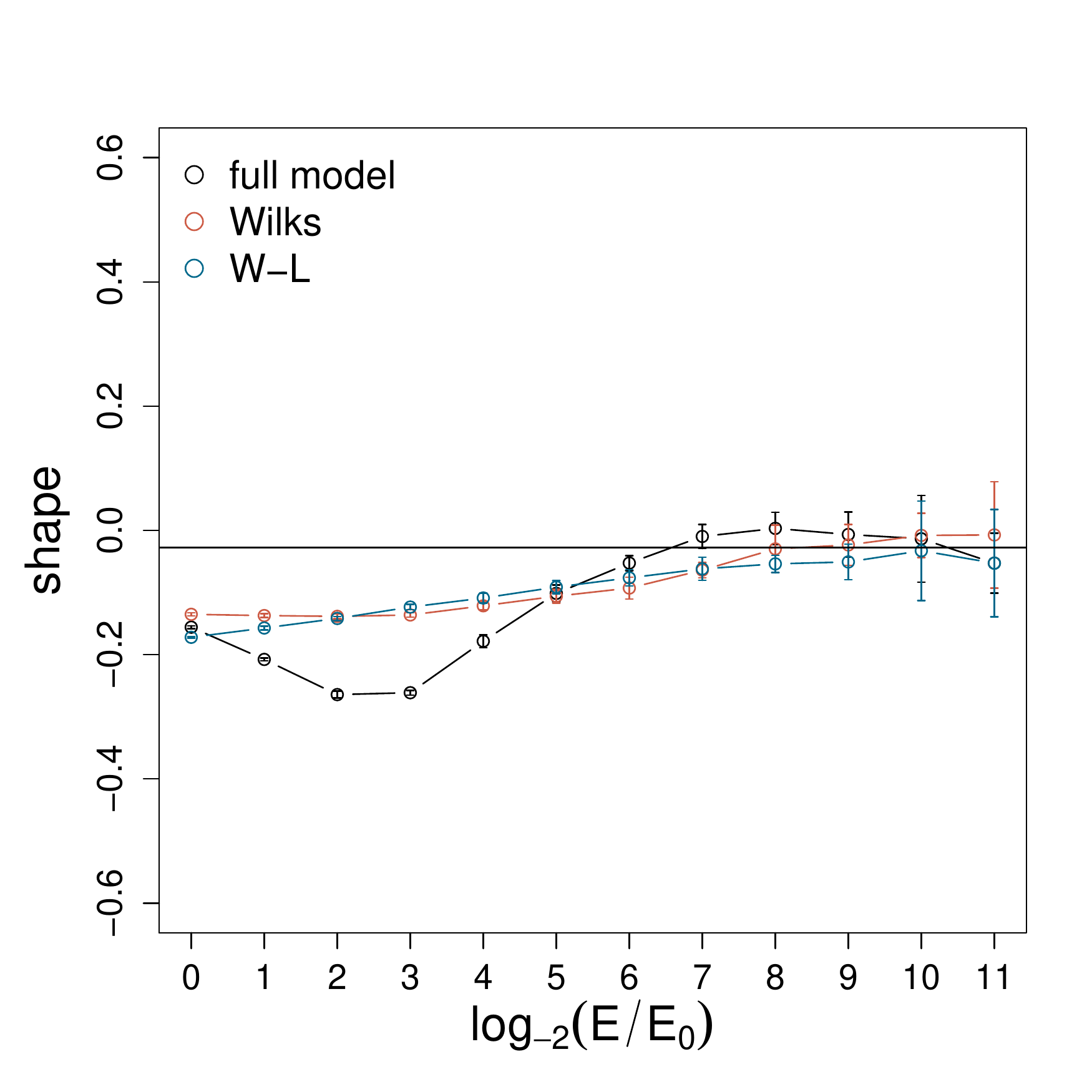}
	\includegraphics[width=0.30\textwidth]{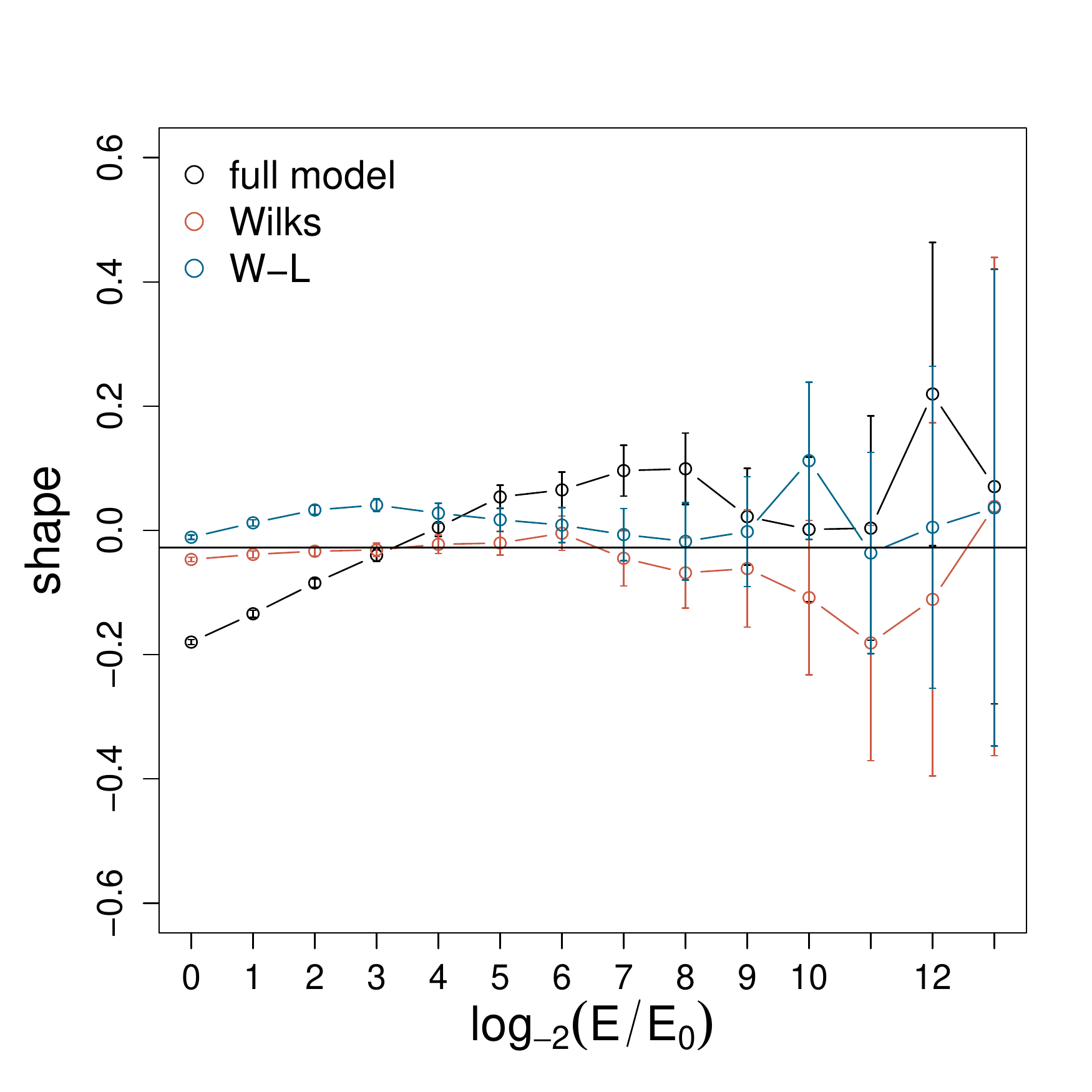}
	\includegraphics[width=0.30\textwidth]{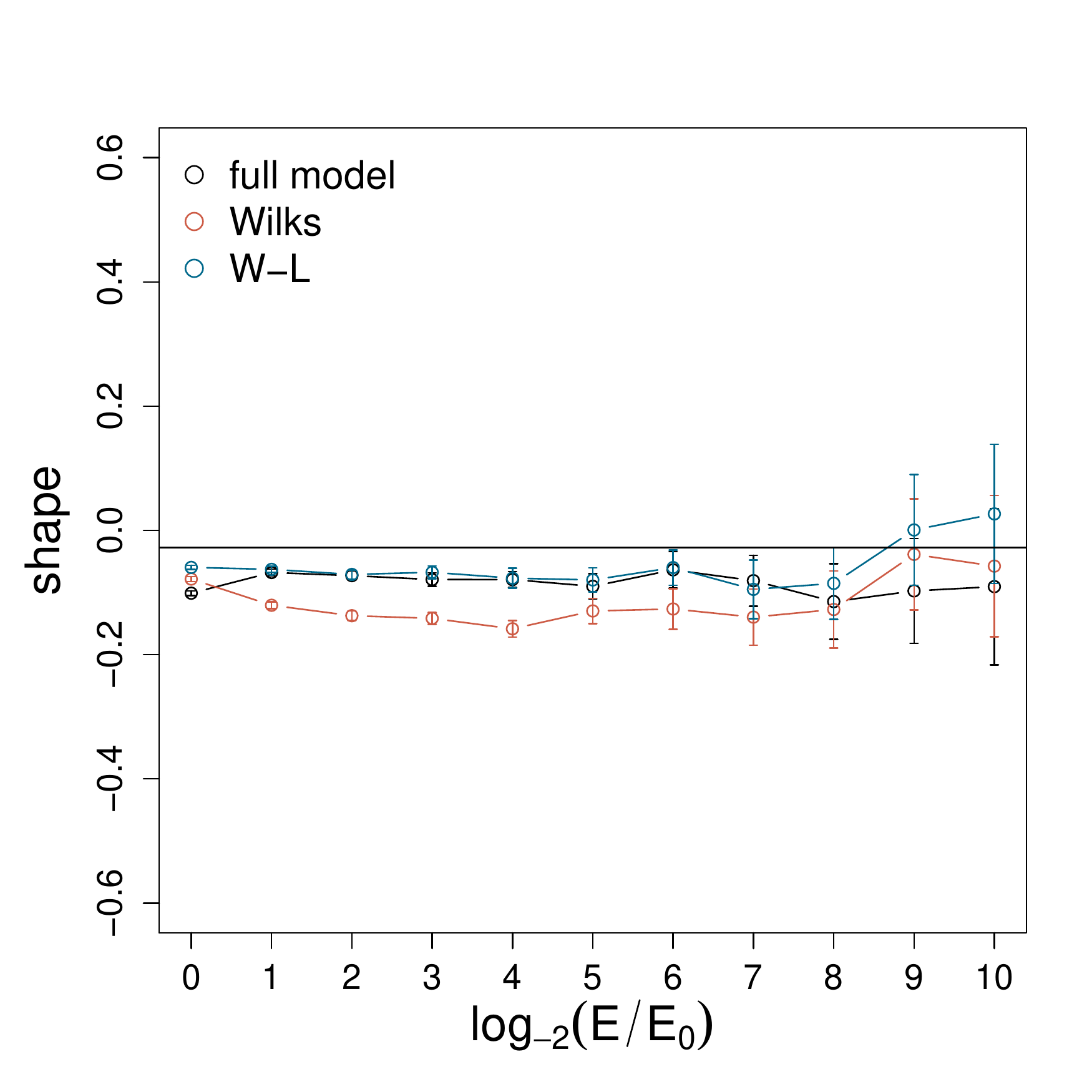}
 	\caption[]{\label{fig:compare_shape_POT}Comparison of the estimates of the GP shape parameter for the observable $A_x$ (left), $A_E$ (middle), and $A_p$ (right) over a range of exceedance ratios. The horizontal line shows the theoretical value of the shape parameter for the full model.}
 	\end{center}
\end{figure*}
\begin{figure*}
    \begin{center}
	\includegraphics[width=0.30\textwidth]{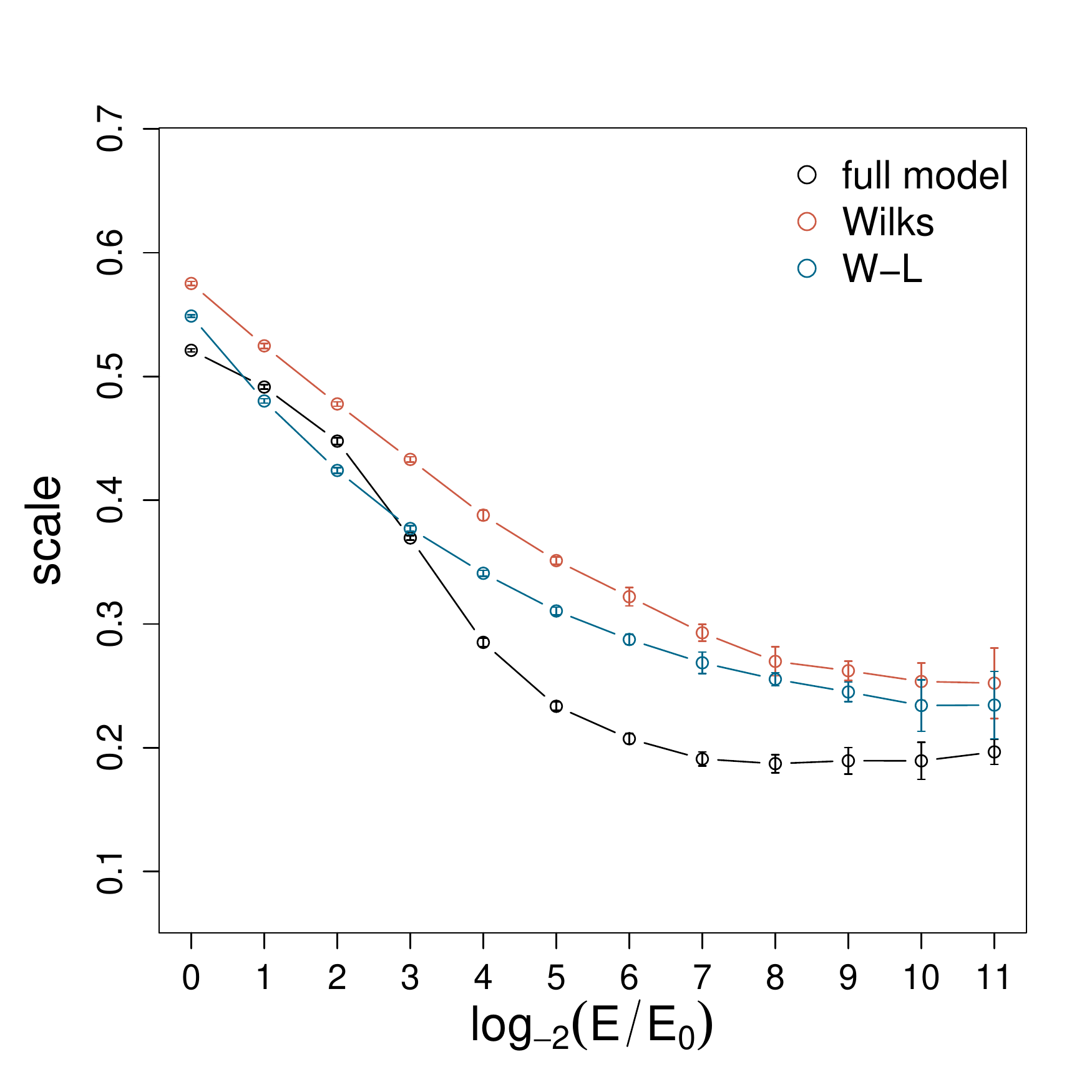}
	\includegraphics[width=0.30\textwidth]{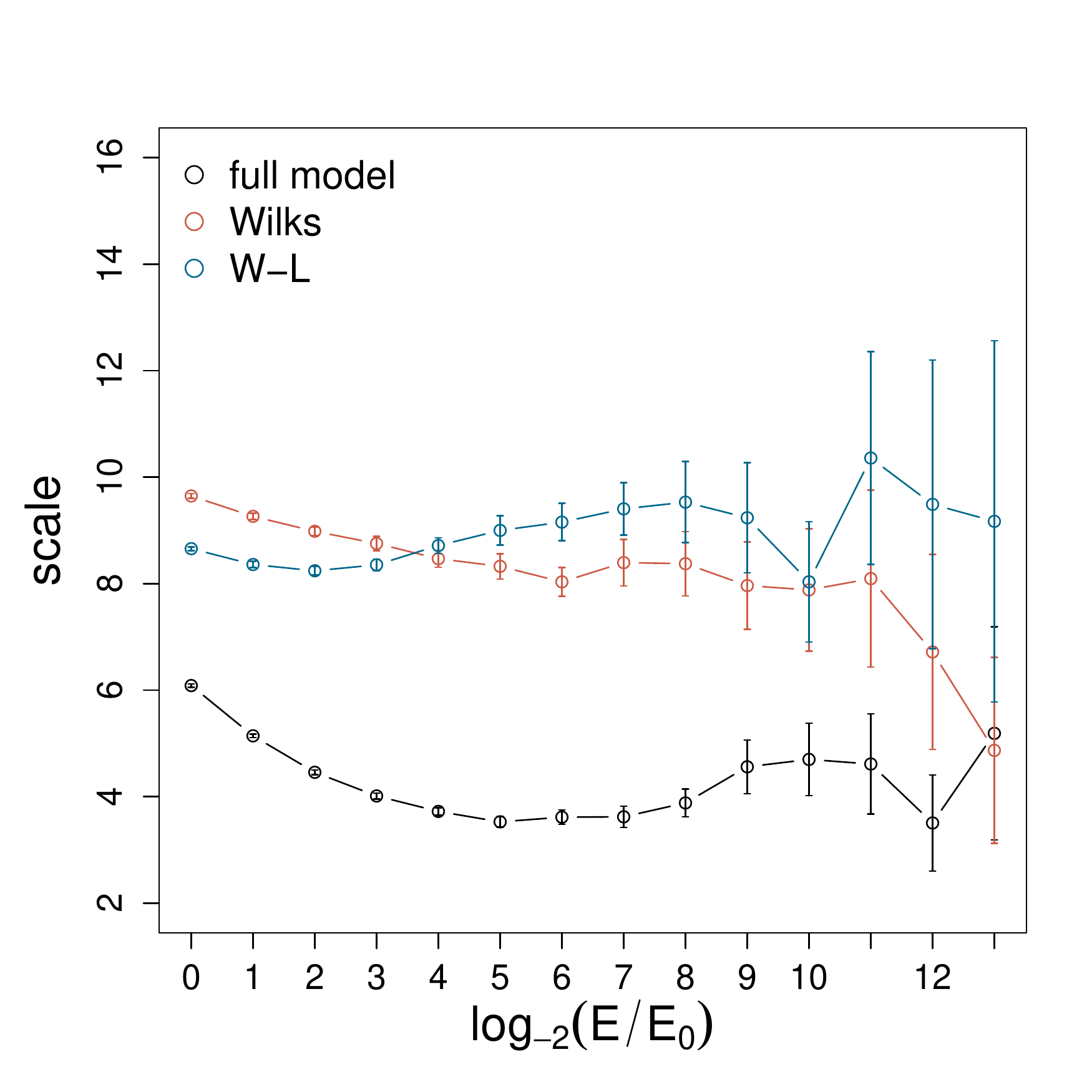}
	\includegraphics[width=0.30\textwidth]{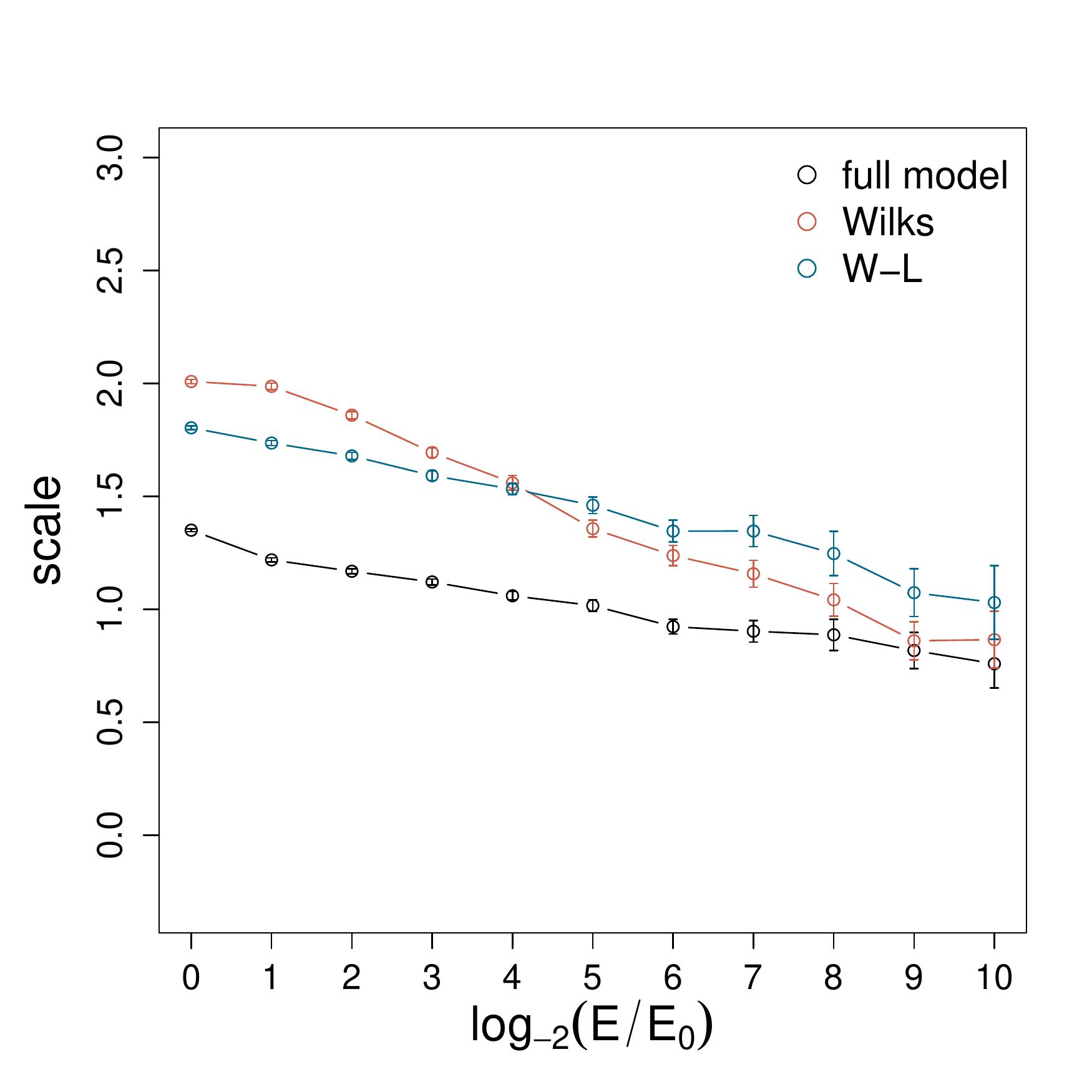}
 	\caption[]{\label{fig:compare_scale_POT}Comparison of the estimates of the GP scale parameter for the observable $A_x$ (left), $A_E$ (middle), and $A_p$ (right) over a range of exceedance ratios.}
 	\end{center}
\end{figure*}

Fig.~\ref{fig:compare_shape} compares the estimates of the GEV shape parameter for the three observables of the full and parametrized models over a range of block sizes. The horizontal line represents the theoretical value of the shape parameter for the full model, which is not applicable to the parametrized models, because they obey a stochastic and not deterministic dynamics. The shape parameter determines the tail behaviour of the distribution: a larger value indicates a slower decay of the tail; and, on the contrary, a smaller value indicates a faster decay of the tail. \\
\indent We first focus on the behaviour of estimates of the shape parameter for the full model. We observe nonmonotonic changes of the estimates with an increase of block size. The estimates for the observable $A_x$ decrease with the block size increasing from $B_0$ to $B_0 \times 2^3$, then it turns to increase and is seen to cross the theoretical value at $\log_2 (B/B_0)=7$, and after that the theoretical value is always within the increasingly large confidence intervals of the estimates. The estimates for the observable $A_E$ cross the theoretical value at $\log_2 (B/B_0)=4$ and then seems to come back at $\log_2 (B/B_0)=11$, however, the confidence interval is very big by then. The estimate for the observable $A_p$ increases at the first three data points, and then the estimates are always within the confidence intervals of any latter estimates. The estimates for the observable $A_p$ seem to reach a value smaller than the theoretical value, however, it should be noted that the last three data points show an upward trend, toward the theoretical value. As the block size increases, we observe to some extent an approximation of the estimates of the shape parameter to the theoretical value instead of a steady convergence. The estimates for the three observables show different erratic behaviours with the increase of the block size, and so none of these behaviours is asymptotic. \\
\indent We now focus on comparing the estimates of the shape parameter given by the parametrized models to that given by the full model. When we look at the observable $A_x$ and $A_E$, the estimates given by the parametrized models are apparently different from that given by the full model over most of the block sizes examined. An interesting result is that when we look at the observable $A_p$, the W-L parametrized model gives very similar estimates compared to the full model. We note, however, that this good correspondence could be just a coincidence; the results may change if we use another system or a different parameter setting. Moreover, this good correspondence should not last as the stochastic model surely has a different asymptotic shape parameter from that of the deterministic model.\\
\indent Fig.~\ref{fig:compare_scale} compares the estimates of the GEV scale parameter for the three observables of the full and parametrized models over a range of block sizes. The estimates for the observable $A_x$ and $A_p$ given by all the three models monotonically decrease as the block size increases, while the estimate for the observable $A_E$ nonmonotonically changes. The parametrized models generally give larger estimates than the full model for all the three observables over all the block sizes considered. We observe that the W-L parametrized model gives more accurate estimates for the local observable over all considered block sizes than the Wilks parametrized model. Similarly to the estimates of the shape parameter for the observable $A_p$, this result could also be just a coincidence.\\
\indent Fig.~\ref{fig:compare_location} compares the estimates of the location parameter for the three observables from the full and parametrized models over a range of block sizes. The location parameter, only appearing in GEV fittings, determines where the center of the distribution is located, and a larger value means that the distribution is shifted to the right, so that we have extremes of higher magnitudes. We observe that both of the parametrized models give larger estimates for all the three observables than the full model. Additionally, the W-L parametrized model gives more accurate estimates for the observable $A_x$ and $A_E$ than the Wilks parametrized model. 

\subsection{The GP Parameters}

\indent Fig.~\ref{fig:compare_shape_POT} compares the estimates of the GP shape parameter for the three observables of the full and parametrized models over a range of exceedance ratios. Fig.~\ref{fig:compare_scale_POT} compares the estimates of the GP scale parameter for the three observables of the full and parametrized models over a range of exceedance ratios. The behaviours of the estimates of the GP parameters are similar to those of the GEV parameters. We shall not comment on these further. The BM and POT approaches give consistent results of extreme value statistics as they are fundamentally equivalent.

\section{Empirical Comparison of Extremes}\label{sec:comparison_direct}

\subsection{Histograms of Block Maxima}

\begin{figure*}
 	\begin{center}
	\includegraphics[width=0.29\textwidth]{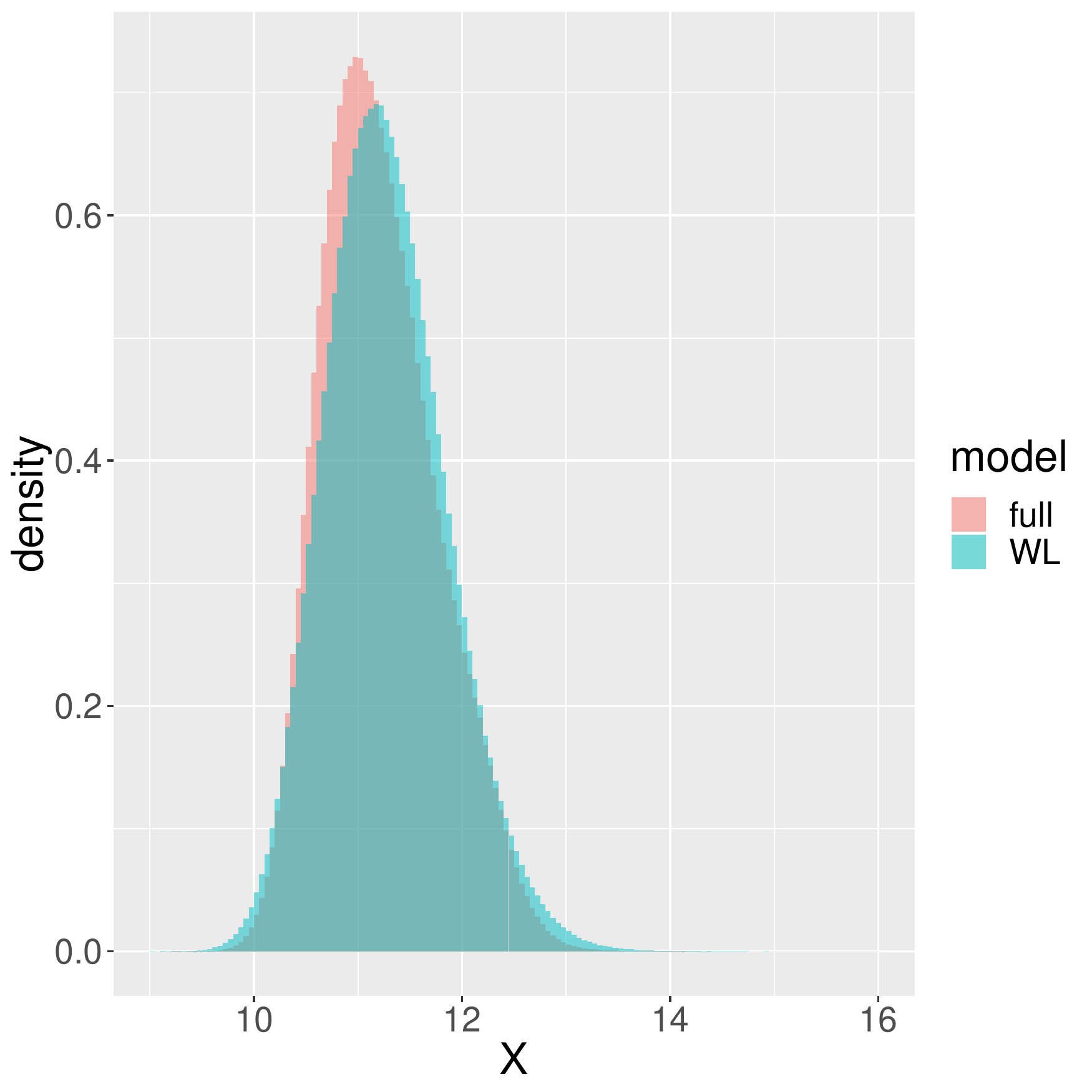}
	\includegraphics[width=0.29\textwidth]{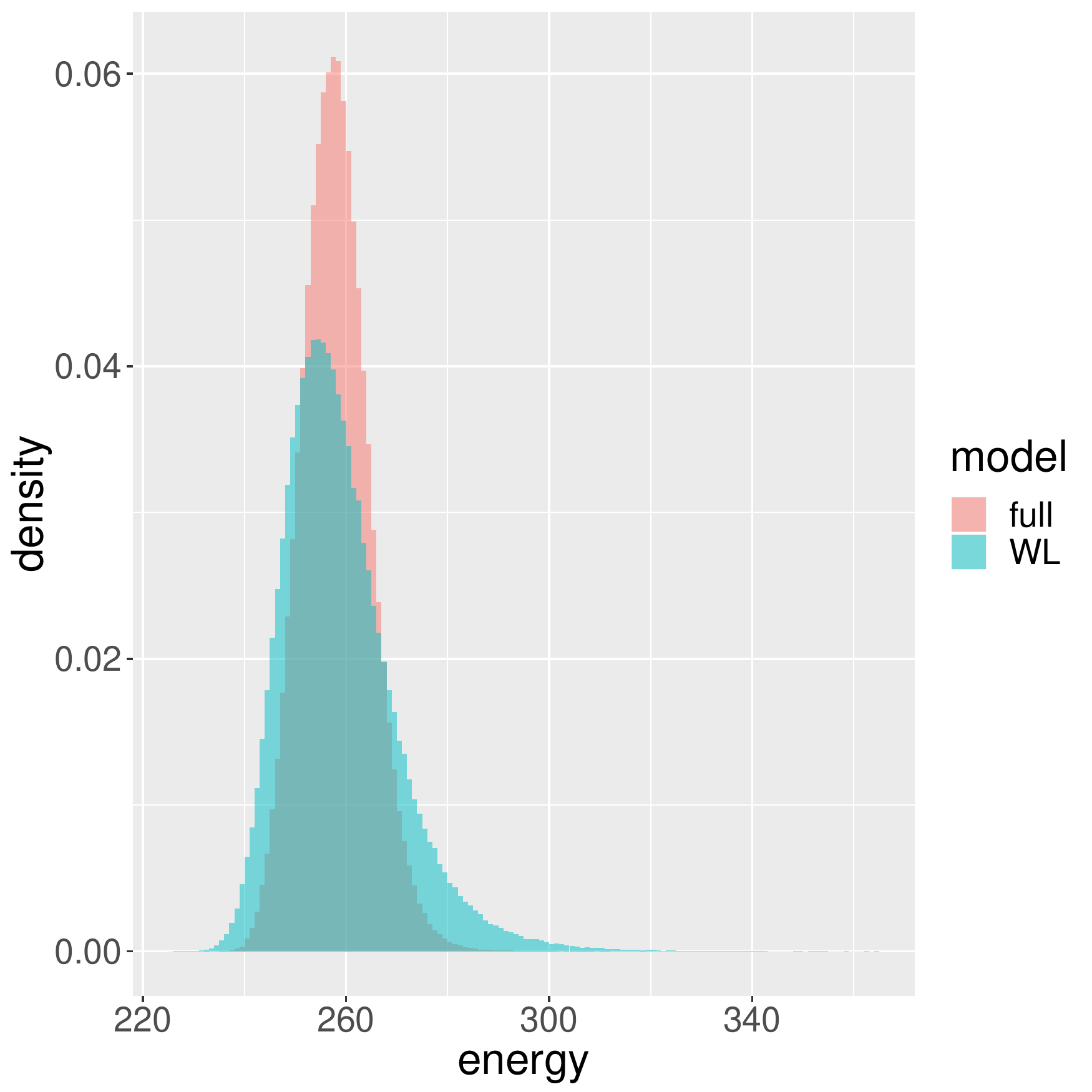}
	\includegraphics[width=0.29\textwidth]{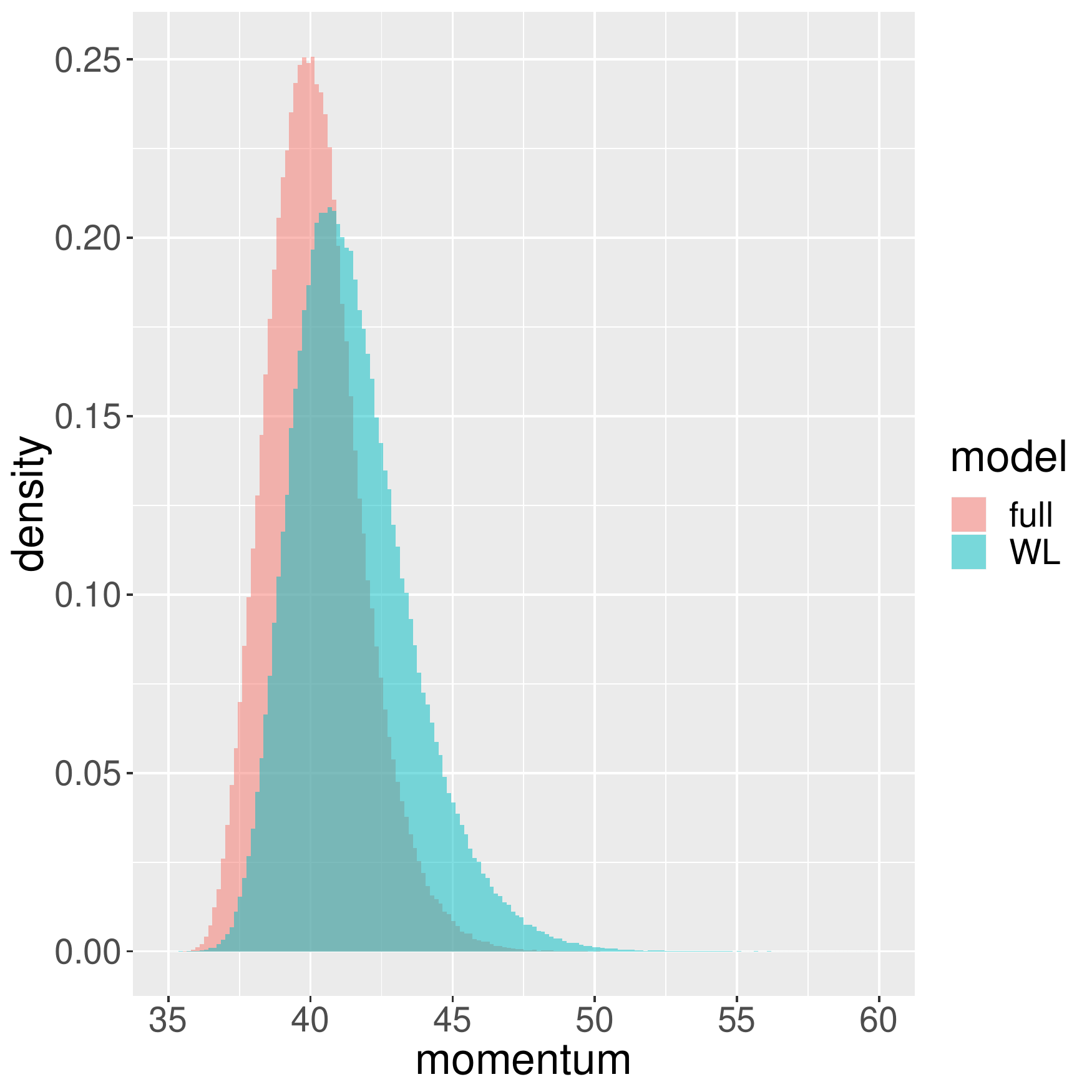}

	\includegraphics[width=0.29\textwidth]{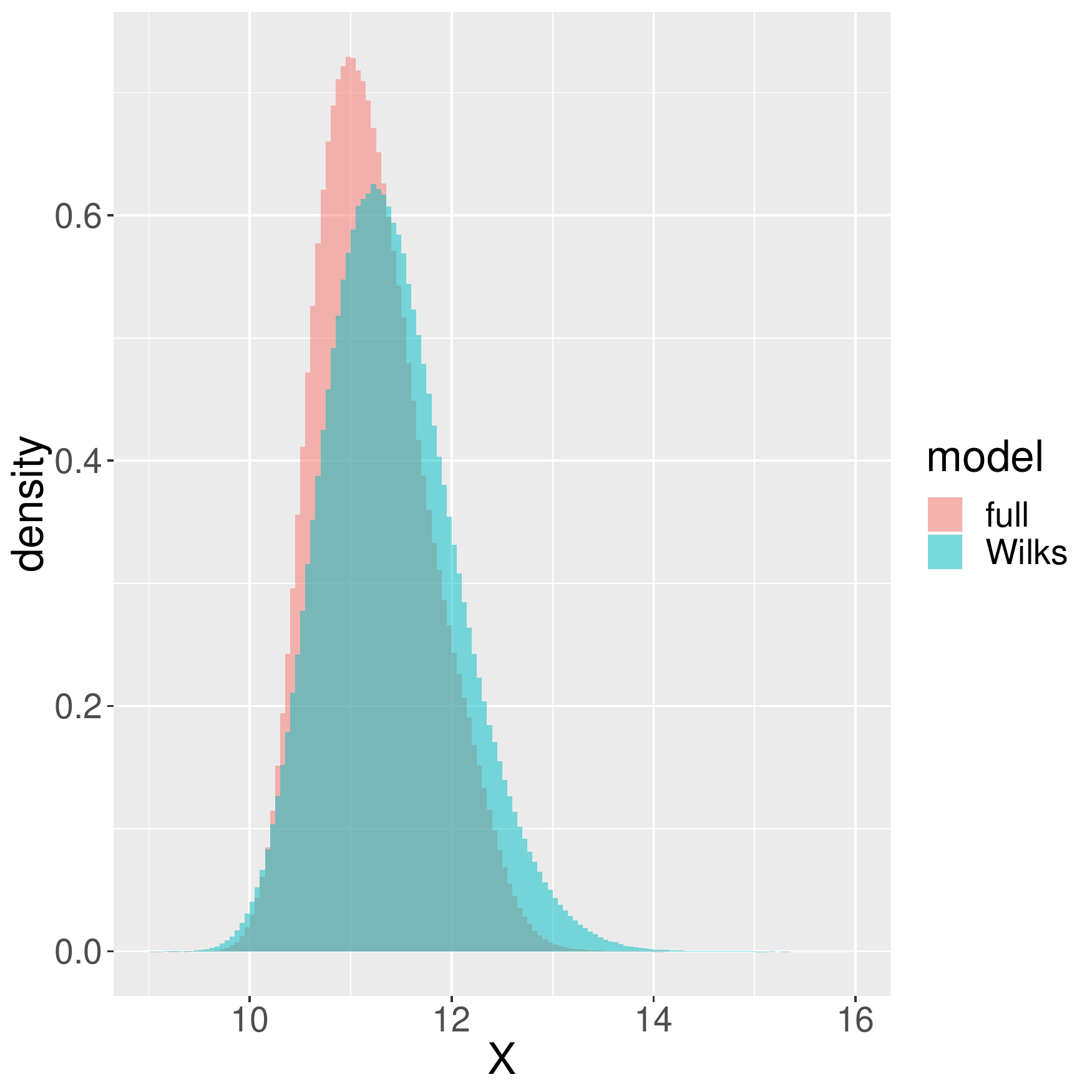}
	\includegraphics[width=0.29\textwidth]{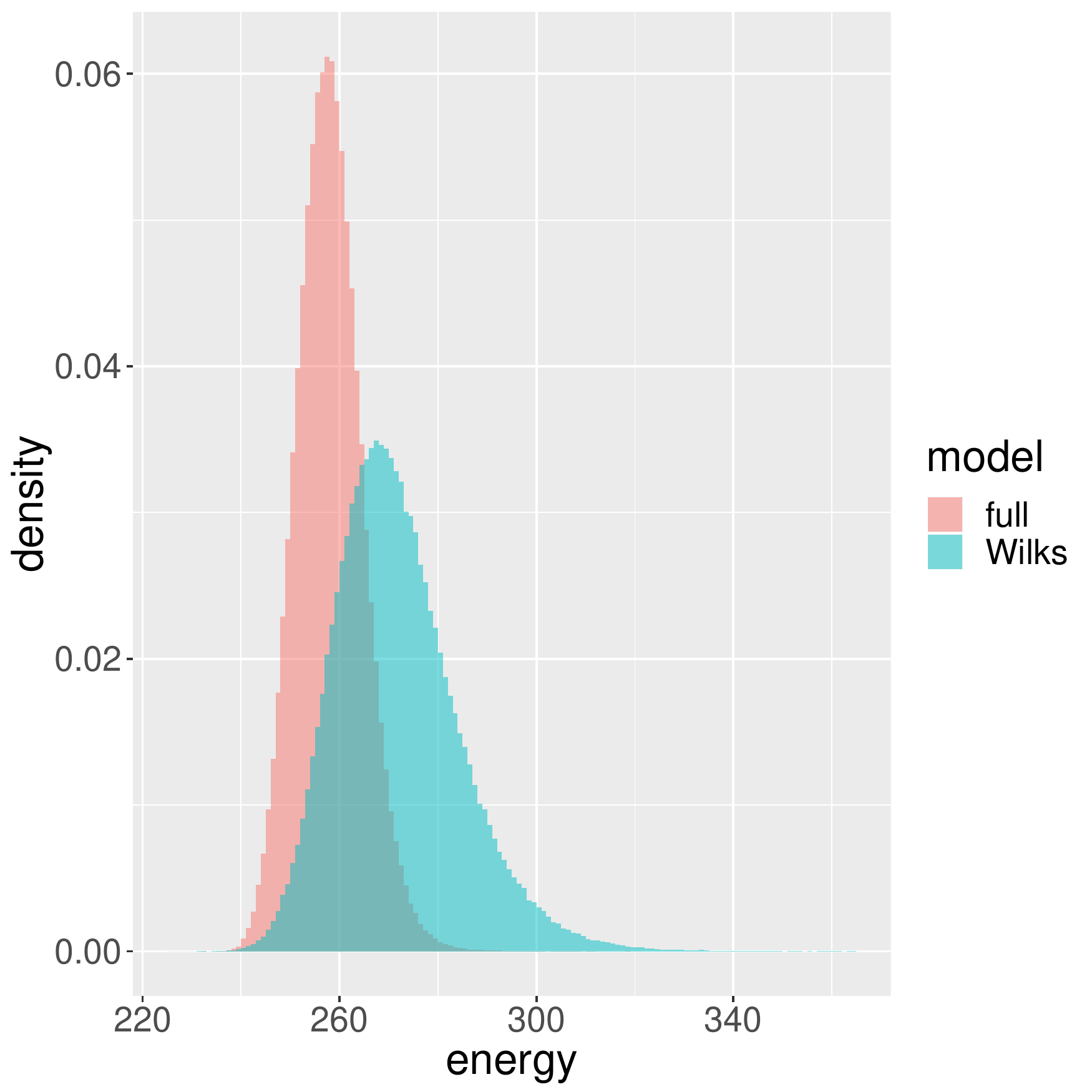}
	\includegraphics[width=0.29\textwidth]{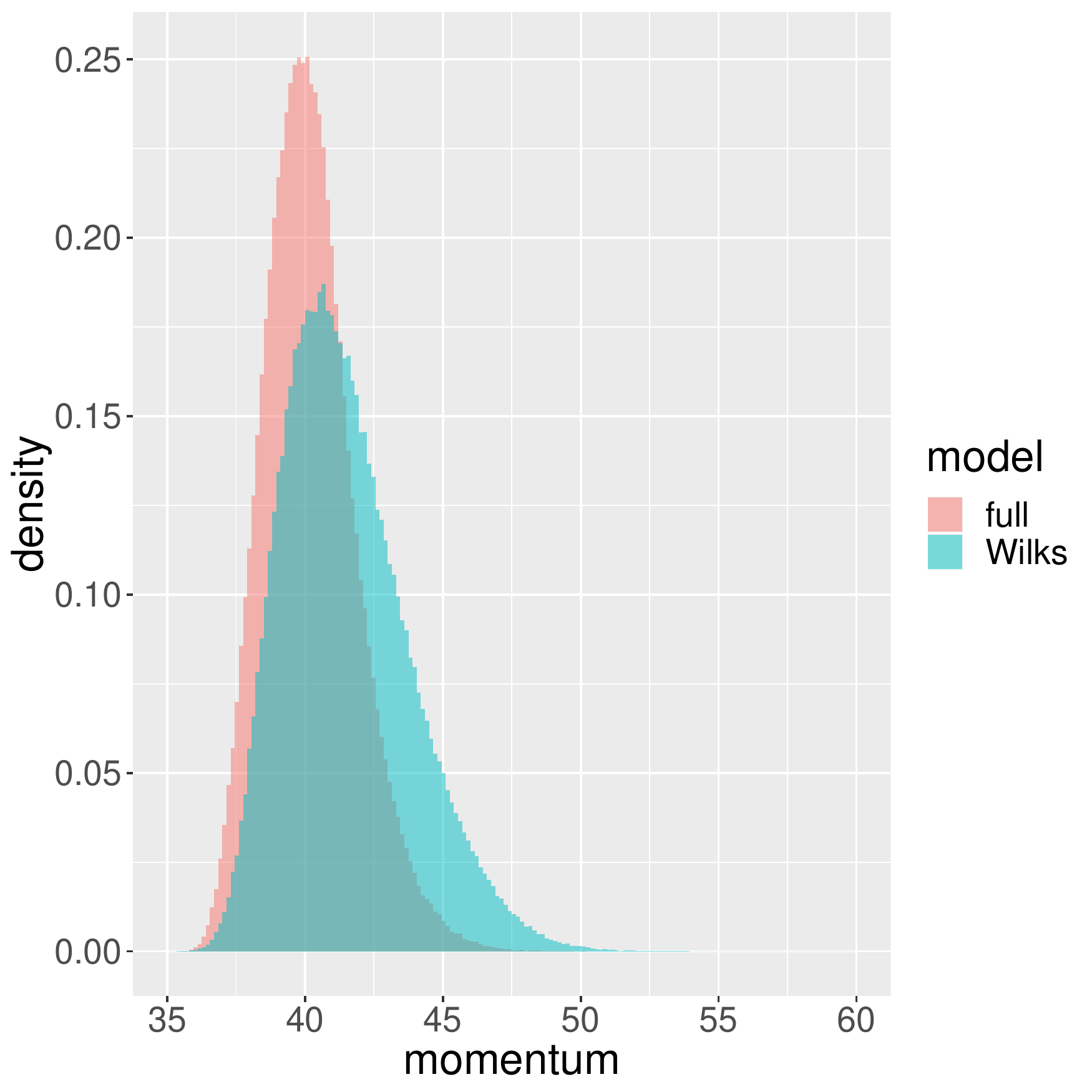}
 	\caption{\label{fig:compare_pdf_ex_bm}Comparison of the histograms of the block maxima of the observable $A_x$ (left panels), $A_E$ (middle panels), and $A_p$ (right panels) between the full and W-L parametrized models (upper panels), and between the full and Wilks parametrized models (lower panels). The block size is $B_0$.}
  	\end{center}
\end{figure*}
\begin{figure*}
 	\begin{center}
	\includegraphics[width=0.29\textwidth]{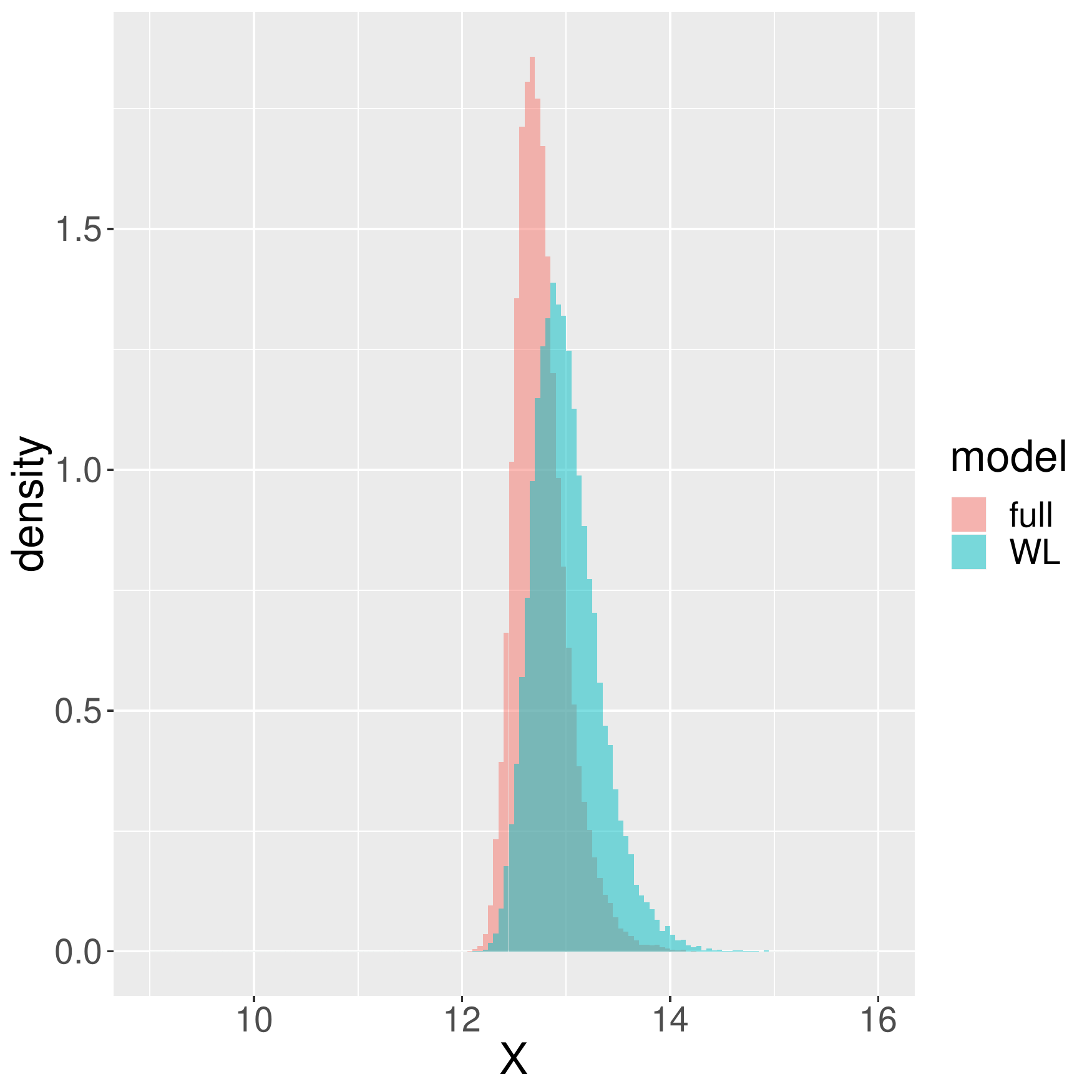}
	\includegraphics[width=0.29\textwidth]{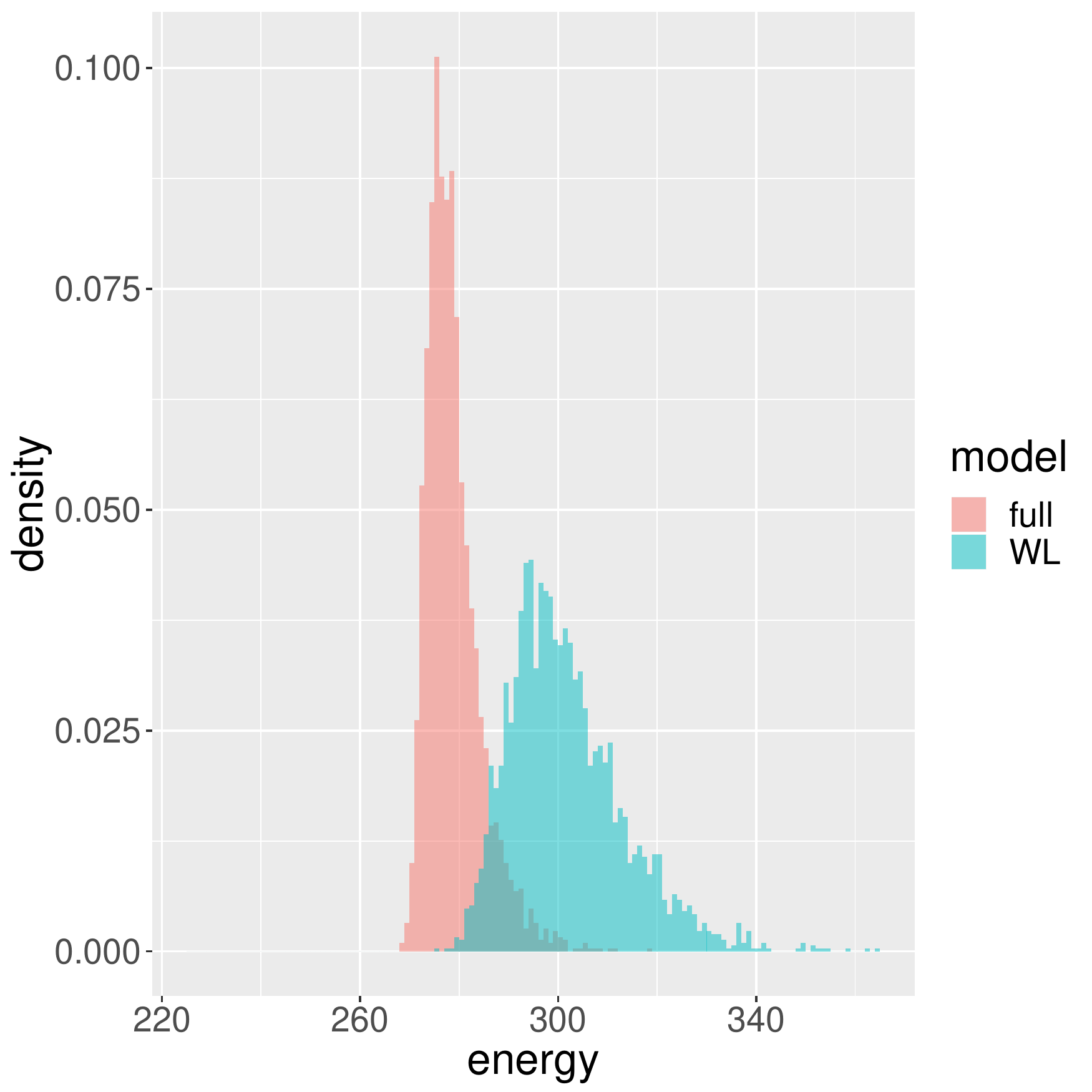}
	\includegraphics[width=0.29\textwidth]{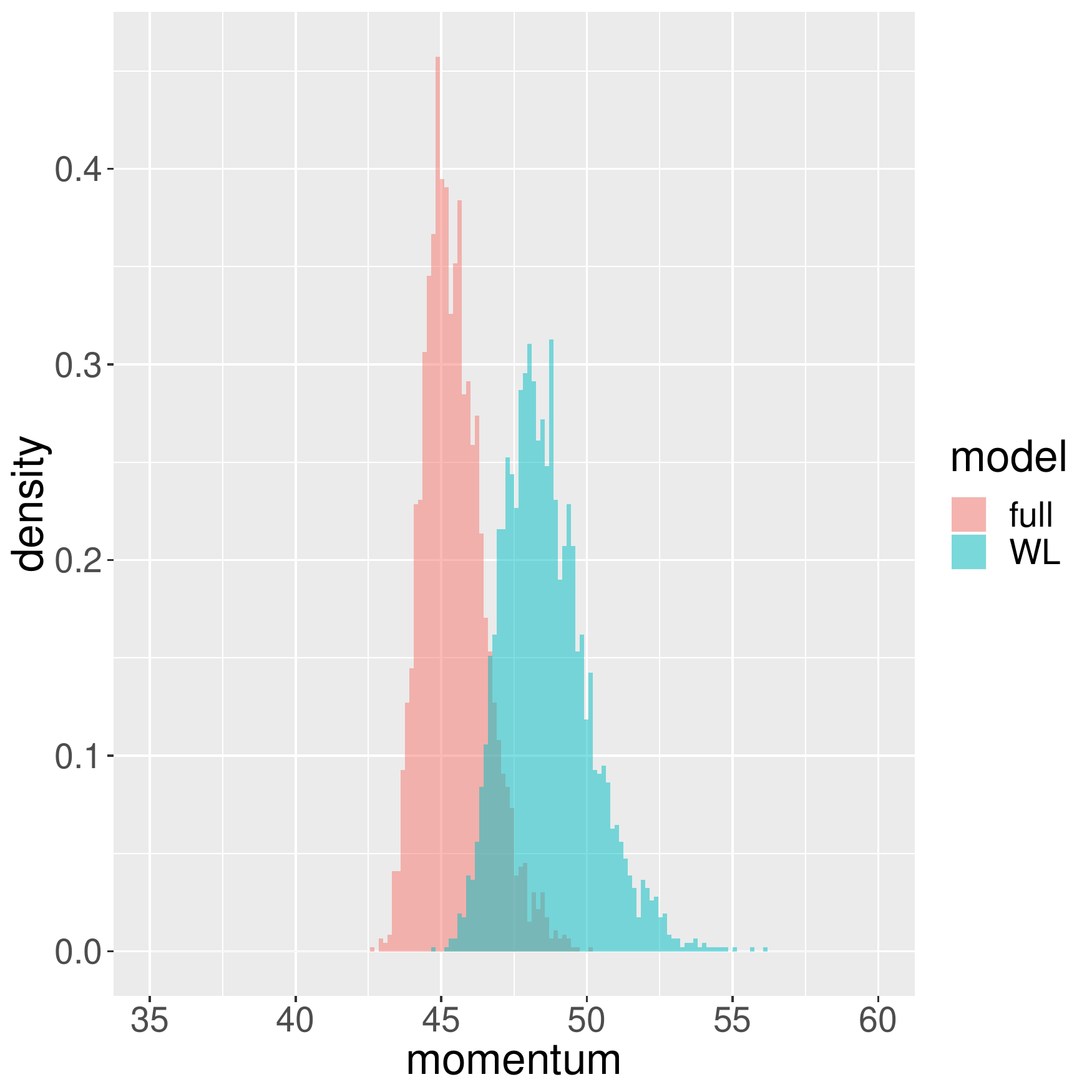}

	\includegraphics[width=0.29\textwidth]{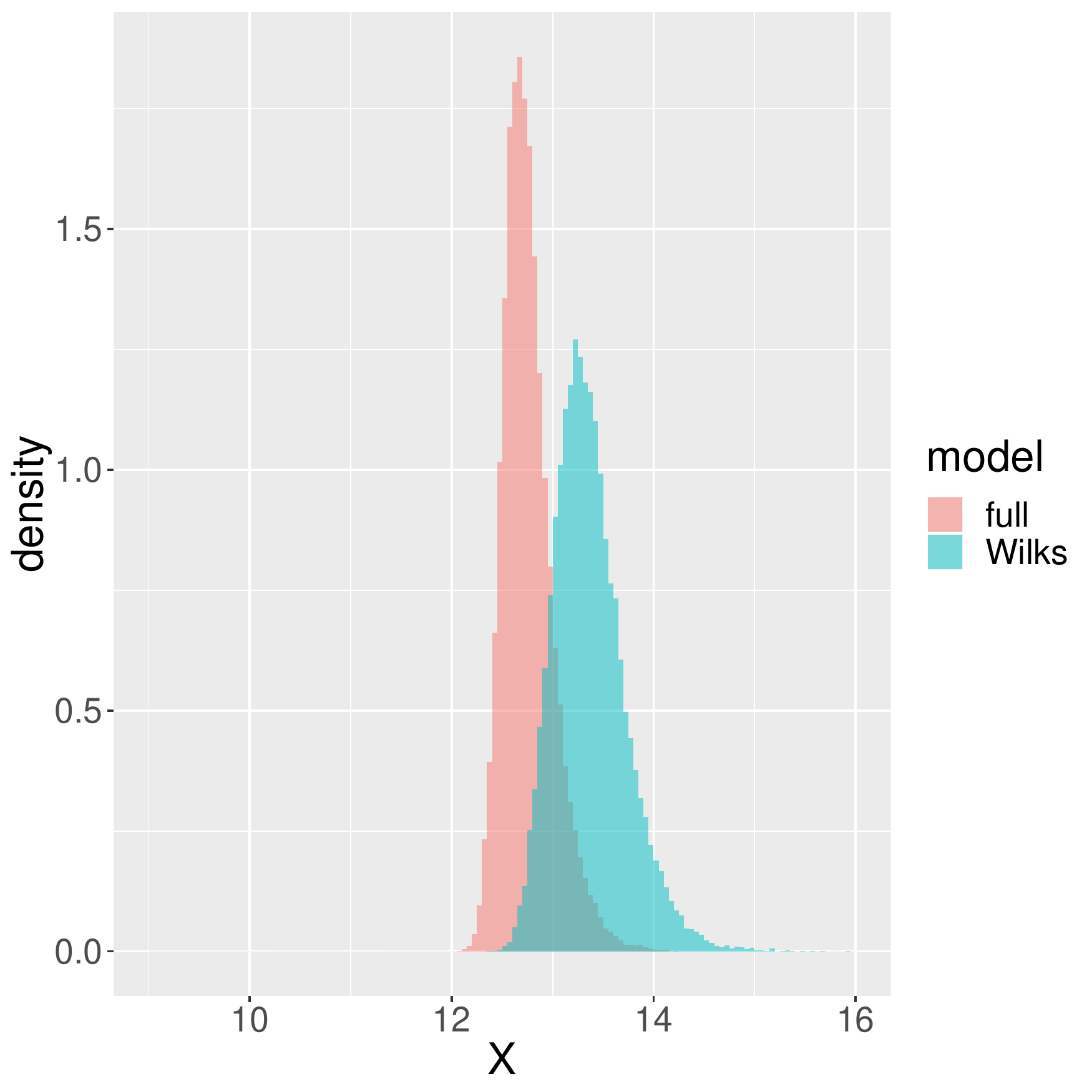}
	\includegraphics[width=0.29\textwidth]{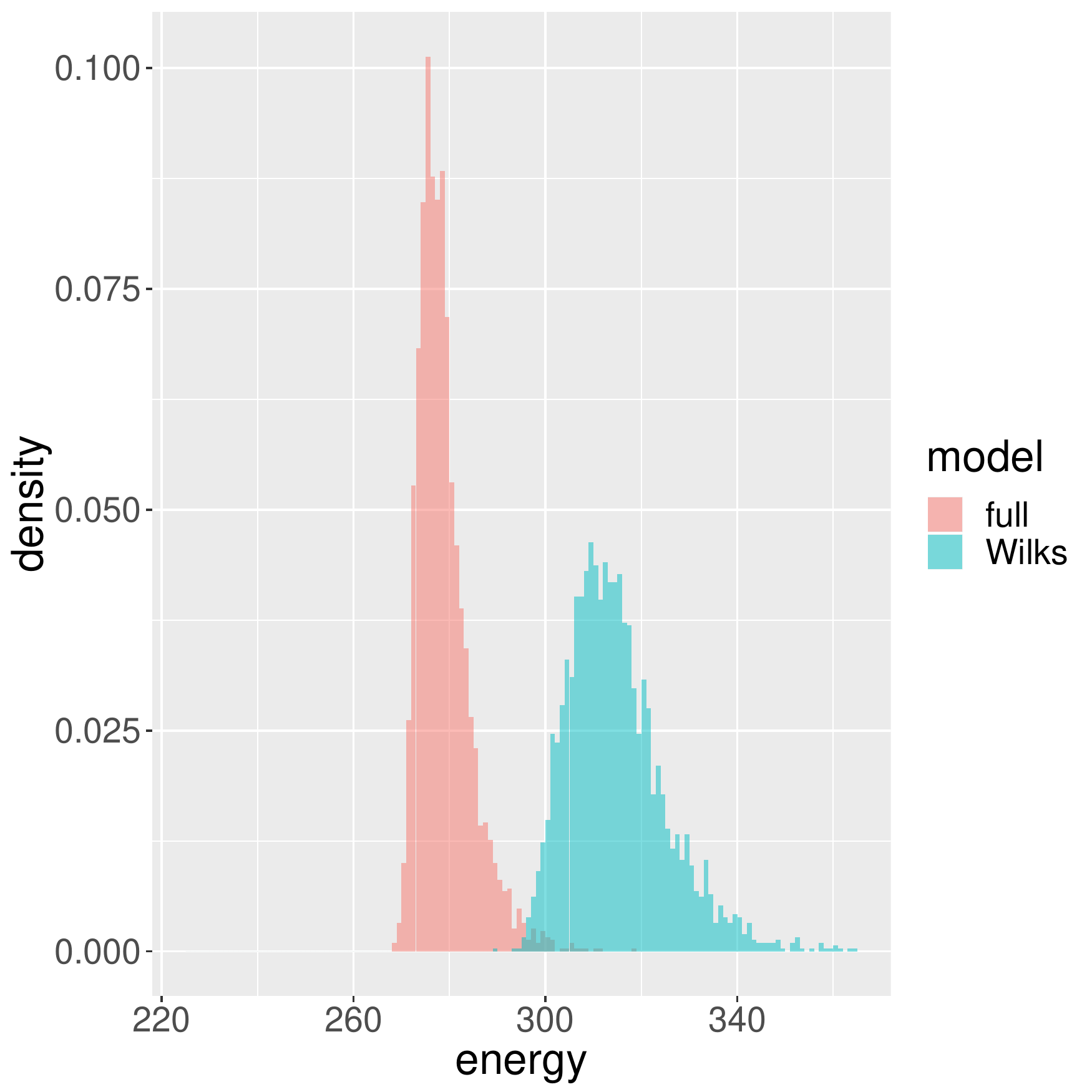}
	\includegraphics[width=0.29\textwidth]{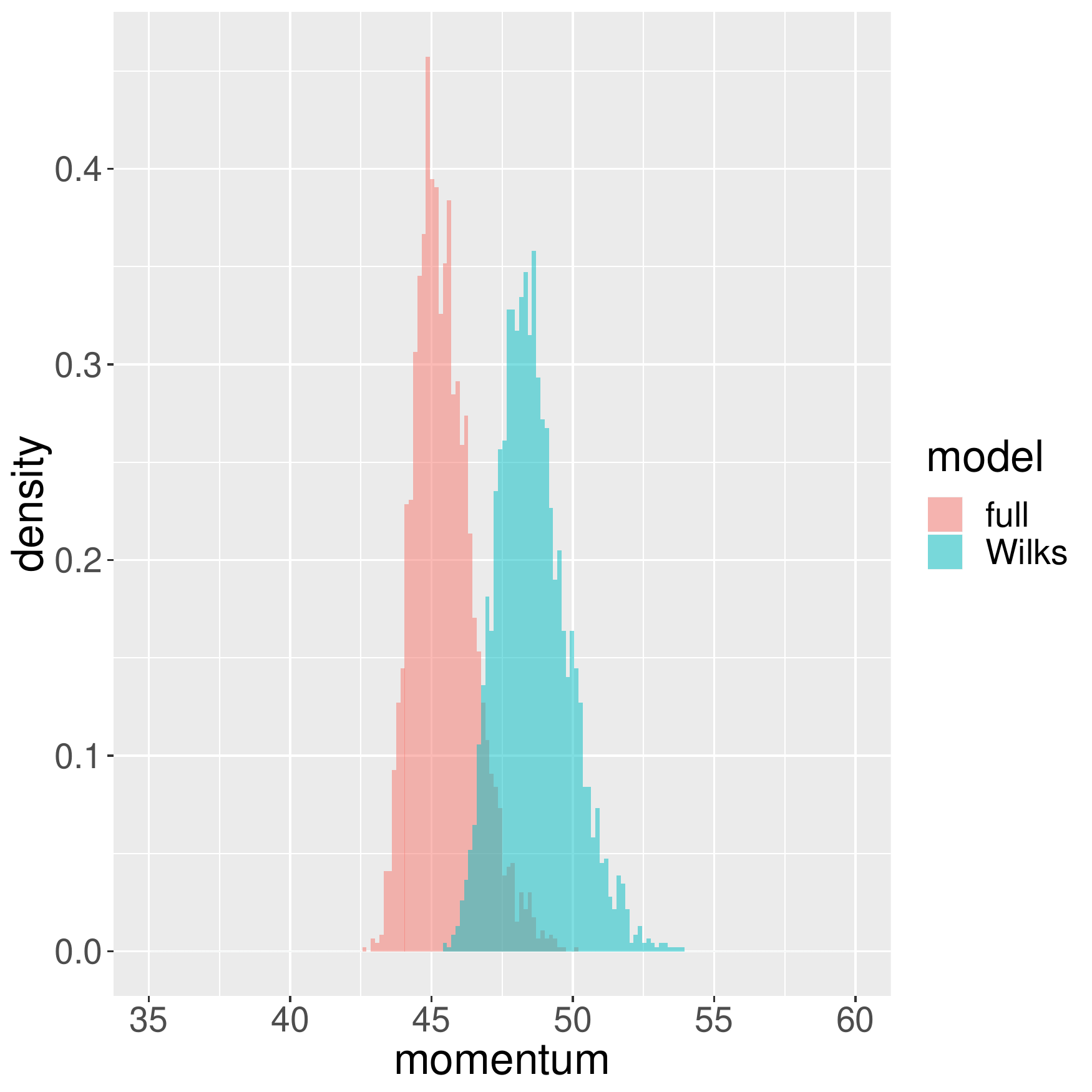}
 	\caption{\label{fig:compare_pdf_ex_bm2}Comparison of the histograms of the block maxima of the observable $A_x$ (left panels), $A_E$ (middle panels), and $A_p$ (right panels) between the full and W-L parametrized models (upper panels), and between the full and Wilks parametrized models (lower panels). The block size is $B_0 \times 2^7$.}
  	\end{center}
\end{figure*}

Fig.~\ref{fig:compare_pdf_ex_bm} compares the histograms of the block maxima of the three observables between the parametrized and full models. The total area of each histogram is normalized to one, and so the vertical axes of the histograms show relative probability density. The histograms give a rough estimate of the PDF of the block maxima. Since we have a large number of data points and small bin sizes, the graphs look very smooth. The patterns of the histograms are unimodal and approximately symmetric; they are slightly right skewed like indeed a GEV distribution. The disagreements between the histograms of the parametrized models and the full model have two main features: 1) the block maxima from the parametrized models are more widely distributed; and 2) the histograms of the parametrized models are right-shifted. These empirical results have been already indicated by the estimates of the scale and location parameter in Sec.~\ref{sec:comparison_EVT}. Fig.~\ref{fig:compare_pdf_ex_bm2} shows the same figures as Fig.~\ref{fig:compare_pdf_ex_bm} but with the block maxima selected by a larger block size. As the block size increases, the block maxima have larger magnitudes and the overlapping area between the histograms of the parameterized and full models becomes smaller. If we look at the observable $A_E$, we can see that the probability density of the block maxima given by the Wilks parametrized model is totally different from that given by the full model. In summary, the histograms of the W-L parametrized model appear to be in somewhat better agreement with the histograms of the full model than that of the Wilks parametrized model. Moreover, the local observable of the full and parametrized models show better agreement than the global observables. 

\subsection{Number of Threshold Exceedances}

\begin{figure}
 	\begin{center}
	\includegraphics[width=3.3in]{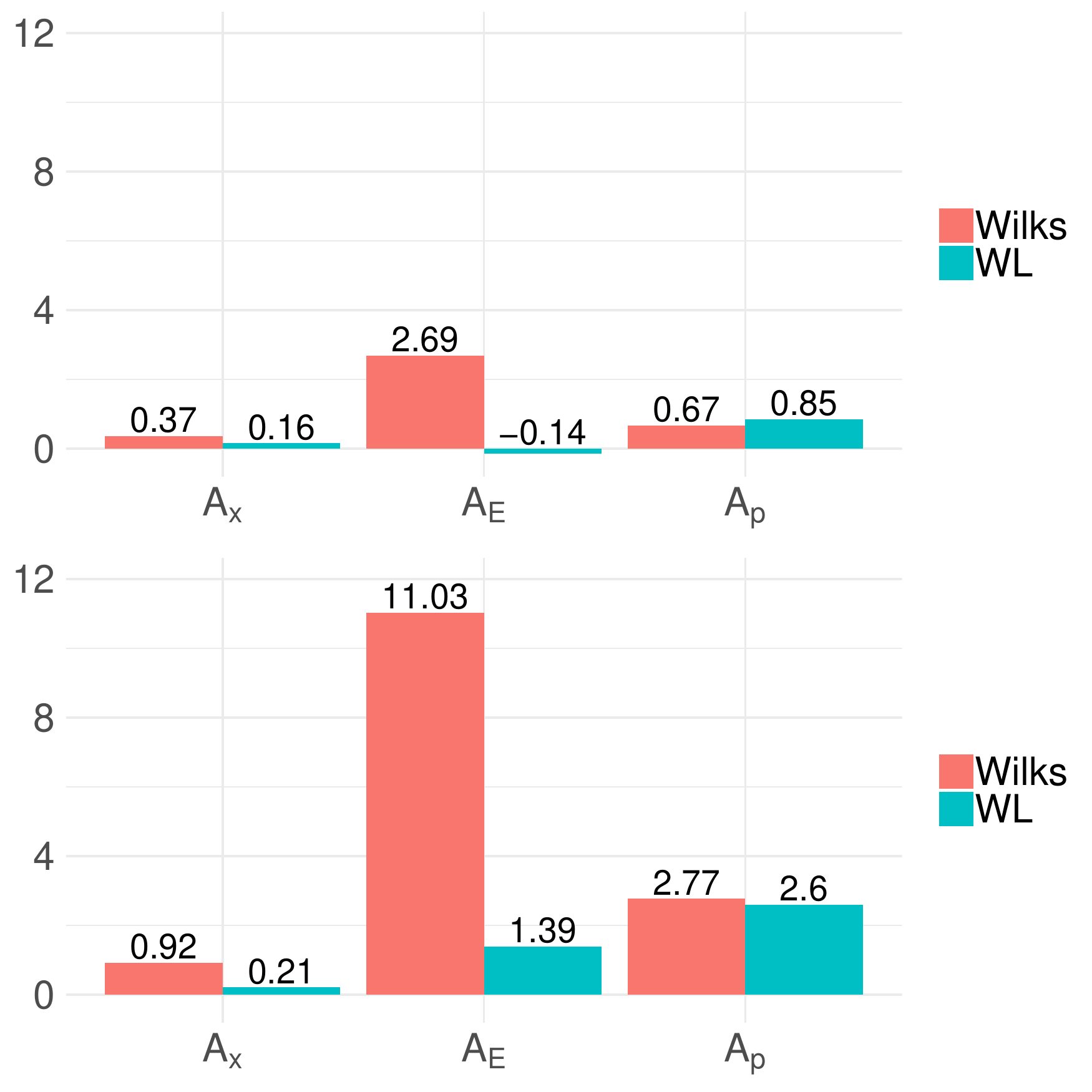}
	\caption{\label{fig:compare_num}The fractional difference of the number of threshold exceedances from the parametrized models compared to that from the full model. In the upper (lower) panel, the threshold is the $99.9$th ($99.99$th) percentile of the corresponding observable of the full model.}
	\end{center}
\end{figure}

When applying the POT method, a main issue is to compare the number of the threshold exceedances produced by the full and parametrized models for a given threshold. Fig.~\ref{fig:compare_num} shows the fractional difference of the number of the threshold exceedances given by the parametrized models compared to that given by the full model.
We calculate the fractional difference using the formula given as: 
\begin{equation}\label{eqn:fractional_difference}
    \frac{n_{par}(u)-n_{full}(u)}{n_{full}(u)},
\end{equation}
where $n_{full}(u)$ denotes the number of exceedances above a threshold $u$ from the full model, and $n_{par}(u)$ denotes that from the parametrized models. We consider two sets of thresholds for the three observables of the three models. The first set of thresholds is chosen as the $99.9$th percentile of each observable of the full model, and the second set of thresholds is more stringent, which is the $99.99$th percentile of each observable of the full model. As shown in Fig.~\ref{fig:compare_num}, both of the parametrized models give more threshold exceedances than the full model, with the exception of the observable $A_E$ of the W-L parametrized model in the upper panel. Moreover, the global observables demonstrate larger differences than the local observable with the same exception. When considering the observable $A_x$ and $A_E$, the W-L parametrized model has a substantially better performance than the Wilks' one. Note that if we look at the observable $A_E$ of the Wilks parametrized model in the lower panel, we can see a huge discrepancy; the number of threshold exceedances given by the Wilks parametrized model is more than $11$ times greater than that given by the full model. Instead, the performance of the two parametrizations is comparable when looking at the observable $A_p$. We also observe that the fractional difference becomes larger when a higher threshold is given.

\subsection{Quantile-Quantile Plots}

\begin{figure*}
 	\begin{center}
	\includegraphics[width=0.29\textwidth]{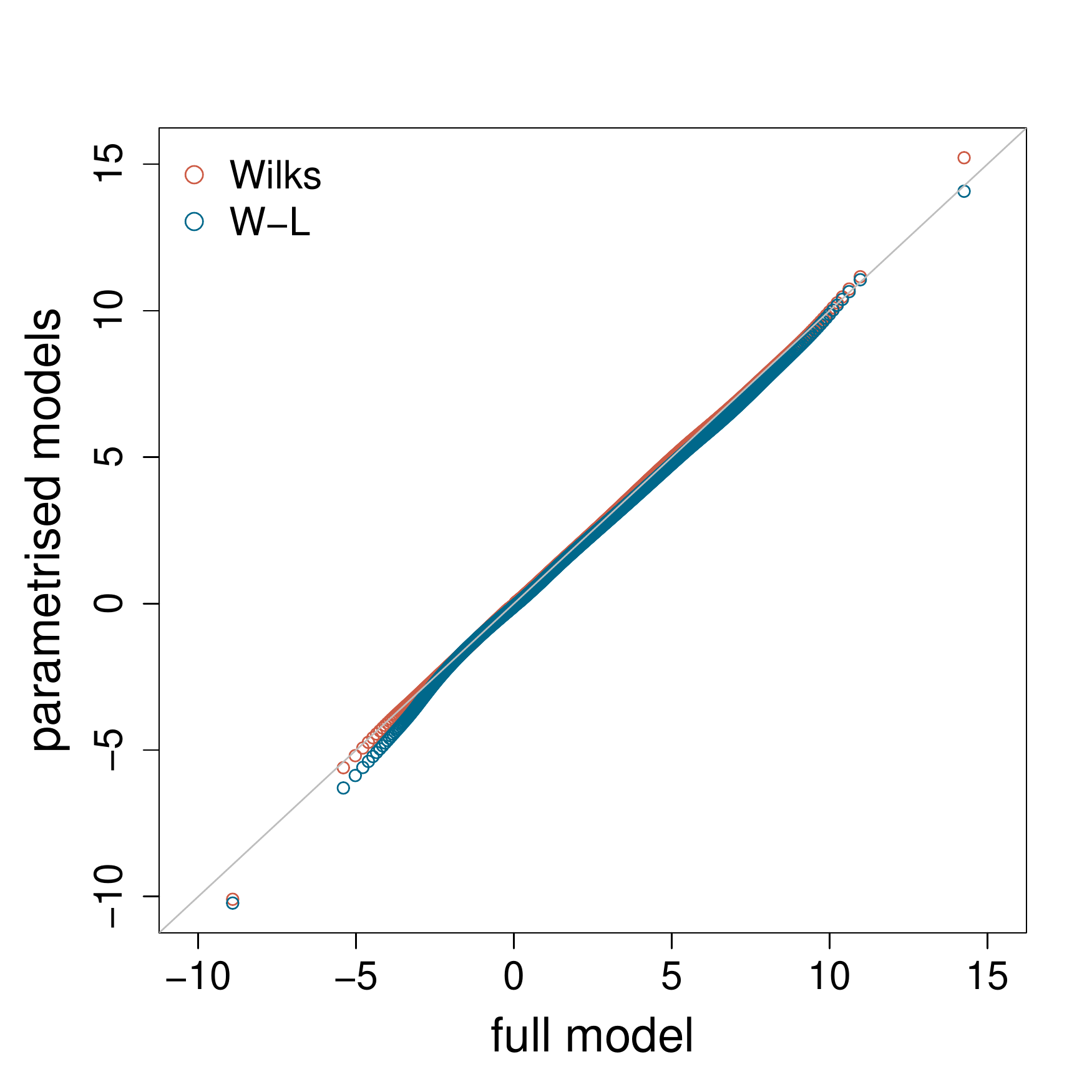}
	\includegraphics[width=0.29\textwidth]{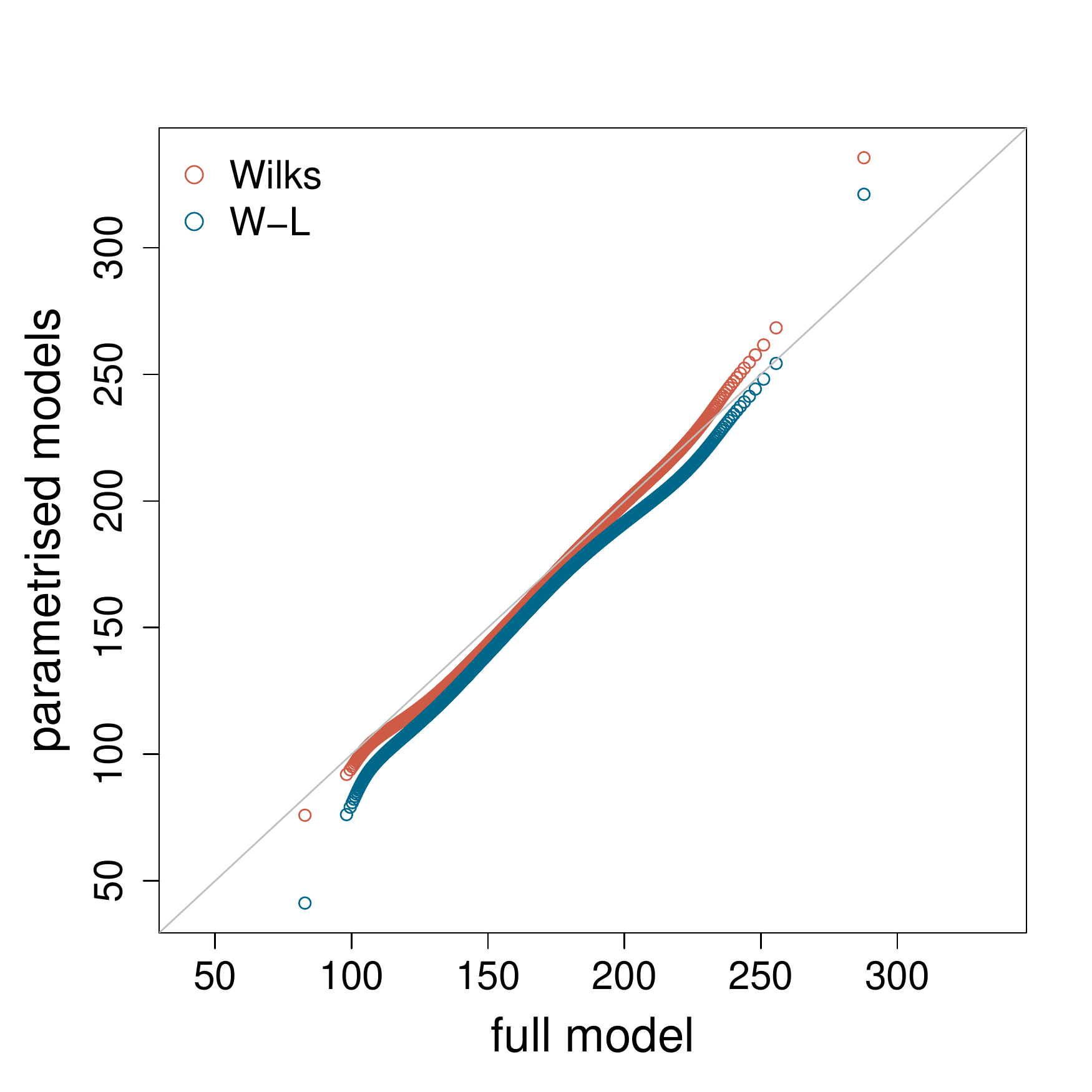}
	\includegraphics[width=0.29\textwidth]{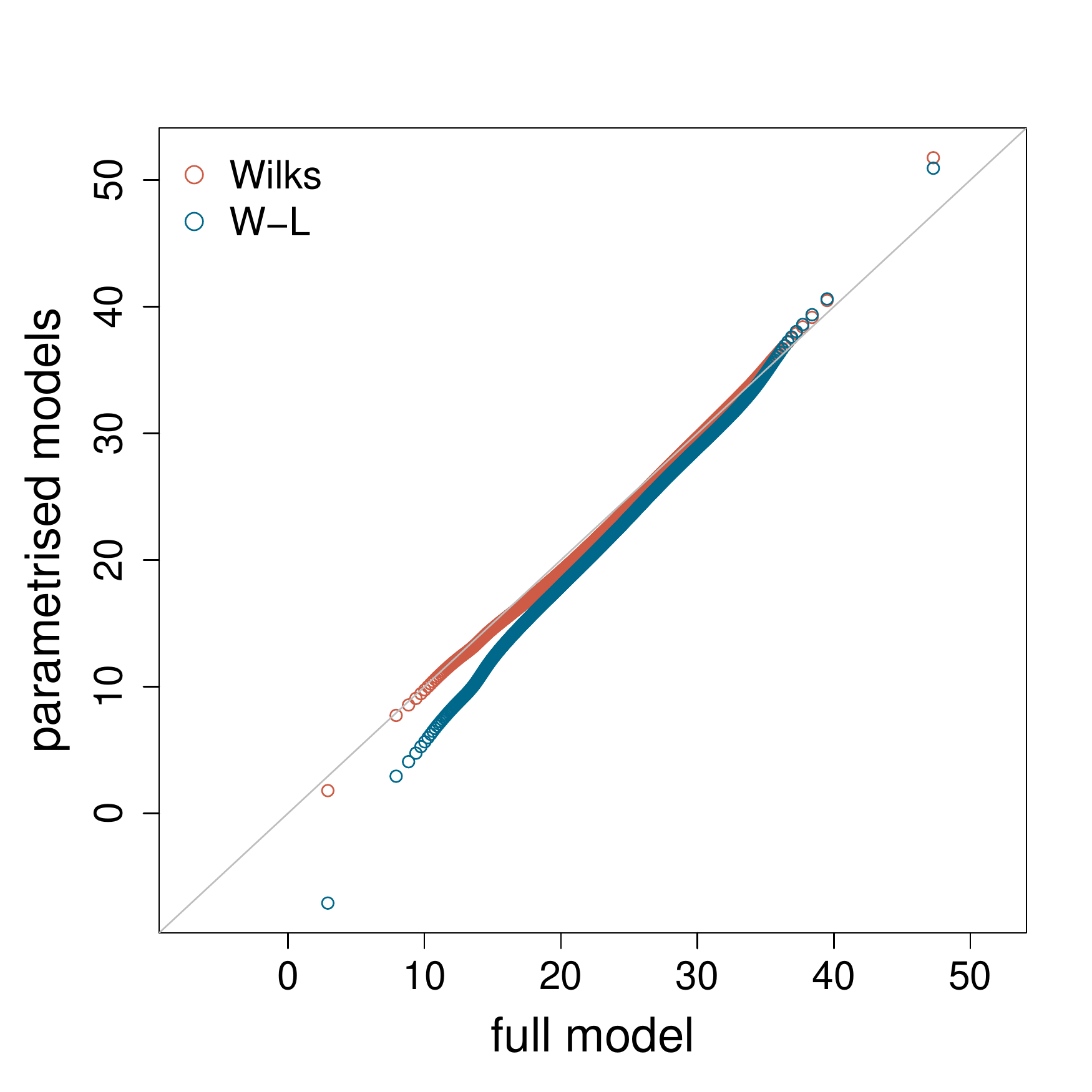}
 	\caption{\label{fig:compare_qq_all}The empirical quantiles of the observable $A_x$ (left), $A_E$ (middle), and $A_p$ (right) from the parametrized models against that from the full model.}
 	\end{center}
\end{figure*}

We further compare the empirical quantiles of the three observables from the parametrized models to those from the full model. We compare $1001$ quantiles, starting with the smallest element of the sample, up to the largest element, and 999 equidistant quantiles in between these two elements. Fig.~\ref{fig:compare_qq_all} shows the empirical quantiles of the three observables from the parametrized models against those from the full model.
\begin{figure*}
 	\begin{center}
    \includegraphics[width=0.29\textwidth]{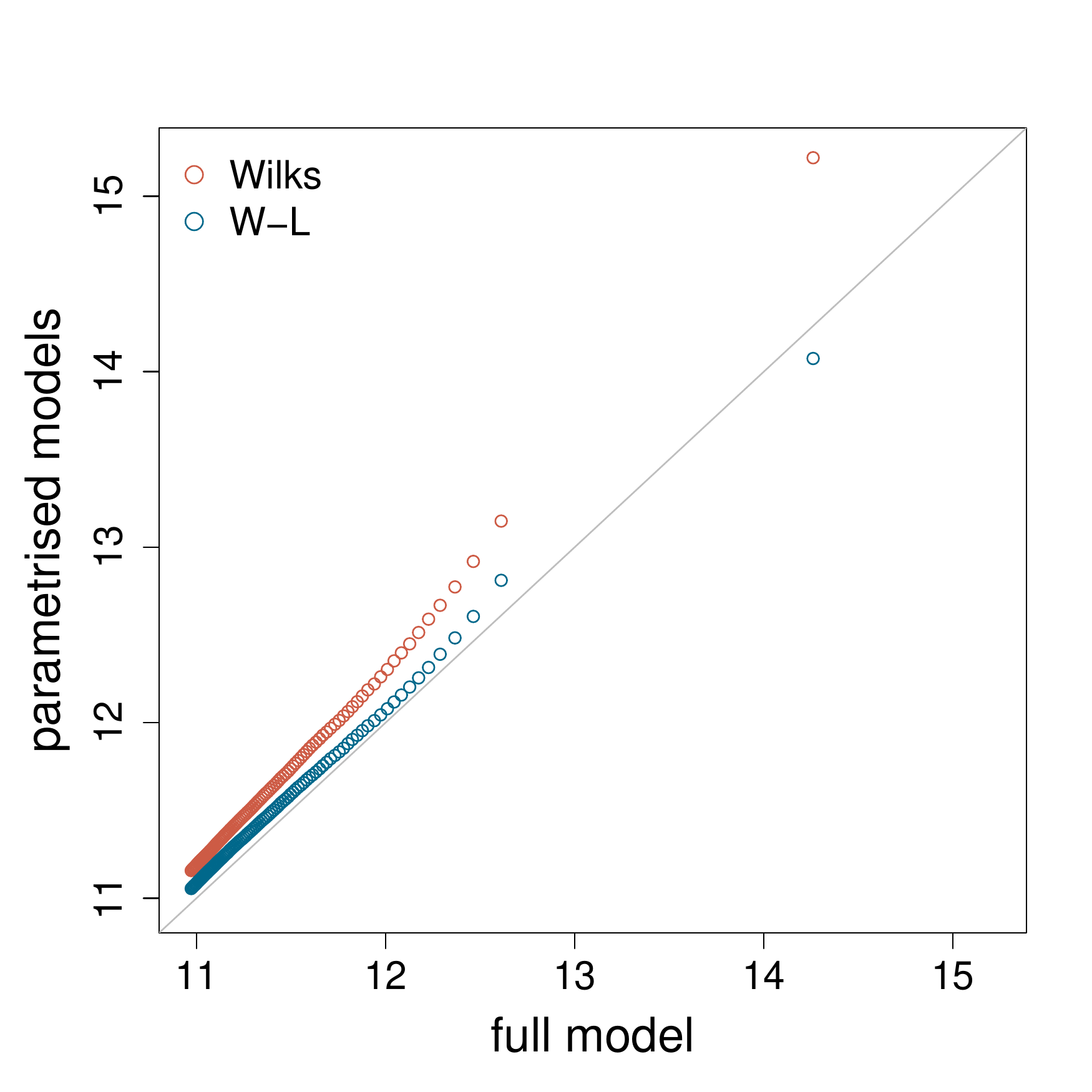}
    \includegraphics[width=0.29\textwidth]{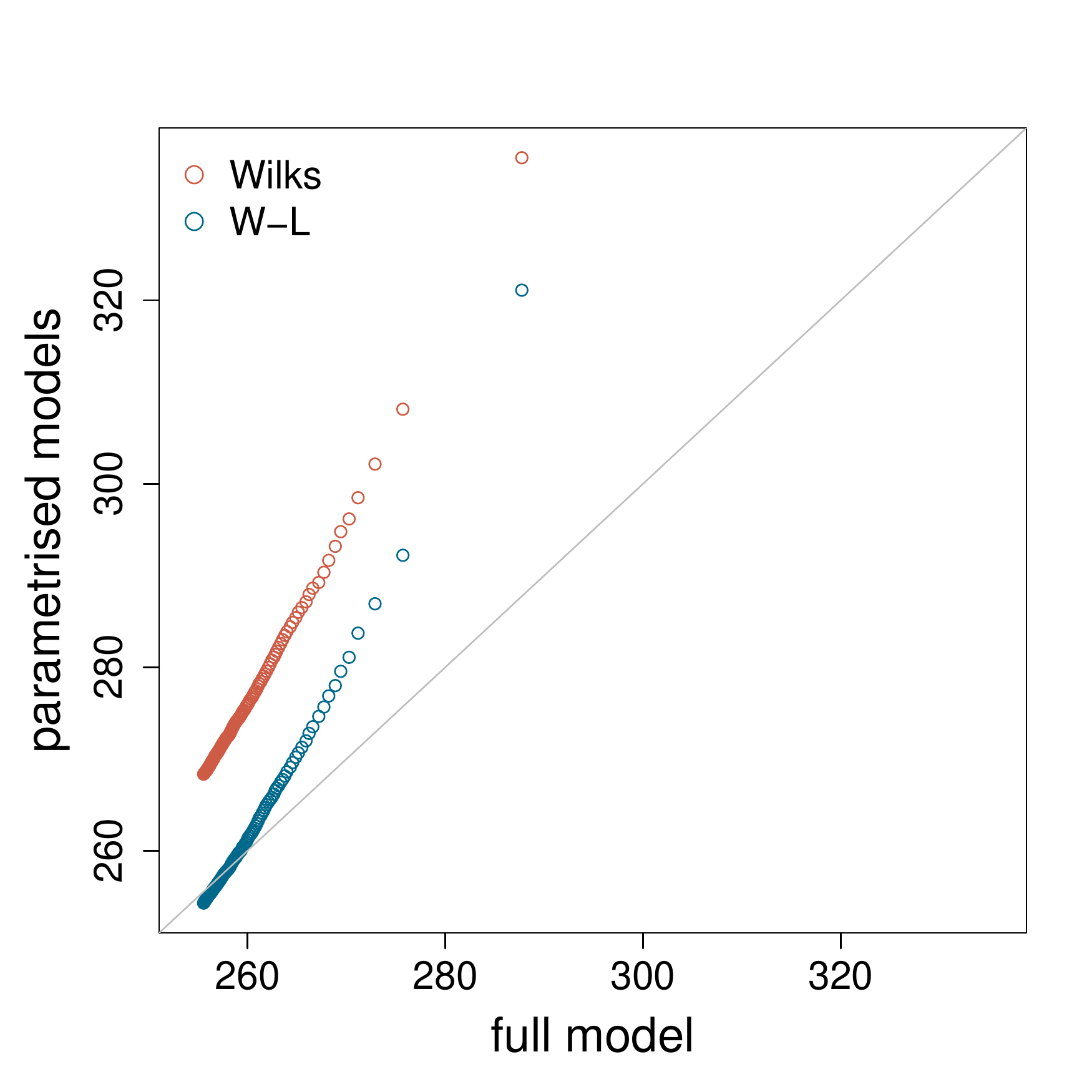}
    \includegraphics[width=0.29\textwidth]{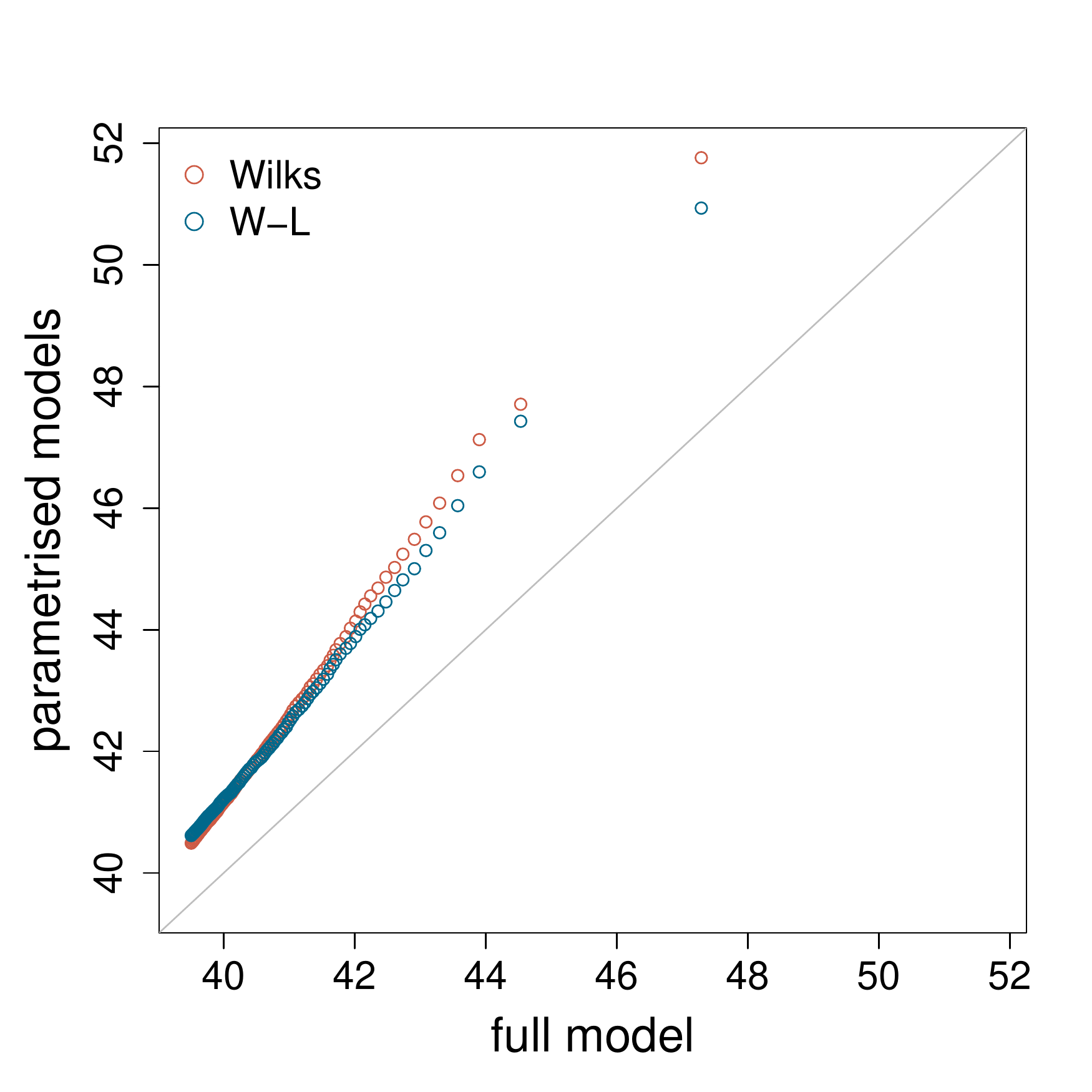}
 	\caption{\label{fig:compare_qq_extremes}The empirical extreme quantiles of the observable $A_x$ (left), $A_E$ (middle), and $A_p$ (right) from the parametrized models against that from the full model. }
 	\end{center}
\end{figure*}
In the left panel, except for several very high-level and low-level quantiles, the other moderate quantiles from the parametrized models agree well with those from the full model, indicating that the parametrized model can reproduce well the bulk statistics of the observable $A_x$ of the full model. In the middle panel, the points are slightly farther away from the diagonal, indicating that the quantiles of the observable $A_E$ from the parametrized models are somewhat in disagreement with those from the full model. In the right panel, we observe that the quantiles of the observable $A_p$ from the Wilks parametrized model agree well with those from the full model, whereas the low quantiles from the W-L parametrized model show an obvious disagreement. In summary, the parametrized models give better quantiles of the local observable than the global observables. \\
\indent The data between the last two quantiles are the largest one thousandth of the sample data. These data are the extreme values we want to investigate. Fig.~\ref{fig:compare_qq_extremes} shows a zooming onto the extreme quantiles that are higher than the $99.9$th percentile. We consider $101$ extreme quantiles, including the $99.9$th percentile, the largest element of the sample, and $99$ equidistant quantiles between them. As shown in the left panel, the extreme quantiles of the observable $A_x$ from the W-L parametrized model are slightly larger than those from the full model, and, in comparison, the extreme quantiles from the Wilks parametrized model have a larger disagreement. In the middle panel, the extreme quantiles of the observable $A_E$ from the Wilks parametrized model totally miss that from the full model and the approximation of the extreme quantiles from the W-L parametrized model to those from the full model breaks down after the $99.95$th percentile. In the right panel, we observe that the extreme quantiles of the observable $A_p$ from both of the parametrized models are far away from those from the full model.

\section{Return Periods of Extremes }\label{sec:return_time}

\begin{figure*}
  \begin{center}
  \includegraphics[width=0.29\textwidth]{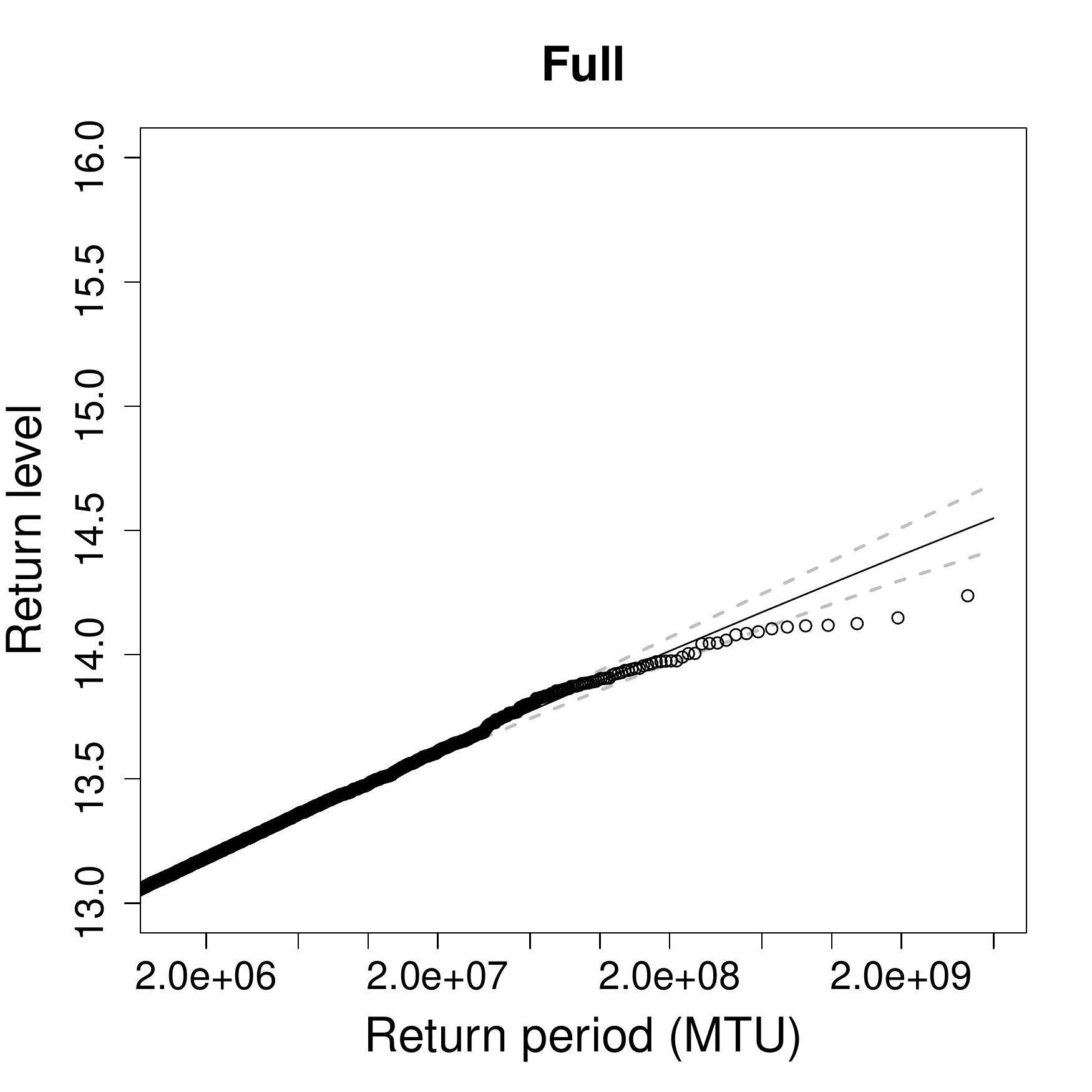}
  \includegraphics[width=0.29\textwidth]{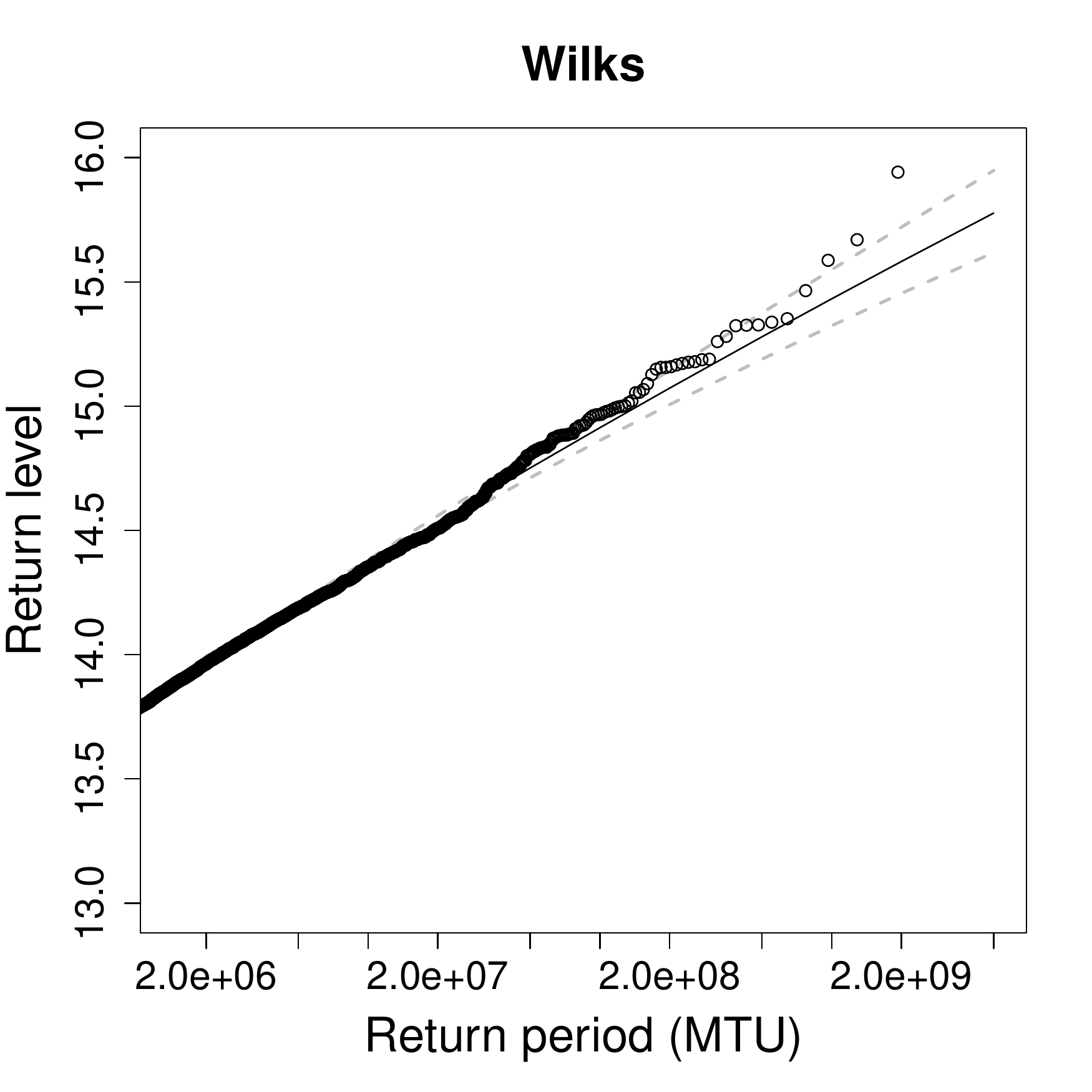}
  \includegraphics[width=0.29\textwidth]{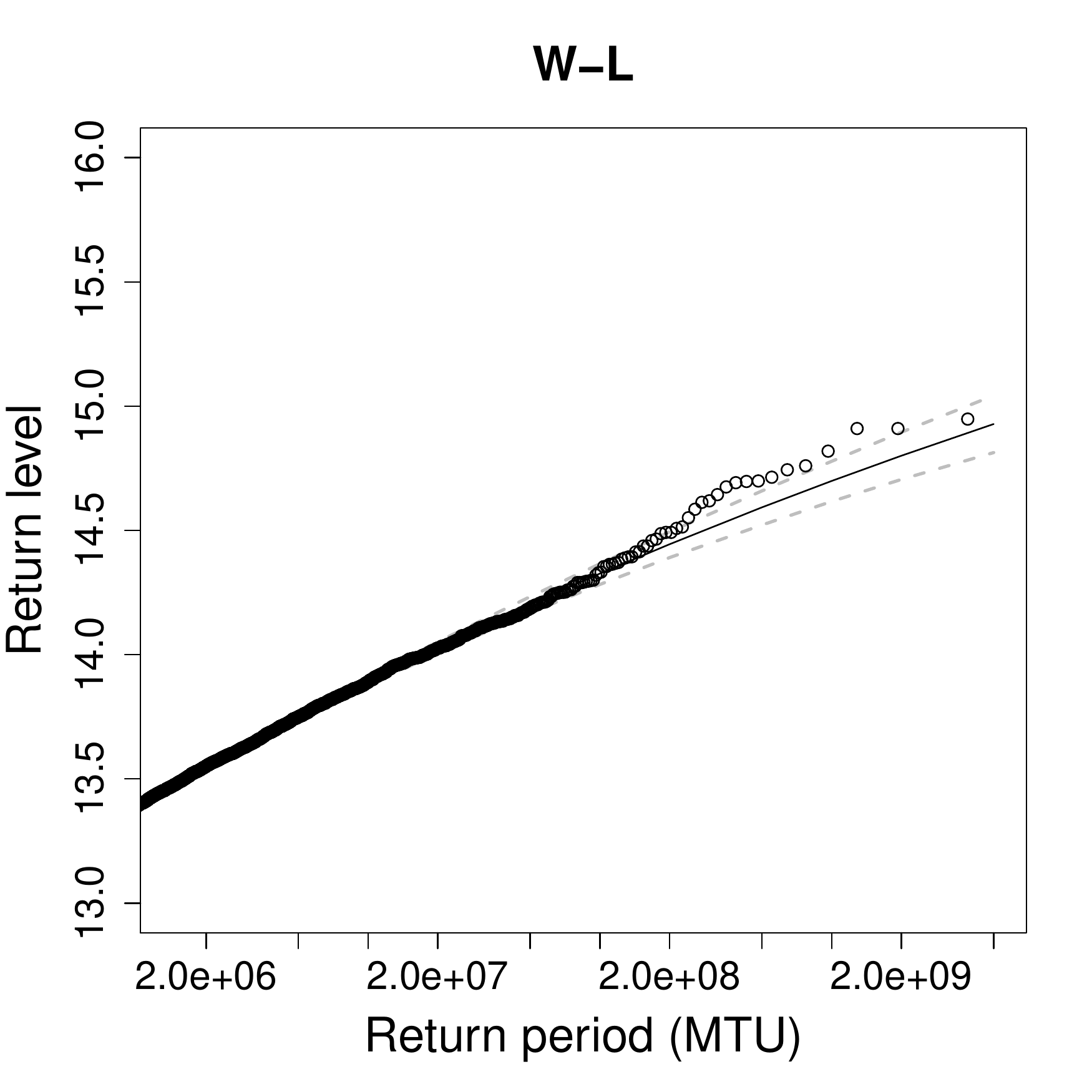}	
  
  \includegraphics[width=0.29\textwidth]{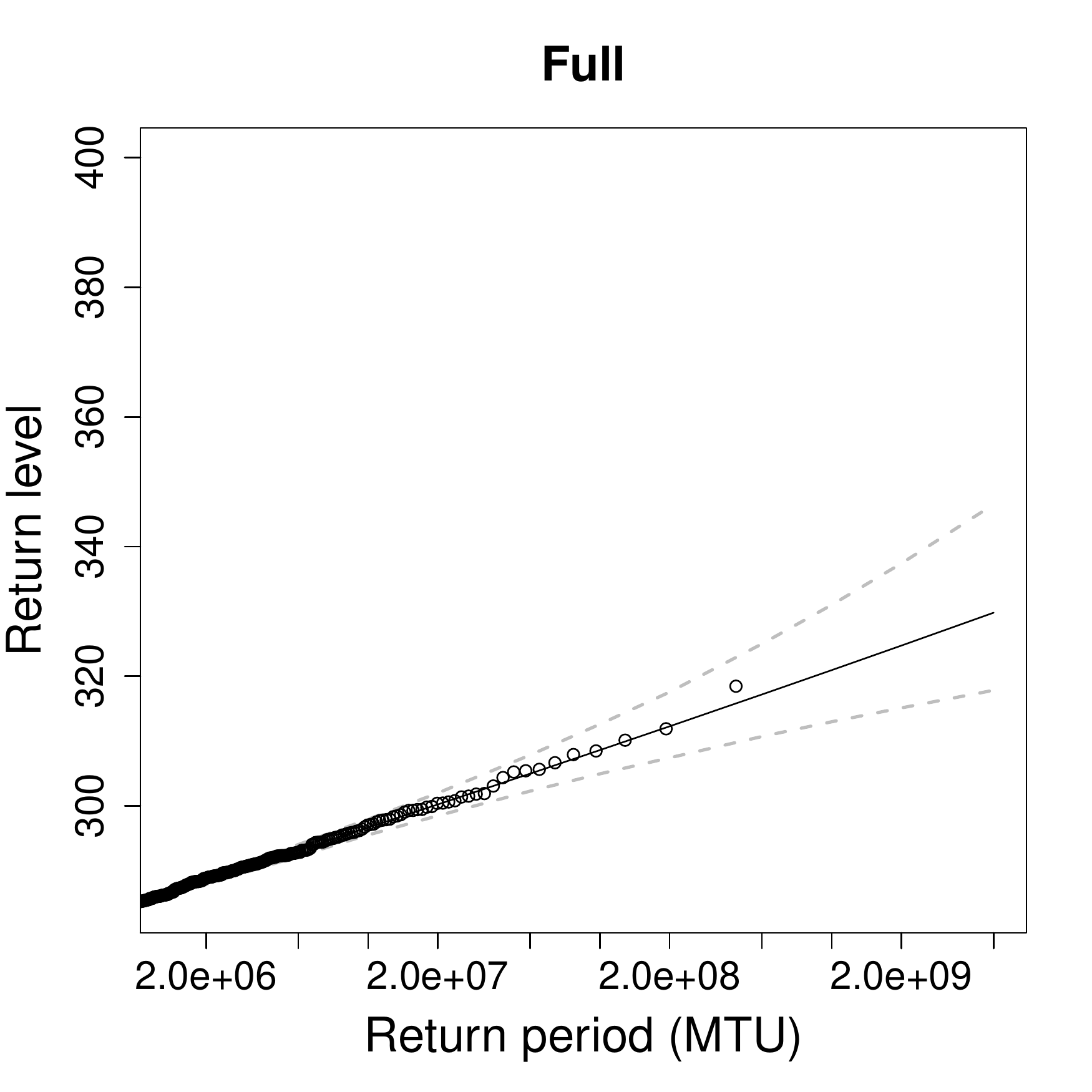}	
  \includegraphics[width=0.29\textwidth]{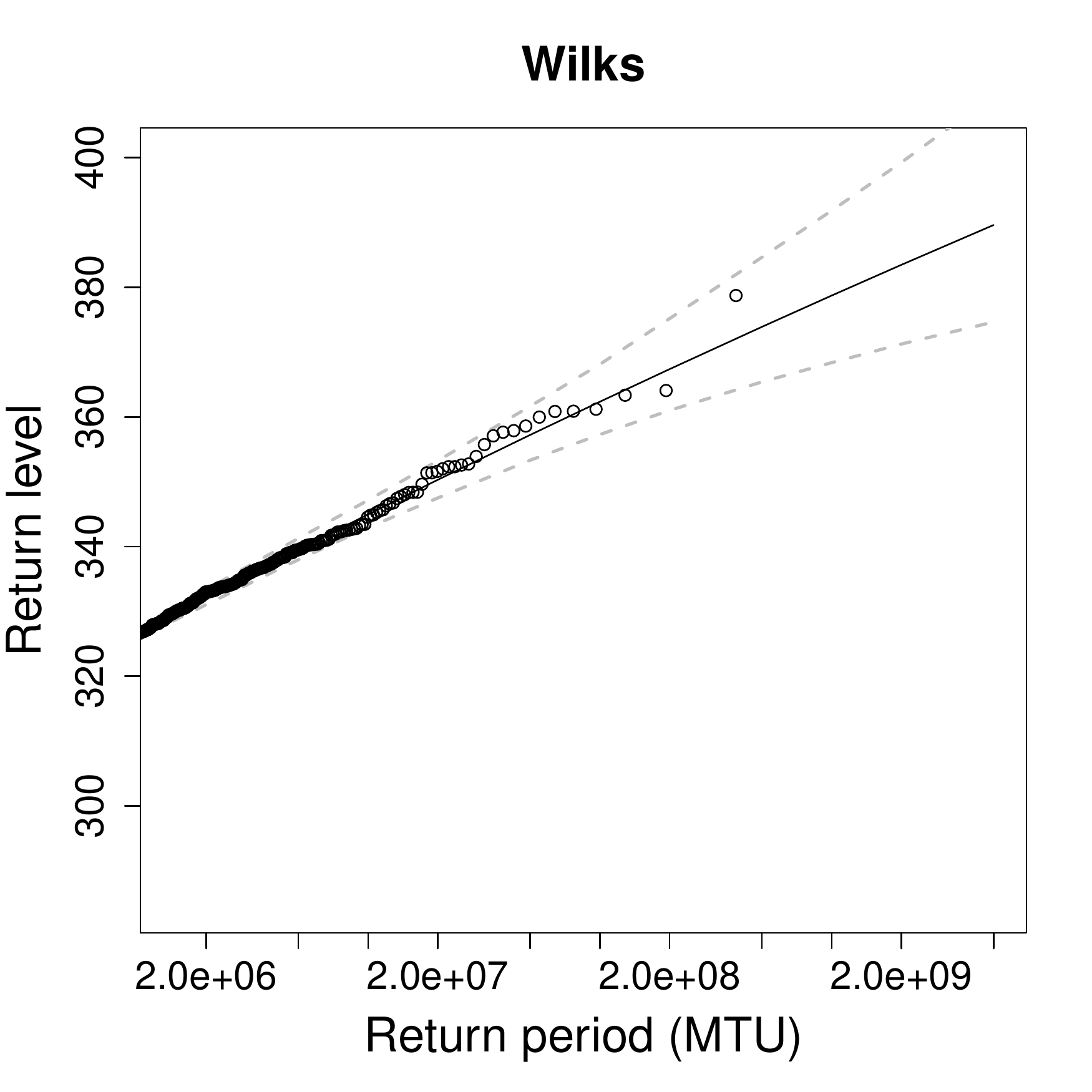}
  \includegraphics[width=0.29\textwidth]{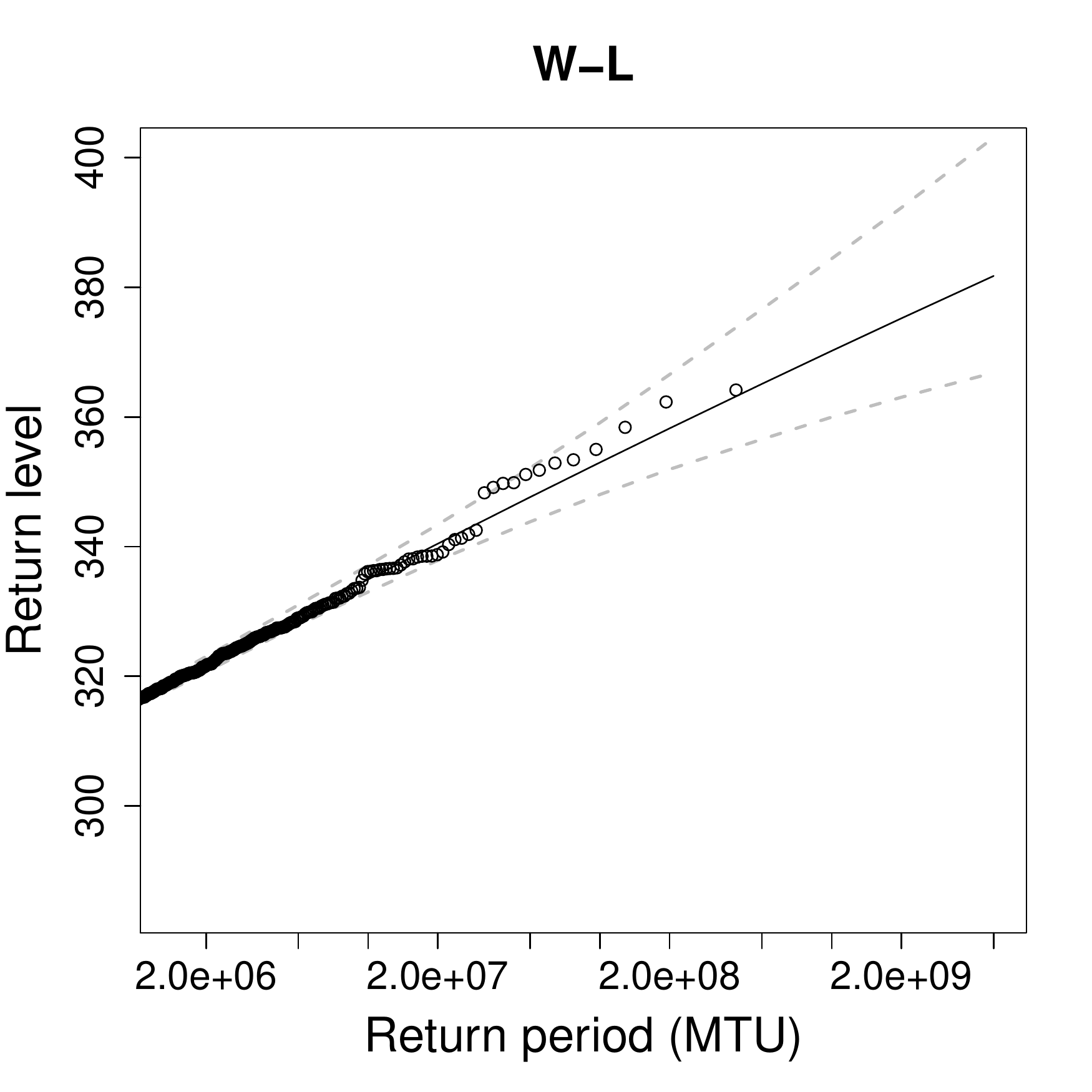}
  
  \includegraphics[width=0.29\textwidth]{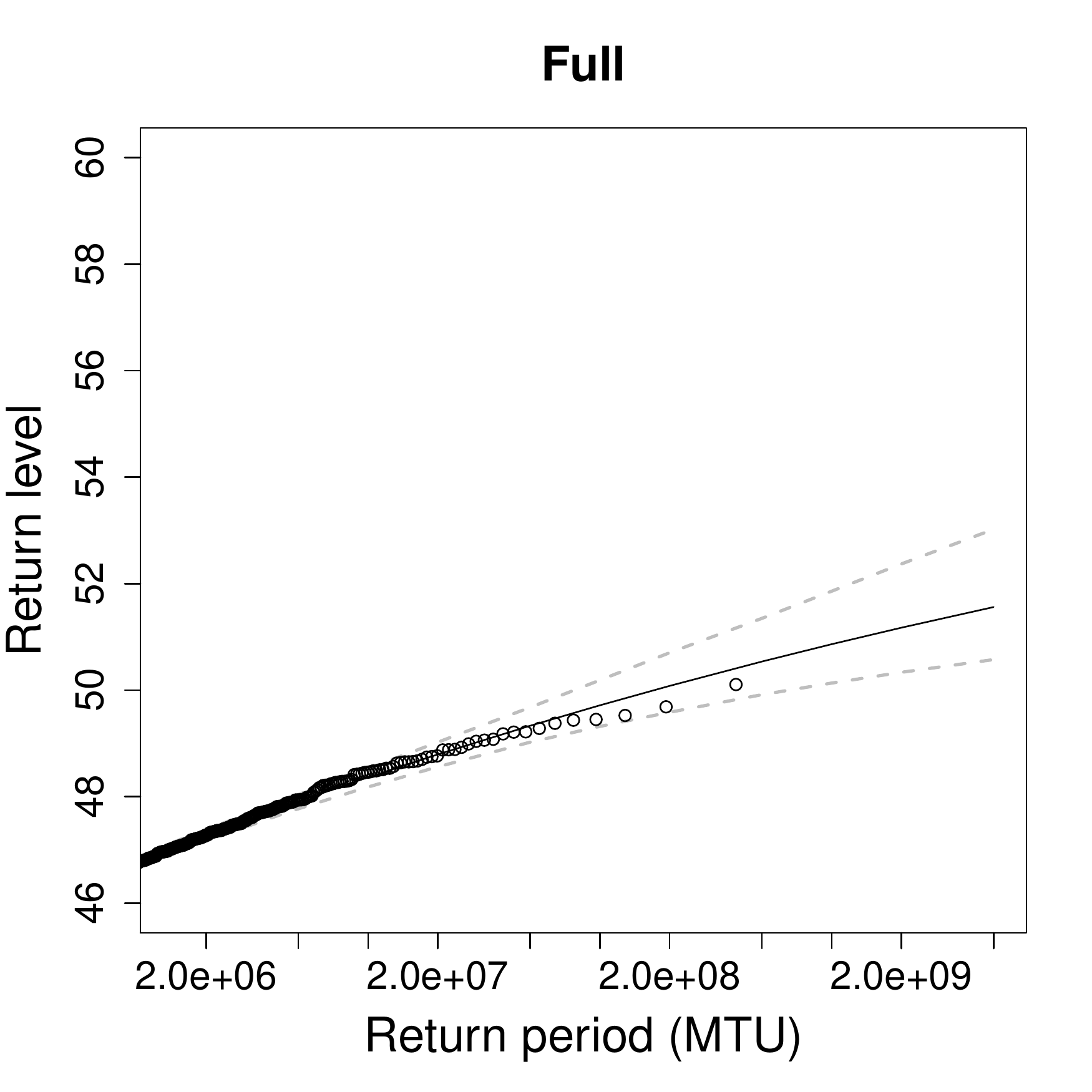}	
  \includegraphics[width=0.29\textwidth]{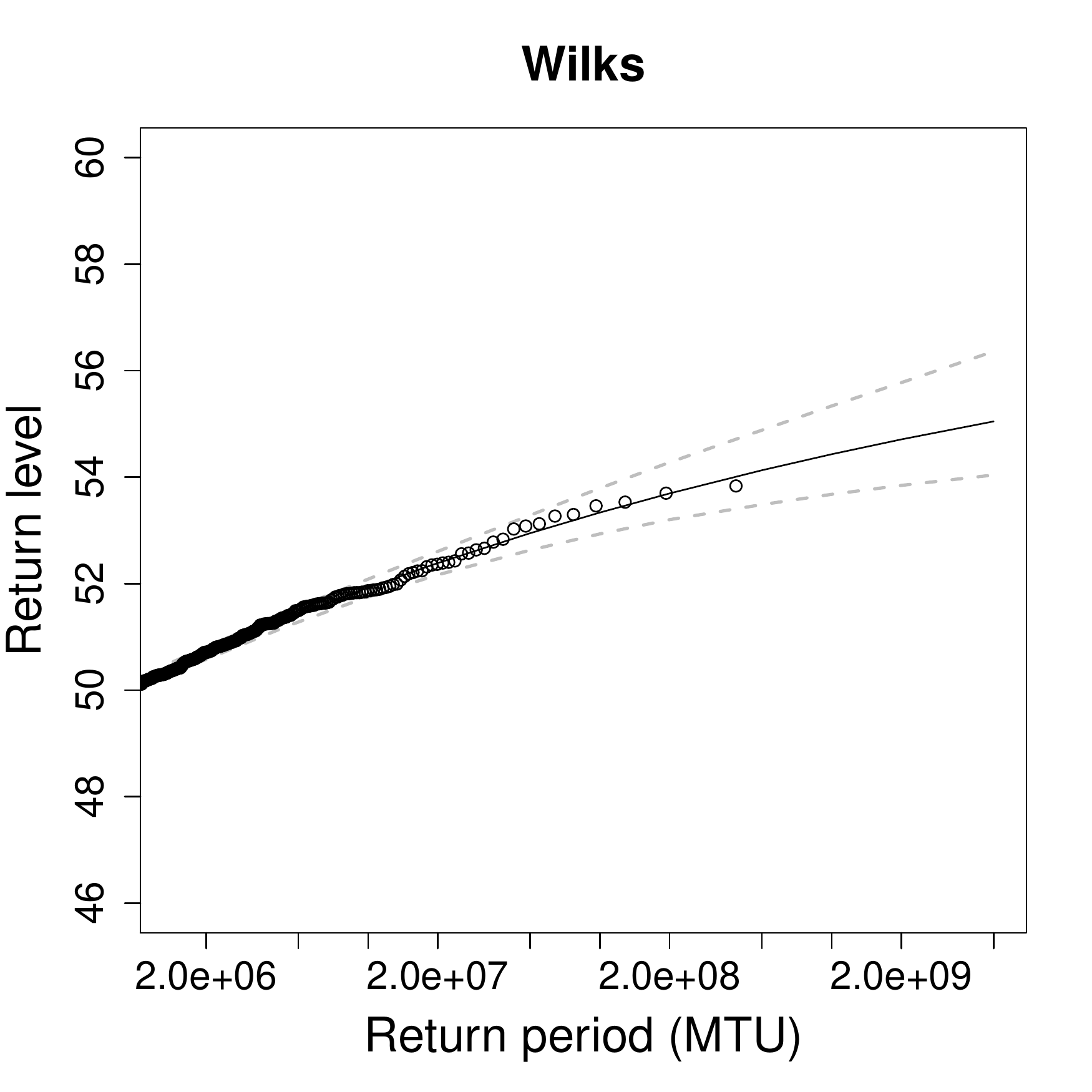}
  \includegraphics[width=0.29\textwidth]{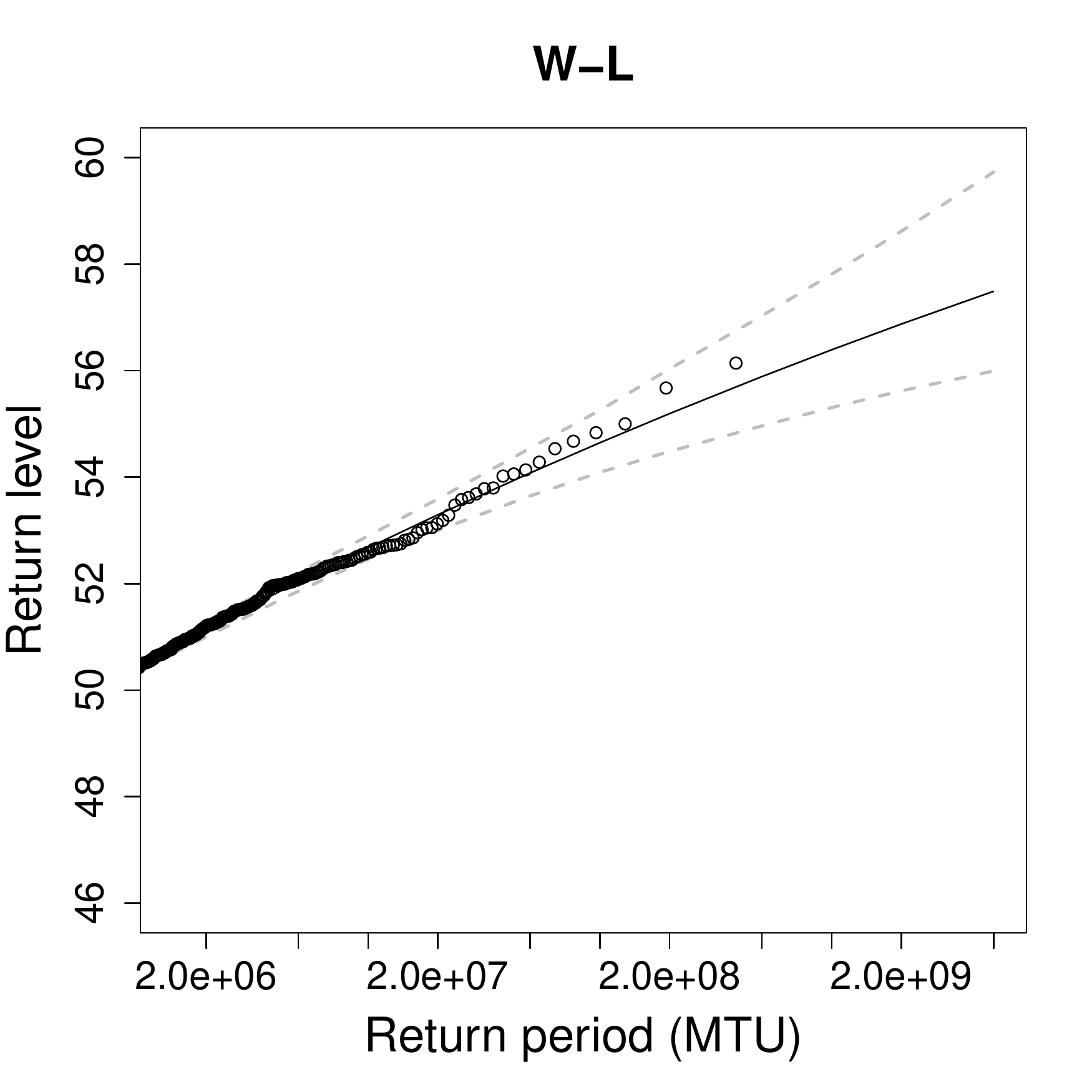}
  \caption{Return levels against return periods for the block maxima of the observable $A_x$ (upper panels), $A_E$ (middle panels), and $A_p$ (lower panels) from the full and parametrized models. The black dots signify the empirical return levels, the solid black lines show the predicted return levels computed based on the fitted GEV distributions, and the dashed grey lines present the $95\%$ confidence intervals of the estimates.}\label{fig:rl}
  \end{center}
\end{figure*}
\indent Fig.~\ref{fig:rl} presents the mean return periods against the return levels of the extremes for the three observables of the three models. The extremes are selected by the BM approach using a block size of $B_0 \times 2^{10}$. It can be seen that the GEV mean return time estimates using the chosen block size fit the empirical data quite well, except for very high return levels of the observable $A_x$, at which the empirical estimates are out of the confidence interval of the GEV estimates.\\
\indent Comparing the parametrized models to the full model, we find that for all the three observables the parametrized models always produce higher-level extremes in the same return periods. In other terms the return period of the extremes with the same magnitude is shorter in the parametrized models than in the full model. Moreover, in comparison with the Wilks parametrized model, the W-L parametrized model gives better estimates of return levels, which are closer to the estimates given by the full model. One exception is observed in momentum observable; the Wilks parametrized model gives slightly better estimates of return levels than the W-L parametrized model. 

\section{Sensitivity to coupling strength}\label{sec:coupling_strength}
\begin{figure*}
 	\begin{center}
	\includegraphics[width=0.29\textwidth]{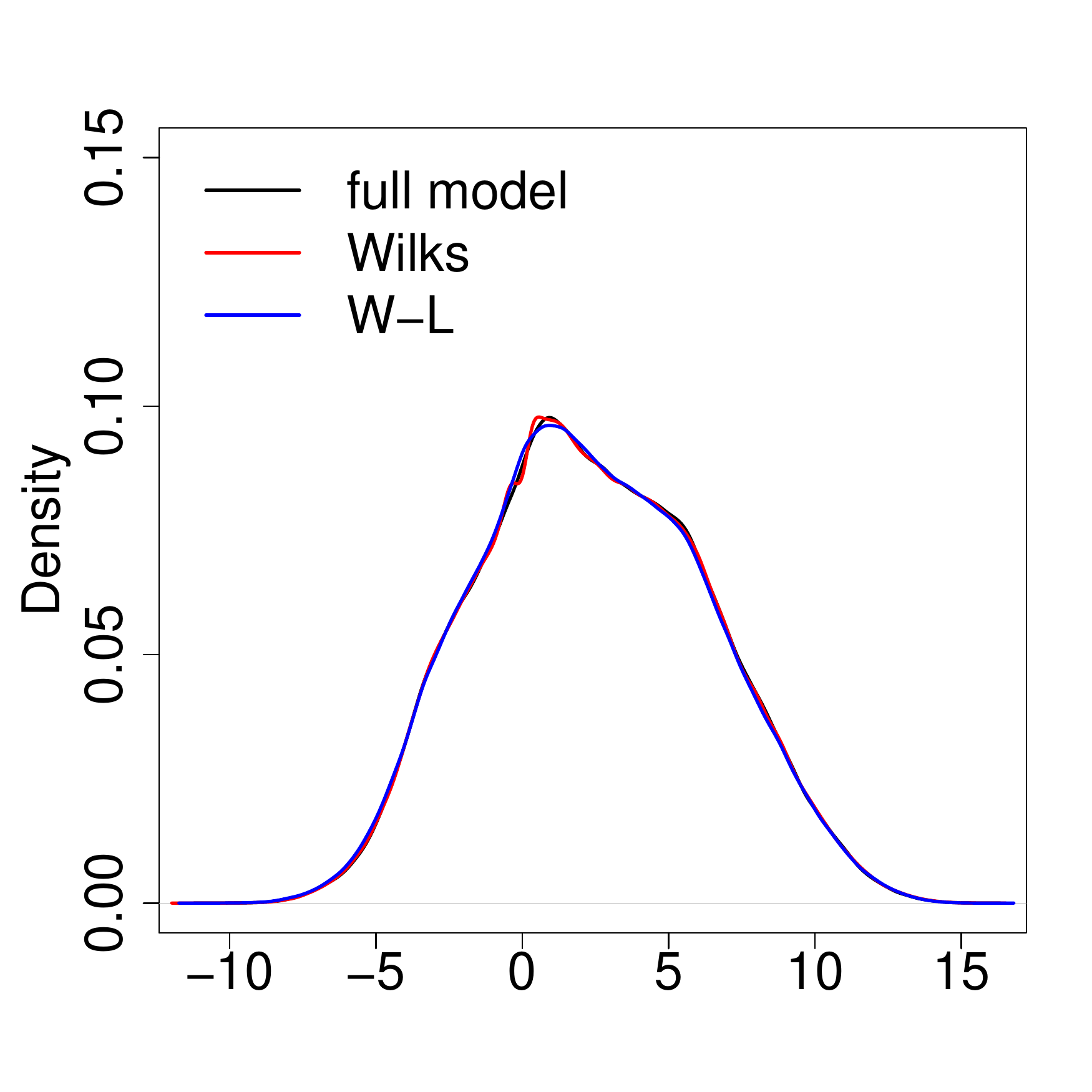}
	\includegraphics[width=0.29\textwidth]{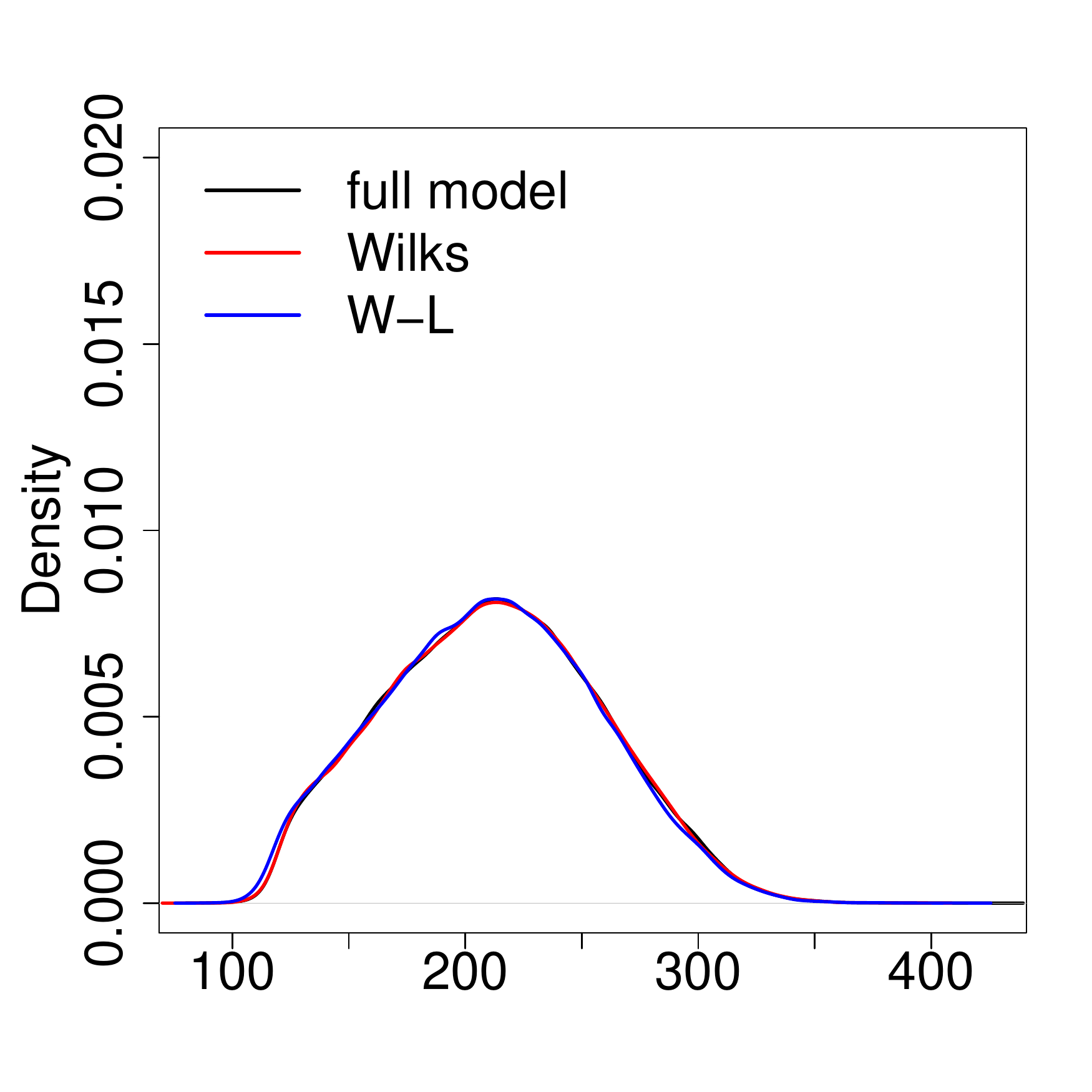}
	\includegraphics[width=0.29\textwidth]{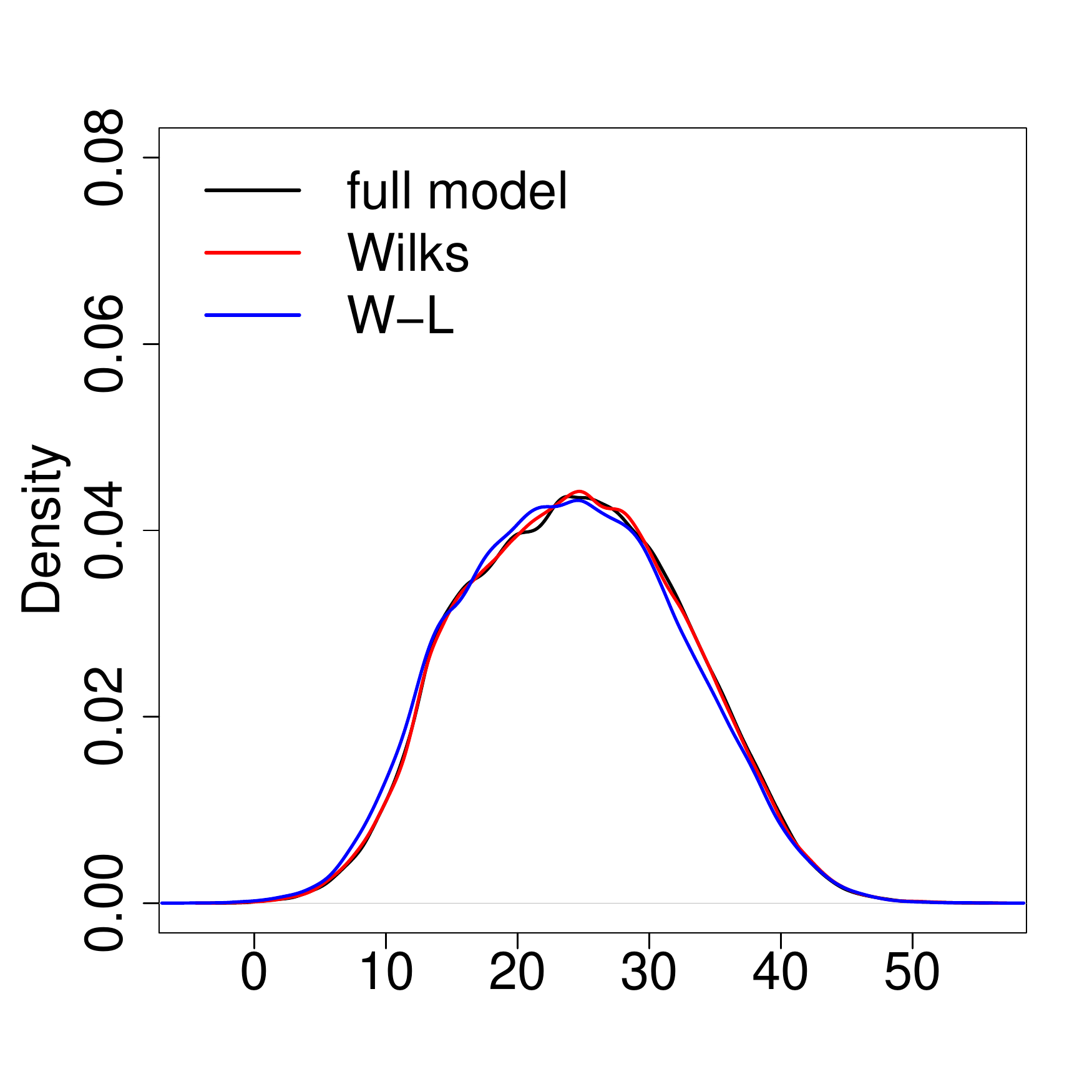}
 	\caption{\label{fig:pdf_h05}Probability density functions of the observable $A_x$ (left), $A_E$ (middle), and $A_p$ (right) from the full and parametrized models in a weak coupling case ($h=0.5$).}
 	\end{center}
\end{figure*}

\begin{figure}
 	\begin{center}
	\includegraphics[width=0.29\textwidth]{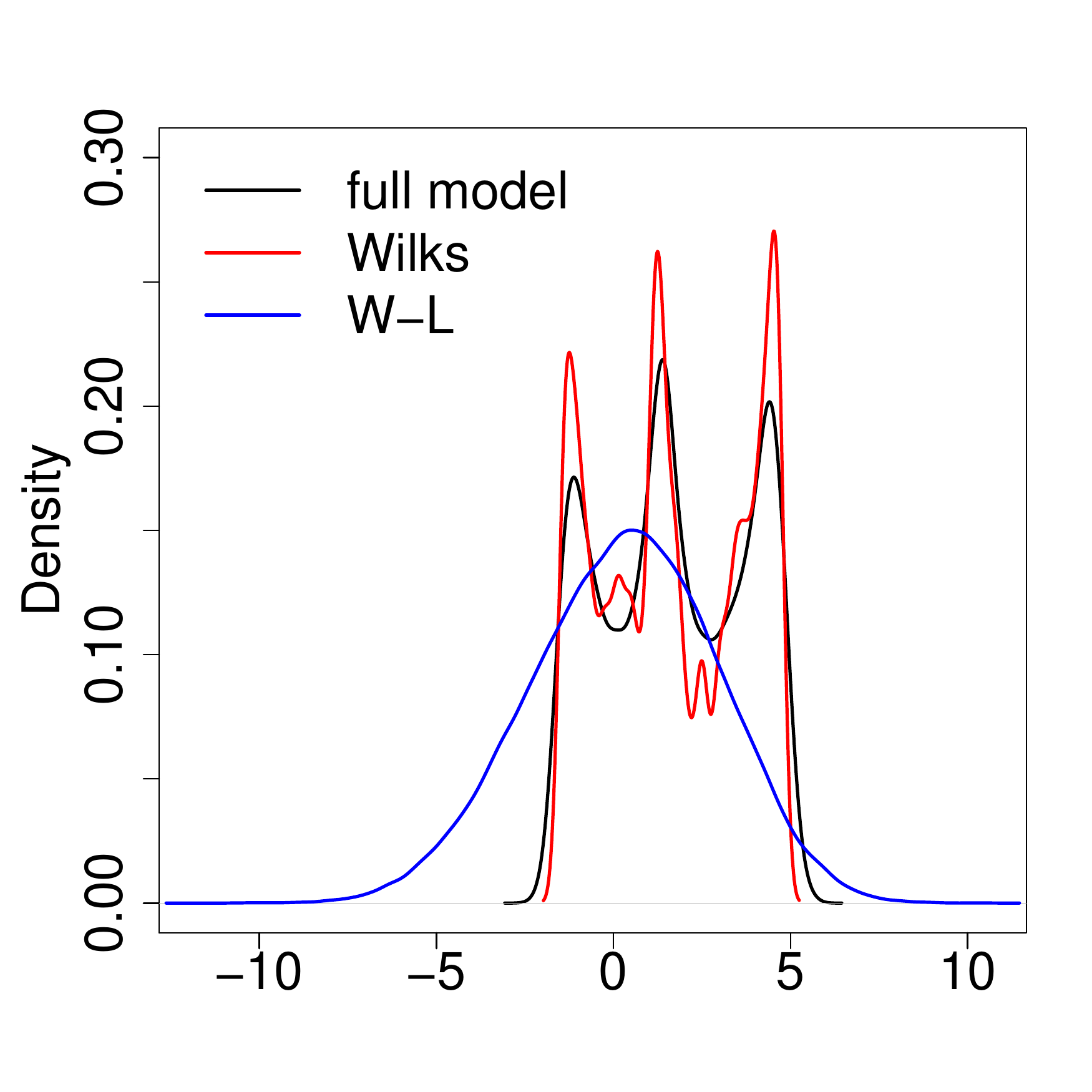}

 	\caption{\label{fig:pdf_h2}Probability density functions of the observable $A_x$ from the full and parametrized models in a strong coupling case ($h=2$).}
 	\end{center}
\end{figure}

\begin{figure*}
 	\begin{center}
	\includegraphics[width=0.29\textwidth]{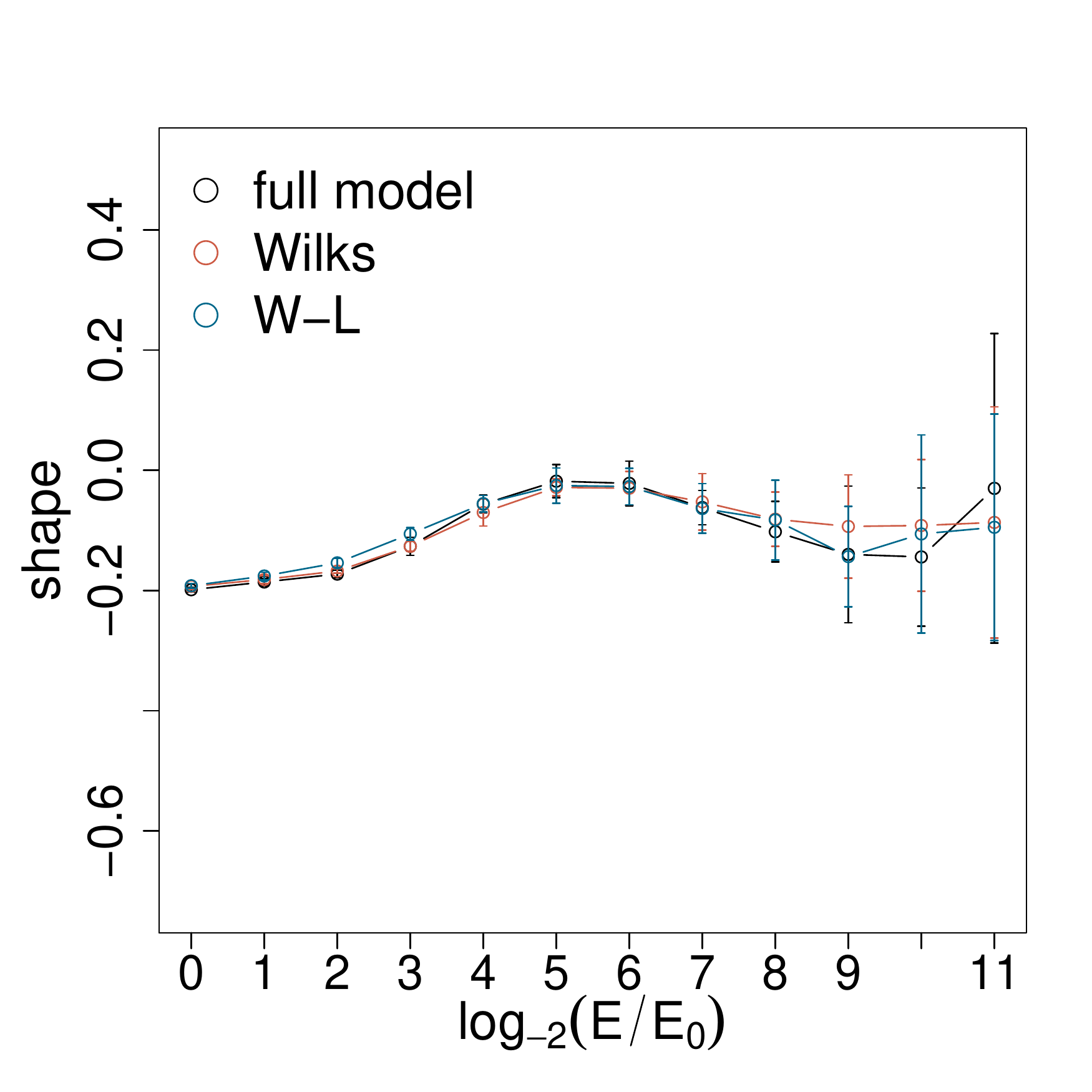}
	\includegraphics[width=0.29\textwidth]{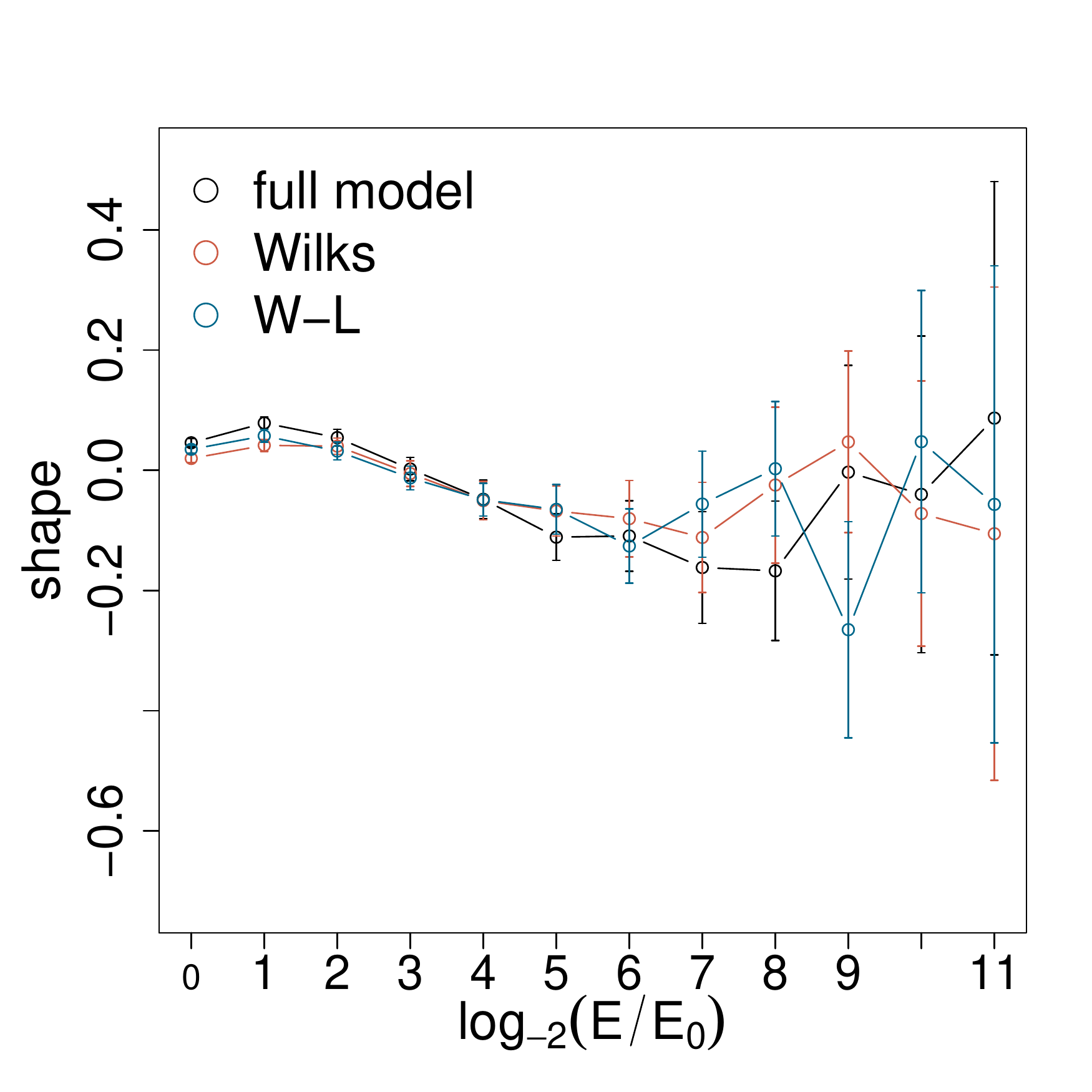}
	\includegraphics[width=0.29\textwidth]{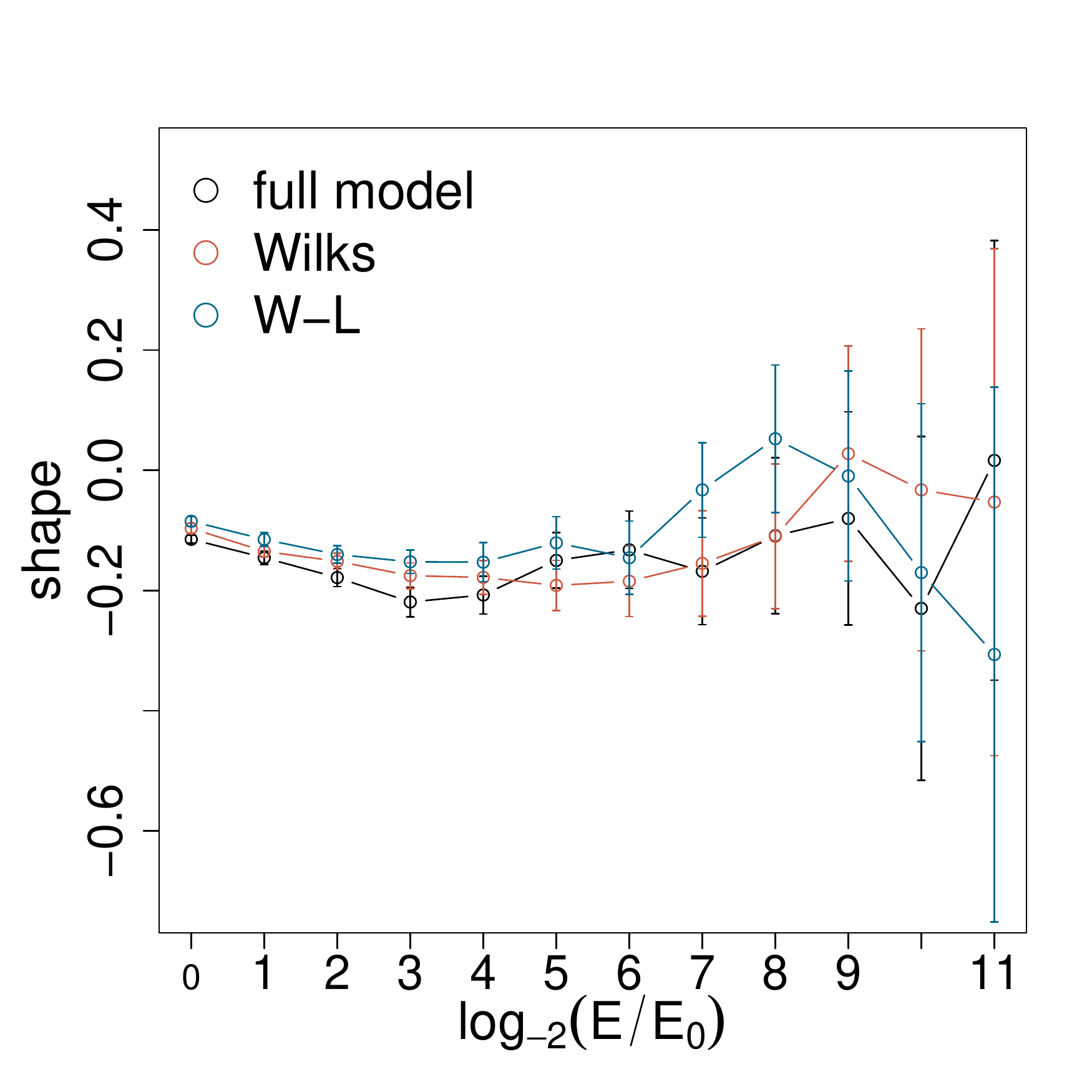}
 	\caption{\label{fig:shape_other_h}Comparison of the estimates of the GP shape parameter for the observable $A_x$ (left), $A_E$ (middle), and $A_p$ (right) over a range of exceedance ratios in a weak coupling case ($h=0.5$).}
    \end{center}
\end{figure*}

\begin{figure}
 	\begin{center}
	\includegraphics[width=3.3in]{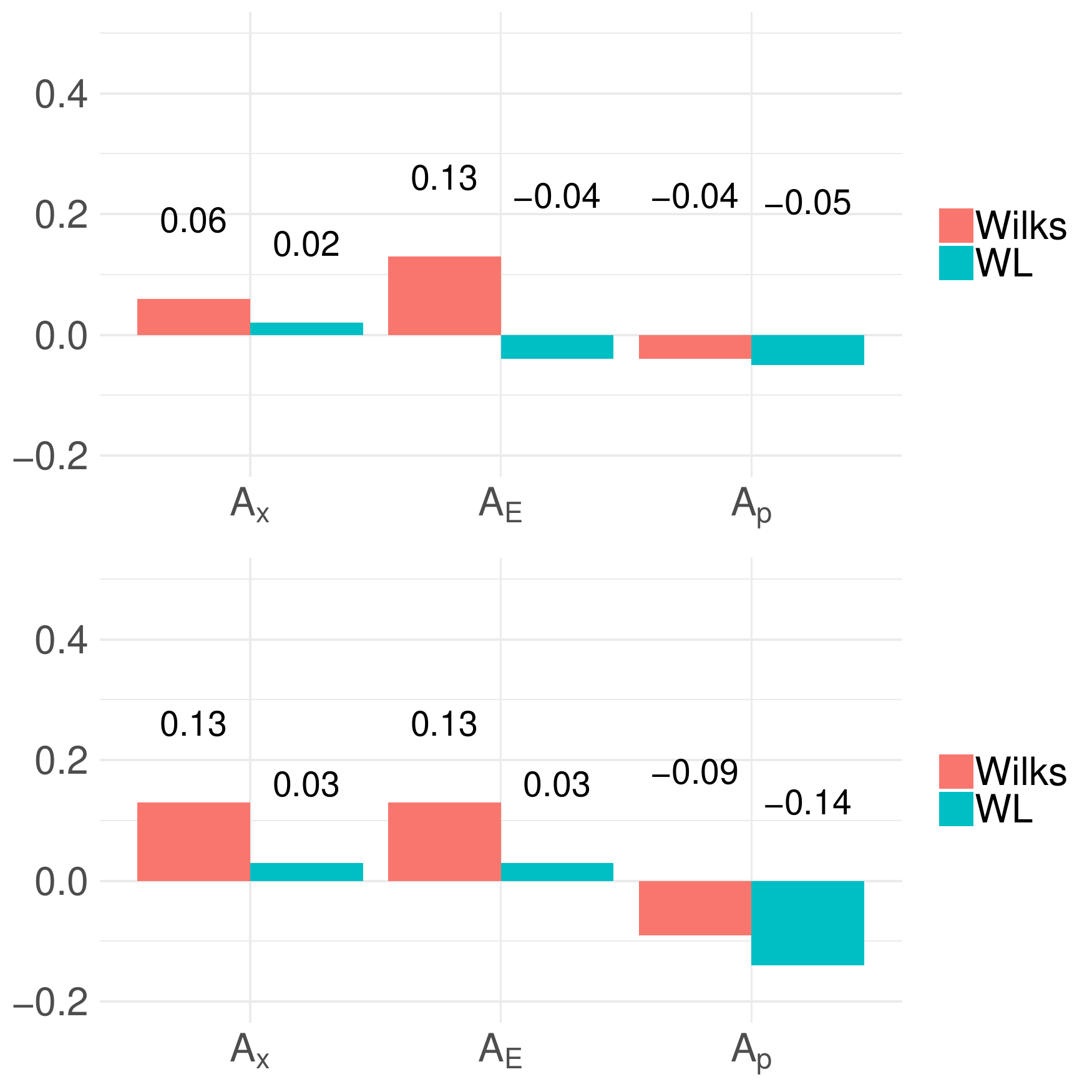}
	\caption{\label{fig:compare_num_h05}The fractional difference of the number of threshold exceedances from the parametrized models compared to that from the full model. In the upper (lower) panel, the threshold is the $99.9$th ($99.99$th) percentile of the corresponding observable of the full model. Coupling strength $h=0.5$.}
    \end{center}	
\end{figure}

In previous numerical experiments, we applied the Wilks and W-L parametrizations to the two-level L96 with the standard value of the coupling strength $h=1.0$. We now test and compare the performances of the two parametrizations as we consider both a factor of $2$ weaker and a stronger coupling strengths, i.e., $h=0.5$ and 2, respectably.\\
\indent We need to recompute the parameter values for the Wilks parametrization through fitting the residual time series as we vary the coupling strength, but this is not the case for the W-L parametrization, the three terms of which can be simply rescaled as follows (see Eq.~($13$) of Vissio and Lucarini (2018)): $\tilde{D} = (\tilde{h}/h)D$, $\tilde{S} = (\tilde{h}/h)S$ and $\tilde{M} = (\tilde{h}/h)^2M$, where the symbols with tilde represent the case when we have a new value of $h$, and the symbols without tilde represent the case when we have the standard value of the coupling strength. The parameter values of the Wilks parametrization for the weaker and stronger coupling strength are presented in Table~\ref{tab:parameter}. Note that the Wilks parametrization lacks stability in the case of $h=2.0$ if we include the stochastic term in the parametrization, therefore, we use a deterministic parametrization, which does not contain the first-order autoregressive model.\\
\indent Fig.~\ref{fig:pdf_h05} shows the PDF of the three observables from the full and parametrized models in the case of $h=0.5$. We observe a better agreement for all the three observables between the full and parametrized models in comparison with the case of $h=1.0$. Because of the way that the W-L parametrization is constructed, the error in the W-L parametrized model with respect to the full model is a function of the coupling strength, which is $\mathcal{O}(\epsilon^3$), as mentioned before, and this applies to all observables in the system \cite{luc14,vis18}. Therefore, it is unsurprising that the W-L parametrized model demonstrates a better performance in the case of a weaker coupling strength. The Wilks parametrization is constructed based on the correlation between the $X_k$ variables and the coupling term, which is the sum of the $Y_{j,k}$ variables. As the coupling strength becomes weaker, this correlation also becomes smaller, however, in the meanwhile, the difference between the uncoupled dynamics and the coupled dynamics reduces, therefore, although the quality of the parametrization scheme is not automatically constrained to improve, we still observe smaller difference between the parametrized model and the full model.
\indent Fig.~\ref{fig:pdf_h2} shows the PDF of the observable $A_x$ from the full and parametrized models in the case of $h=2.0$. In this case, the statistics of the observable $A_x$ in the full model are greatly changed; the PDF changes from unimodal to multimodal. The Wilks parametrization is able to capture the multimodality of the PDF, since the correlation between the $X_k$ and the $Y_{j,k}$ variables still exist. However, we believe that this multimodality prevents us to see a cubic scaling for the difference between the full and W-L parametrized models. The qualitative difference shown in the PDFs suggests that we are outside of a radius of convergence with respect to the coupling strength $h=2.0$, which means
that the scale adaptivity is only valid for a range of the coupling strength. This is unsurprising, because the W-L parametrization can only be applied to weakly coupled systems. The PDFs for the two global observables, $A_E$ and $A_p$, have not been shown, because both parametrizations perform very poorly. \\ 
\indent We now look at the extreme value statistics given by the two parametrized models in the case of $h=0.5$. Fig.~\ref{fig:shape_other_h} compares the estimates of the GP shape parameter for the three observables between the full and parametrized models. In the case of a weaker coupling, the parametrized models display extreme value statistics that matches more closely that of the full model in comparison with the case of the original coupling strength. Furthermore, we compare the number of threshold exceedances for the parametrized models to those for the full model. The thresholds are chosen as the $99.9$th and $99.99$th percentile of the data produced by the full model. Fig.~\ref{fig:compare_num_h05} shows the fractional difference calculated by Eq.~(\ref{eqn:fractional_difference}) in a weak coupling case. The W-L parametrized model gives a closer number of the extremes of the observable $A_x$ and $A_E$ to the full model than the Wilks parametrized model, and the Wilks parametrized model gives a closer number of the extremes of the observable $A_p$. These are features unchanged with respect to the standard parameter setting. Otherwise, both the parametrized models give less extremes of the observable $A_p$ than the full model.  
\begin{table}
\begin{center}
\caption{\label{tab:parameter}Parameter values of the Wilks parametrization for different values of the coupling strength $h$. }
\begin{tabular}{cccccccc}
\hline
$h$ & $a_0$ & $a_1$ & $a_2$ & $a_3$ & $a_4$ & $\phi$ & $\sigma_e$ \\
\hline
$0.5$ & $-0.95$ & $-0.0349$ & $-0.0008$ & $-0.0002$ & $0.00002$ & $0.967$ & $0.3548$ \\
$2.0$ & $-3.54$ & $-0.3292$ & $-0.0548$ & $-0.0593$ & $0.0132$  & $0.0$   & $0.0$ \\
\hline
\end{tabular}
\end{center}
\end{table}

\section{Summary and Conclusions}\label{sec:summary}

This paper addresses the problem of how well stochastic parametrizations perform, in a simple yet informative case; we conduct numerical experiments in a conceptual atmospheric model, the two-level L96 \cite{L96}. Instead of investigating the bulk of the statistics, as what has been usually done, here we focus on examining extreme value statistics. The problem has been explored using EVT, which provides us with a mature statistical framework to define and analyze extremes. Under this framework, probability distributions obey a limit law and so some universality, as opposed to the bulk statistical properties. We considered two kinds of parametrization schemes for the two-level L96: the widely applied Wilks parametrization scheme \cite{wilks05} and a recently proposed scale-adaptive parameterzation scheme \cite{vis18}. The latter one is constructed by the methodology proposed by Wouters and Lucarini (2012, 2013, 2016) and hence denoted as W-L parametrization. The parametrized models constructed by these two schemes can reasonably reproduce the bulk, but not the extreme value statistics of the large-scale variables of the full model. The differences are mainly found in the following aspects:
\begin{itemize}
    \item The extreme values from the parametrized models have different shape parameters, scale parameters and location parameters than those from the full model; 
    \item The extreme values selected by the BM approach have a larger variance in the parametrized models than in the full model;
    \item The parametrized models give different numbers of extreme events than the full model when we apply the same threshold to them, and in most cases, the parametrized models produce more extreme events.
\end{itemize}
From a practical point of view, these differences will affect the mean return times of extreme events; the extreme events from the parametrized models have typically shorter return times than those from the full model. Our results suggest that stochastic parametrizations should be accurately tested against their performance on extremes, because a good performance on typical events cannot ensure a good performance on untypical events.\\
\indent The parametrization schemes we use are constructed in two fundamentally different ways, and so they have different advantages and disadvantages. As discussed in Sec.~\ref{sec:model}, the Wilks parametrization is an empirical one, which is constructed based on the observed fact that in the two-level L96 the small-scale tendency strongly and nonlinearly depends on the value of the large-scale variables \cite{wilks05}. The W-L parametrization is constructed through a statistical mechanical approach, where the small-to-large-scales coupling is treated as perturbations of strength $\epsilon$ added to the uncoupled large-scale dynamics. It is based on a series expansion of the invariant measure of the uncoupled system with respect to the small coupling parameter $\epsilon$. Including first- and second order terms in the W-L parametrization implies that the difference between the expectation value of all observables of the large-scale variables in the full model and in the parametrized model scales as $\mathcal{O}(\epsilon^3$). In fact, this scaling applies to tail probabilities of our interest too, because they are observables~\citep{luc14,lucarini2016extremes}. \\ 
\indent As shown by Vissio and Lucarini (2018), the Wilks parametrization is more accurate than the W-L parametrization in representing the bulk of the statistics of the full model. This is not surprising because the Wilks parametrization is constructed through an empirical fitting procedure that aims at providing an optimal representation of the typical events of the system, Note that this does not ensure a good representation of rare events. Indeed, our new results show that there are many cases where the W-L parametrized model reproduces the extreme value statistics better than the Wilks parametrized model. Figs. 9, 10, 11 are especially clear in this regard. However, we must emphasize that the performance of any parametrization scheme should depend on the observable of interest via a prefactor for the scaling $\mathcal{O}(\epsilon^3$) of errors. We are yet to develop an understanding why the W-L scheme outperforms Wilks with respect to the local observable $A_x$ of the site variables and the global energy $A_E$ observable but not the global momentum $A_p$ observable.\\
\indent An important advantage of W-L that sets it apart from Wilks is that if an observable of interest is not reconstructed accurately enough for a purpose, then one can opt -- within the perturbative framework -- to work out higher-order correction terms. Within a radius of convergence of the series expansion with respect to $\epsilon$, the performance can be improved arbitrarily in principle.
The other major advantage of the W-L parametrization is the scale adaptivity; when the coupling strength $\epsilon$ is changed, we can adapt the obtained parametrization schemes by a suitable rescaling procedure. In contrast, we need to recompute the parameter values of the Wilks parametrization every time when the coupling strength is changed. But if the coupling strength is too strong, the W-L parametrization fails; it is outside of the the radius of convergence, while the Wilks parametrization might still work (although even that was found to be not the case for $h=2$, only the deterministic parametrization). This shows advantages and disadvantages of the two strategies.\\
\indent In this paper, we also examined the convergence (or approximation, with a finite residual error possibly) of finite size estimates of the GEV and GP shape parameters to the theoretical value of the deterministic model, which is computed by the partial dimensions of the attractor of the system. In the two-level L96, the estimates of the shape parameter non-monotonically approach the theoretical value as the block size or threshold increases. The estimates show erratic behaviours rather than an asymptotic behaviour in the observed ranges of the block maxima and thresholds. The approximation of the estimates to the theoretical value takes place very slowly, and they have different rates for different observables. Different rates of convergence of the estimates of the shape parameter for different observables had also been reported in G{\'a}lfi, B{\'o}dai, and Lucarini (2017).

\bigskip

\noindent Statement on computer code availability: All code used for this study and written by us is available on request.

\section*{Acknowledgments}
GH is funded by China Scholarship Council (CSC). TB and VL acknowledge financial support from the Horizon 2020 projects CRESCENDO (under grant No. 641816) and Blue-Action project (under grant No. 727852). We thank Gabriele Vissio for providing the code for the W-L parametrized model. We thank two anonymous reviewers for their effort with reviewing our manuscript and for providing feedback.

\nocite{*}
\bibliographystyle{plainnat}
\bibliography{ref}

\end{document}